\documentclass[usenatbib,twocolumn]{mn2e}
\usepackage{graphicx,amsmath,amssymb,enumitem,verbatim,sidecap}
\begin{document}

\title[Two galaxy SFR history approximations]{Approximations to galaxy
  star formation rate histories: properties and uses of two
examples}

\author[Two galaxy SFR history approximations]{J.D.Cohn${}^{1,2}$\thanks{E-mail: jcohn@berkeley.edu}\\
${}^1$Space Sciences Laboratory 
  University of California, Berkeley, CA 94720, USA\\
${}^2$Theoretical Astrophysics Center,
  University of California, Berkeley, CA 94720, USA}
\maketitle
\begin{abstract}
Galaxies evolve via a complex interaction of
numerous different physical processes, scales and components.  In spite of this,
overall trends often appear.
Simplified models for galaxy histories can be used to search for and
capture such emergent trends,
and thus to interpret and compare
results of galaxy formation models to each other and to nature.
Here, two approximations are applied to galaxy integrated star
formation 
rate histories, 
drawn from a semi-analytic model grafted onto a dark
matter simulation.  Both a lognormal functional form and
principal component analysis (PCA) approximate the
integrated star formation rate histories fairly well.
Machine learning, based upon
simplified galaxy halo histories, is somewhat
successful at recovering both fits.  The fits
to the histories give fixed time star formation
rates which have notable scatter from their true final time rates,
especially for quiescent and ``green valley'' galaxies, and
more so for the PCA fit.
For classifying galaxies into subfamilies sharing similar integrated histories,
both approximations are better than using final stellar mass or specific
star formation rate.
Several subsamples from the simulation 
illustrate how these simple parameterizations provide points of
contact for comparisons between different galaxy formation samples, or
more generally, models.
As a side result, the halo masses of simulated galaxies with early
peak star formation rate (according to the lognormal fit) are bimodal. 
The galaxies
with a lower halo mass at peak star formation rate appear to stall in their
halo growth, even though they are central in their host halos. 
\end{abstract}

\section{Introduction}
Many galaxy properties are now observed and measured in samples
extending over huge volumes of sky, reaching back to earlier and earlier
times.  
Several
trends have been discovered to emerge from all the interrelated
complexities of galaxy formation.  These include the fact that small isolated galaxies tend to be
star forming, central\footnote{A satellite galaxy, as compared
to a central galaxy, is a galaxy which has fallen into the dark matter
halo of a larger galaxy.} galaxies in large dark matter halos tend to be
quiescent, and galaxies of a certain stellar mass often inhabit host
dark matter halos of a certain mass.  Finding
these and other trends can help identify and understand
physical causes and effects in galaxy formation.  For instance, several 
such trends are thought to originate from self-regulation
of physical processes, so that tracking one process implies the
behavior of others (for example, \citet{Sch09,HopQuaMur11}).
Simple models can be used to try to identify such trends.  
These trends can also help to guide
the construction of simple models, especially when they have
simple physical interpretations, such as the stellar mass-halo mass
relations.

Here, the focus is on simple descriptions of (integrated) galaxy histories rather
than fixed time properties.  These descriptions
can provide a point of contact between results of detailed models (arising
from the interplay of all the model processes and components) and
observations, or between two different models.  Again, these
descriptions can also encode
known trends, and help to search for new ones.
For instance, galaxy halo histories on average can be fit by 
 a simple parameterized form (e.g., \citet{Wec02,Zha03,Tas04,
   McBFakMa09, Zha09, Dek13,RodPriAviFab17} and many
others).  Several of these halo history parameterizations incorporate
the physical insight that galaxy halos often have a quickly
growing phase, dominated by significant mergers, followed by a slower
accretion dominated phase.   That is, the functional form of the
simplified models allow a
physical interpretation as well.

In the following, two simplified descriptions of integrated galaxy star
formation rate histories are applied to several
samples constructed from the
L-galaxies semi-analytic model \citep{Hen15}.  The 
N-body Millennium simulation \citep{Spr05, Lem06,AngWhi10,AngHil15} provides
the underlying halo and subhalo histories.  
One description is based upon an integrated lognormal fit, following
the proposal studied in detail
 in \citet{Gla13,Abr16,Die17}. A specific physical shape is assumed.
The second description follows \citet{CohvdV15,Spa15}, applying principal
component analysis (PCA),  not
to the instantaneous star formation rate histories (as in
those works) but instead to
the integrated star formation rate histories.  PCA uses fluctuations
around the sample average
history, determined by the sample.  
PCA thus
incorporates all of a sample's galaxy histories in
its definition, in addition to assigning parameters to each galaxy's
individual history.   Using
integrated rather than instantaneous star
formation rate histories was proposed as key to
reducing scatter in \citet{Die17}, these integrated histories are
taken as the
main quantities of interest here.

This work can be considered as a natural combination and extension of
that of 
\citet{Die17} and \citet{Pac16}.  The
relations among the lognormal fit parameters, and between them and several galaxy and star formation rate properties were
explored in \citet{Die17}.  The integrated star formation rate was
also introduced therein as a basic quantity. 
In \citet{Pac16}, average
histories were found for star formation rates.  In detail, individual
galaxy star formation rate histories were sorted into subfamilies
according to whether they were quiescent or star forming, their final
stellar mass, and their time of observation, and then stacked within
each subfamily.
The properties of the scatter around each of the history subfamilies
studied by \citet{Pac16} is
measured below in an analogous sample, and compared to the scatter of subfamilies created using the
lognormal and PCA fits.

In \S\ref{sec:setup}, galaxy samples and methods are described. 
The integrated star formation rate histories are analyzed 
using both descriptions in \S\ref{sec:param}, and the accuracy of using
the fits as approximations is measured.
In \S\ref{sec:otherprops},
correlations between the two descriptions and between
them and final time properties or other galaxy histories are quantified.
Machine learning is used to investigate
how well several galaxy properties, including 
the history of the largest halo only at each
time, can directly predict the fit parameters.
Different ways of sorting the integrated star formation rate histories into subfamilies are considered in
\S\ref{sec:split}.
A summary and discussion are found in
 \S\ref{sec:discussion}, and the appendix has more details of the
 machine learning results and of splitting up galaxy samples into
 subfamilies using the
 history-defined (fit) parameters.

\section{Samples and methods}
\label{sec:setup}
Star formation rate histories are taken from
the \citet{Hen15}
L-galaxies model, built upon the Millennium Simulation
\citep{Spr05,Lem06}.  The simulation is dark matter only, and the
histories were downloaded from
the German Astrophysical Virtual Observatory.\footnote{At http://gavo.mpa-garching.mpg.de/Millennium/. I thank G. Lemson for his
  patient assistance.}  The underlying MRscPlanck1 simulation is the original Millennium
simulation, rescaled via the method in
\citet{AngWhi10,AngHil15} to the Planck parameters
$\Omega_m = 0.317$, $h=0.673$, $\sigma_8 = 0.826$
and side $470.279$ Mpc/$h$.  

\begin{figure}
\begin{center}
\resizebox{3.3in}{!}{\includegraphics{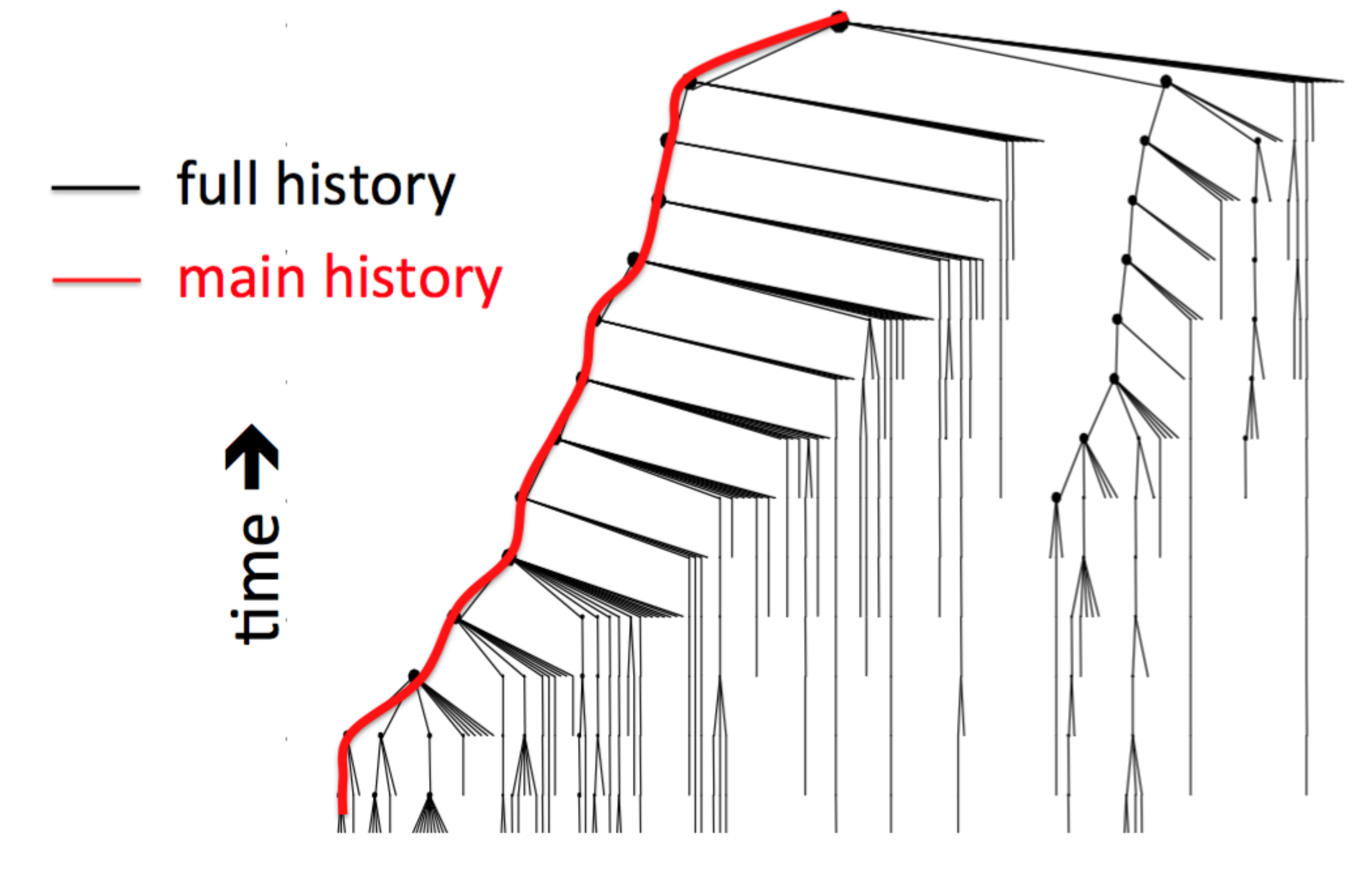}}
\end{center}
\caption{Two types of histories, main (red line) and full (all lines) for a single galaxy (the dot at
  the top).  Dots are galaxies
  and lines connect them between time steps. Time flows upward.  For
  the galaxy at the top (final time), 
  the full star formation rate at any time is the sum of star
  formation rates of 
  all dots (galaxies/halos) at the height corresponding to that time.
  All galaxies in the picture contribute to the full star formation rate.
 The main star formation rate history only includes the star formation
 rate of the single
  galaxy on the main (most massive progenitor) history of the galaxy,
  connected by the red line over time.
A galaxy's spectrum encodes the full
star formation rate history (after dropping the contributions from
stars which have been stripped, and including ageing).  The main history is
the star formation rate of what might be considered as a single galaxy
evolving through time.
Merger tree
courtesy M. White.
}
\label{fig:histtypes} 
\end{figure}

\begin{table*}
\centering 
\begin{tabular}{l|c|c|c|c|c|} 
Sample & method&range& $N_{\rm gal}$&all final galaxies above&history type\\
\hline
$M_h$& 20 bins of $\leq 2000$ &$10.60 < \log M_h/M_\odot < 15.60$&31383&$\log M_h/M_\odot
> 14.30$&main, full\\
& in $\log M_h$& & & & \\
$M^*$&20 bins of $\leq 2000$ &$8.00< \log M^*/M_\odot <12.43$&34246&$\log
                                                                M^*/M_\odot>11.77
                                                                $&main,
                                                                   full
  \\
& in $\log M^*$& & & & \\
{\tt ran} &random& $M^*/M_\odot > 10^9 $&32775& N/A&main, full \\
cen $M_{h,{\rm big}}$&all central galaxies& $M_h/M_\odot\geq
                                      10^{12}$&386919&central
                                                       $M_h/M_\odot \geq
                                                 10^{12}$&main
\\
\hline
\end{tabular} %
\caption{Seven galaxy samples used for measurements.  In the
binned samples chosen from equally spaced logarithmic bins (final
$M^*$ or $M_h$), 2000 galaxies were
  randomly chosen in each bin.  However, in the highest bins, 
  fewer than 2000 galaxies exist in the simulation; all galaxies were
  included above the scales as noted.  Here, $M^*$, $M_h$ refer to
  values at final time.}
\label{tab:galsamps}
\end{table*}
\begin{figure}
\begin{center}
\resizebox{3.3in}{!}{\includegraphics{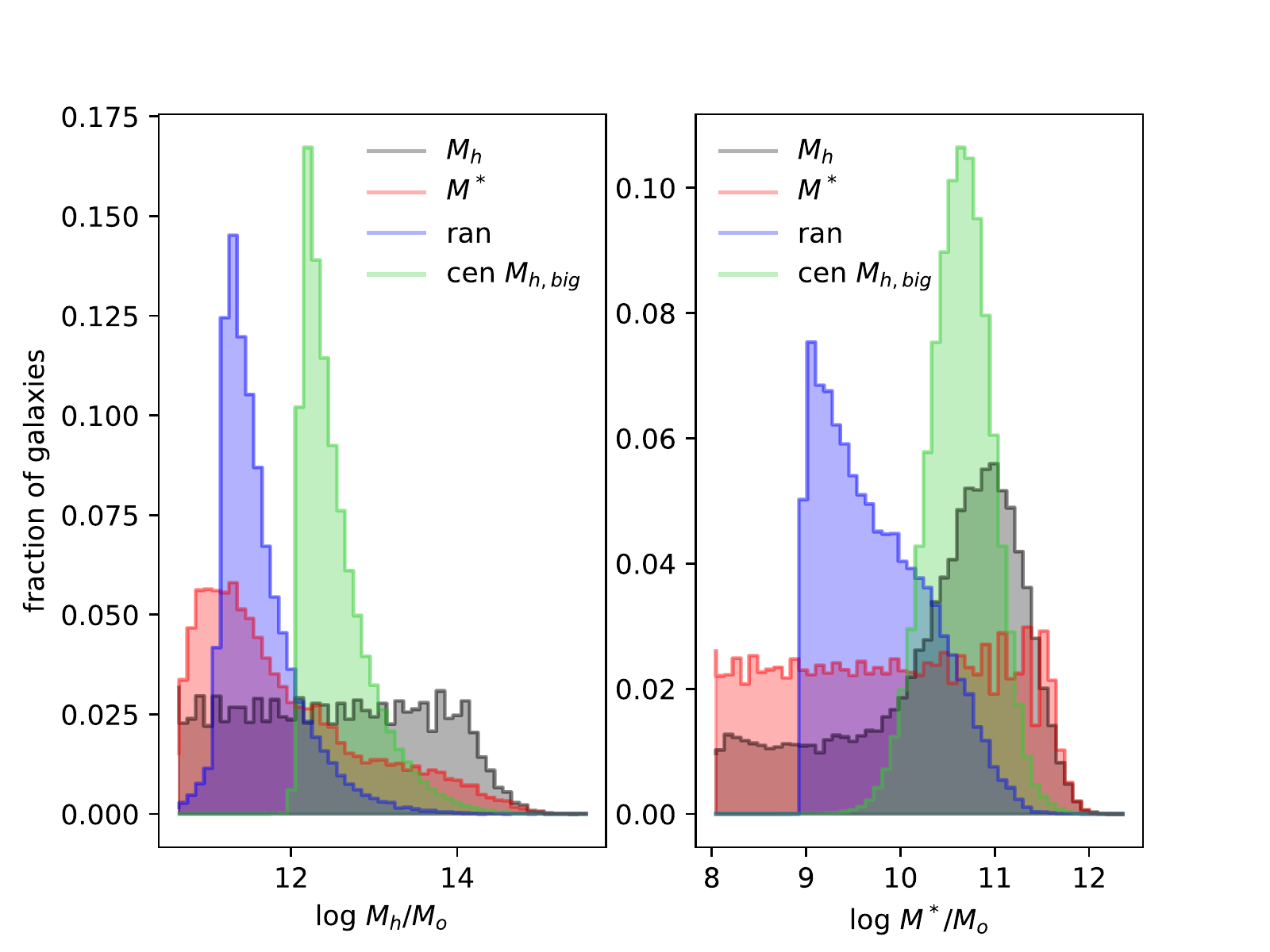}}
\end{center}
\caption{Final time halo mass and stellar mass distributions for the 4 simulation
  samples described in the text and Table \ref{tab:galsamps}.  The
  $M_h$, $M^*$ and {\tt ran} samples are all randomly selected according to
  some criteria, while the cen $M_{h,{\rm big}}$ sample includes all galaxies
  $M_h \geq 10^{12} M_\odot$ which are central at the last two time outputs.
}
\label{fig:massdists} 
\end{figure}

There are two natural definitions of star formation rate histories,
schematically illustrated with a
sample simulated galaxy dark matter history in Fig.~\ref{fig:histtypes}.
All dark matter halos
which eventually merge to form the final galaxy are shown.   Time runs up the
picture, with the single final galaxy at top, and progenitors
appearing at the time when they are first resolved in the simulation (the size of dot is
proportional to dark matter halo mass).  The progenitors of the final
galaxy that exist at any given time are shown on the same row, with
lines connecting them to their descendants in the row above. 

The full star formation rate history specifies the formation time of
all stars in
all galaxies which eventually merge to produce
the final galaxy.
At any give time, this rate is the sum of star formation rates across
the appropriate
row in Fig.~\ref{fig:histtypes}. The full star formation
rate history is encoded in the spectrum of the final
galaxy, measured observationally, although stellar ageing and
stripping can remove stars.  
Every star in the final galaxy was formed as part of the full star
formation rate.
  
In contrast, the main star formation rate
history is composed of the star
formation rate of the largest progenitor galaxy at any time (here shown at left,
with its history traced by the red line).   
The main star formation history considers the final galaxy as
a single object throughout its history, with other galaxies
merging into it.  Such a galaxy could then be described, for instance, as moving
through the blue cloud in a certain way, and onto the red sequence (if
it quenches). 
Stars due to mergers did not form in the main history
and thus are not related to this (``in situ'') star formation
rate.

The full star formation rate history was
considered by \citet{Die17} in the Illustris simulation, and
by \citet{Pac16}, who used spectra to get the star formation rate
histories, and then
 matched them to a semi-analytic post treatment of the Millennium simulation histories. 
Both definitions of star formation rate histories are considered
below.

\subsection{Galaxy history samples}
\label{sec:samples}
In order to study properties of galaxy histories as a function of halo
mass or stellar mass, four galaxy samples are considered from the
$\sim$ 2.26$\times 10^7$ galaxies at the scale factor $a=0.9997$ time step of the
simulation (called the final time hereon).
One sample is 
a random selection
of galaxies with final $M^*
> 10^9 M_\odot$.  This sample is dominated by the
lowest mass galaxies in the sample, due to the shape of the mass
function.  To better identify properties as a function of final halo or
stellar mass, rather than being swamped by properties of low mass galaxies,
random samples with a roughly uniform distribution in
log final  $M_h$ or log final $M^*$ were also created. 
All three subsamples have approximately 33000 galaxies,
and full and main histories are studied for both.   
A fourth sample was taken for comparison with
machine learning work by \citet{KamTurBru16a}, and includes 
all galaxies with final time halo mass above
$10^{12} M_\odot$ which are central at both the last and second to
last time step (17\% were satellites at some point in their histories).  There are 386919 galaxies in this sample, so only the main
star formation rate history was considered.\footnote{Downloading 386919 
  full histories is computationally time intensive and
  not expected to yield significantly more insight.  Comparisons of
  this sample to the exact sample used in \citet{KamTurBru16a} are in
  the appendix, \S\ref{sec:morekam}.}
  These samples are called
$M_h$, $M^*$, {\tt ran}, and cen $M_{h,{\rm big}}$ below (with main or full to
identify the choice of star formation rate history, except for cen
$M_{h,{\rm big}}$ where only the main history was extracted).  More details are
in Table \ref{tab:galsamps}, and the stellar mass and halo mass
distributions at final times are in Fig.~\ref{fig:massdists}.  Besides highlighting 
higher $M_h$ and $M^*$ galaxies, using several
subsamples illustrates how the fits below can be used to compare different
galaxy samples (or different models
built on the same or different simulations).
 
The starting redshift is 9.7 (when the universe is about
450 Myr old), and following \citet{Die17}, star formation rate
histories are integrated to the present day, using the galaxy
formation model star formation
rates 
at each output time.\footnote{The simulation data at each output does
  not include starburst contributions.}
The integrated 
star formation rate
from the initial time to time
$t$ is $\tilde{\cal S}(t)$.
For the seven samples here, there are 48 output times, 
outputs 11 to 58 in the MRscPlanck1
simulation.\footnote{There are star formation rate histories 
  at 20 output times directly available from these simulations as
  well \citep{Sha15}.} 

Note that the final time integrated star formation rate  $\tilde{\cal
  S}(t_f)$ is not the final stellar mass.  For the main histories, stellar
mass gain due to mergers and stripping is not included.  Although the full
histories include all stars formed by galaxies which eventually merge into the
final galaxy, they still do not account for stars which are stripped
off, or those which are added by stripping of other galaxies which don't
eventually merge, and again, for both, starbursts are not available
at the simulation output times.\footnote{To get the main stellar
  mass history as considered in \citet{CohvdV15},  stripping,
 ageing (which happens instantaneously in this simulation, dropping the stellar mass to about 60 percent of
stars formed) and mergers must be combined with the main integrated star
formation rate history.  These added
contributions and subtractions for stellar mass also make it difficult
to use the stellar mass to estimate the
amount of star formation due to starbursts between time outputs.} 
In both cases, the stars age as well.

These integrated histories  $\tilde{\cal S}(t)$ are assigned peak times, using a lognormal
fit (following \citet{Gla13,Die17}) and principal component expansions around the
average history as follows.

\subsection{Lognormal fit}
For the lognormal parameterization,
star formation rate (SFR) histories are taken to have the form \citep{Die17},
\begin{equation}
SFR_{\rm log}(t) = \frac{A}{\sqrt{2 \pi }\tau t} \exp(-\frac{(\ln t - T_o)^2}{2
  \tau^2}),
\label{eq:sfrlog}
\end{equation}
with corresponding integrated star formation rate history 
\begin{equation}
\begin{array}{ll}
\tilde{\cal S}_{\rm log}(t) &= \frac{A}{2}\{1 - {\rm erf}(- \frac{\ln t - T_0}{\tau
    \sqrt{2}})\}  \\
&= \int_{t_{\rm initial}}^t SFR_{\rm log}(t') dt' \; . \\
\end{array}
\label{eq:intsfr}
\end{equation}
Fits are done to this integrated star formation rate history,
following \citet{Die17}, due to its reduced scatter.

This parameterization has
a peak time, width and peak SFR 
\begin{equation}
\begin{array}{lll}
t_{\rm peak} &=& e^{T_0-\tau^2} \\
\sigma_t &=& 2 t_{\rm peak} \sinh(\sqrt{2 \ln(2)}\tau) \\
SFR_{\rm peak}&=& \frac{A}{\sqrt{2 \pi}\tau} e^{-T_0 + \tau^2/2}
\end{array}
\label{eq:logfit}
\end{equation}
The width $\sigma_t$ is the amount of time between the two points in the history
where the star formation rate is above $1/2$ of its peak value.  More
generally \citep{Die17},
the time where the star formation rate reaches $1/\beta$
of its peak value, $SFR(t_\beta) = \frac{1}{\beta} SFR_{\rm peak}$ is
\begin{equation}
t_{\pm 1/\beta} = t_{\rm peak} e^{\pm \tau \sqrt{2 \ln (\beta)}} \; .
\end{equation}
One particular value of interest is $t_{1/2}$, where the star
formation rate drops to half of its peak value (it is part of
$\sigma_t = t_{1/2}-t_{-1/2}$ and can be roughly thought of as a sort of
quenching parameter).

\citet{Die17} applied this lognormal fit to integrated star formation rate histories in the Illustris
simulation, as well to the integral of observed average quiescent
galaxy star formation rate histories stacked by \citet{Pac16}, and
compared to similar fits on observations by \citet{Gla13}.  They used
the 29203 galaxies in Illustris with
$M^* \geq 10^9 M_\odot$, integrating the star formation rates
starting when the universe was 54 Myr old along 100 equally spaced
output times. \footnote{The other fit considered in detail in \citet{Die17} was a double power
law, as used in \citet{BWCz8}.  The resulting fit was often singular
when applied to the histories here, although when non-singular, it
tended to be better according to the criterion Eq.~\ref{eq:goodfitcrit}.
\citet{Die17} suggest the improved fit is 
likely in part due to a double power law having an extra parameter and thus extra
flexibility, but see also, e.g., \citet{Car17}.  Just as in the lognormal
fit, discussed below, some of the bad fits are due to rejuvenating histories.}

For Illustris, the parameters $t_{\rm peak},\sigma_t$ are correlated,
obeying a mean
relation, 
\begin{equation}
\sigma_t \sim 0.83 t_{\rm peak}^{3/2} \; {\rm in \; Gyr} .
\label{eq:sigt}
\end{equation} 
(For example, see Figs.~5 and 6 in \citet{Die17}.)

By construction, this fit is an approximation. They
defined the goodness of fit for their parameterization as
\begin{equation}
D ={\rm max}_t  \frac{|\tilde{\cal S}_{\rm log}(t) - \tilde{\cal S}(t)|} {{\rm
    max}(\tilde{\cal S}(t))} \; .
\label{eq:goodfitcrit}
\end{equation}  
and found that satellites tended to have worse fits than central galaxies.
This goodness of fit measure will be used below for both approximations and all 7 samples.

\subsection{Principal component analysis}
\label{sec:PCA}

Principal component analysis
(PCA) offers another approximation to galaxy integrated star formation rate
histories.  For PCA in general, vectors are decomposed into the
average of the sample, plus coefficients $a_n$ times principal
components $PC_n$.  The $PC_n$, basis vectors for fluctuations
around the average, are eigenvectors of the covariance matrix of the
vector components.
The integrated star formation rate history of one galaxy up to a
particular time is an element of the vector $\tilde{\cal S}(t)$ for
that particular galaxy.
The full ensemble of a
sample's integrated star formation rate histories, for all of its galaxies,
determines the average and the fluctuation vectors $PC_n(t)$.

In more detail, the integrated star formation rate histories are first
normalized by dividing the integrated star formation
rate histories by each galaxy's individual integrated star formation rates at
the final time,
\begin{equation}
{\cal S}(t) = \tilde{\cal S}(t)/\tilde{\cal S}(t_f) \; .
\label{eq:rescale}
\end{equation}
(Again, as mentioned earlier, $\tilde{\cal S}(t_f)$ is not necessarily
the same as final $M^*$.)
Without this normalization, the sample average and fluctuations around it are dominated by
the most massive galaxies, as these tend to have the largest integrated star
formation rates and fluctuations.  
Other candidates for rescaling $\tilde{\cal S}(t)$,
using the final stellar mass or the peak star formation rate, gave much
larger scatters around the resulting average history.

The vector
${\cal S}^\alpha(t)$, the normalized integrated star formation rate history of
any galaxy labeled by $\alpha$, is then written using PCA as
\begin{equation}
{\cal S}^\alpha (t) = \bar{\cal S}(t) + \sum_{n=0}^{N_{\rm times}-1}
a_n^\alpha PC_n(t) \; .
\label{eq:pcdecomp}
\end{equation}
with constant coefficients $a_n^\alpha$.  Here the average $\bar{\cal S}(t)
=\frac{1}{N_{\rm gal}}
\sum_{\alpha} {\cal S}^\alpha (t)  $.
The PCA basis fluctuations $PC_n(t)$ are
the orthonormalized eigenvectors
of the covariance matrix, $C_{ij} = \langle {\cal S}(t_i)
{\cal S}(t_j)\rangle$.
There are as many fluctuation basis vectors $PC_n(t)$ as there are
output times $t_i$, 48 for
the samples under study here, and the expression Eq.~\ref{eq:pcdecomp}
is exact.  The largest contribution to the
sample variance is in the direction $PC_0(t)$, followed by $PC_1(t)$, etc.  (For parameter
counting, to give the unnormalized history there is 
one additional parameter, to undo the rescaling which made ${\cal
  S}(t_f) \equiv 1$ for each galaxy.  Because of this constraint, the
variance in the direction of $PC_{47}(t)$, a vector of zeros except for a 1
at final time, is 0.) 

An approximation to the integrated star formation rate history can be
made by truncating the expansion Eq.~\ref{eq:pcdecomp}, keeping only
some of the $PC_n(t)$.
Hereon, the PCA approximation is taken to be the truncation of the
above expansion to the first three components:\footnote{The
  approximation to the full $\tilde{\cal S}(t)$, when used below, is obtained by multiplying
  ${\cal S}(t)$ by
  $\tilde{\cal S}(t_f)$.  Also note that
  these approximate integrated histories can give a negative
  instantaneous star formation rate.  For fixed time comparisons any
  negative star formation rate is set to zero.  One could introduce
  more complexity by constraining the expansion to give positive star
  formation rates at every time.}
\begin{equation}
{\cal S}^\alpha (t) \approx \bar{\cal S}(t) + 
a_0^\alpha PC_0(t) + a_1^\alpha PC_1(t) + a_2^\alpha PC_2(t) \; .
\label{eq:pcaapprox}
\end{equation}
If this approximate description of average history plus a few
fluctuations is to be useful, a large fraction of the variance of the
sample should
be captured using the first few basis fluctuations, that is, by the
sum of the
first few eigenvalues of the covariance matrix.
Not unrelated, but not automatic, for the approximation to be good for
any particular galaxy labeled by 
$\alpha$, the
$a_n^\alpha$, for $n>2$, should be relatively small, for example, 
in comparison to the variance around the average for the full sample.  
Again, in the PCA decomposition, both the $PC_n(t)$ and the average
integrated history, $\bar{\cal S}(t)$, are properties of the sample,
and depend upon the galaxy histories used.  The sample depends upon
its selection function, and the galaxy histories of course depend upon
the theory used to construct them.

\section{Parameterizing galaxy histories}
\label{sec:param}
The lognormal fit, Eq.~\ref{eq:logfit}, and PCA approximation,
Eq.~\ref{eq:pcaapprox}, were implemented for all seven
galaxy samples in Table \ref{tab:galsamps}.
Some properties of the fits, in particular, the values of the
leading parameters, $t_{\rm peak}$
 and $a_0$, their relation, and measures of goodness of the fits, are
 as follows.
\subsection{Lognormal Fit}
\begin{figure}
\begin{center}
\resizebox{3.5in}{!}{\includegraphics{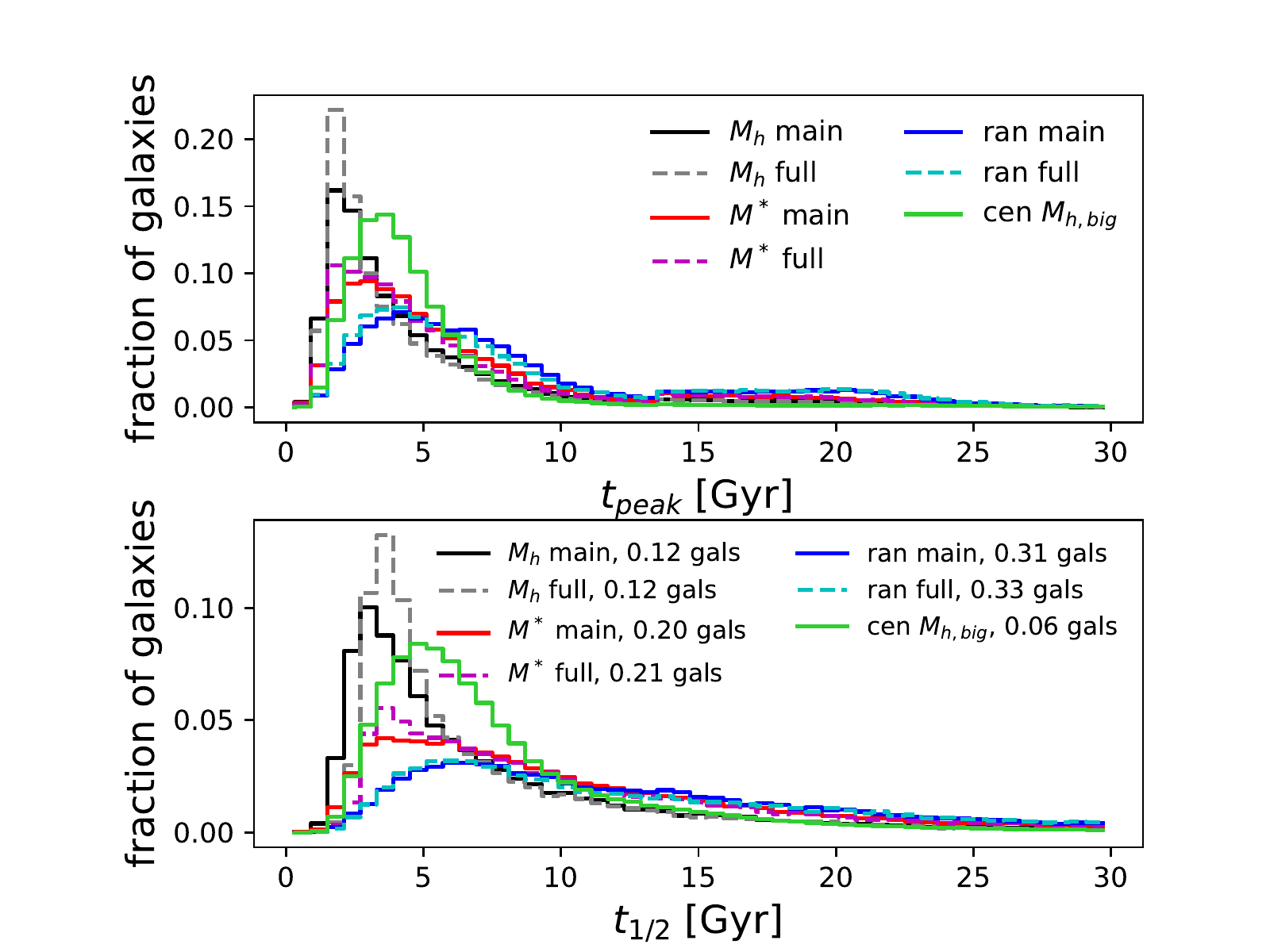}}
\end{center}
\caption{Distribution of (top) peak time $t_{\rm peak}$ and
(bottom) $t_{1/2}$ for integrated star formation
rate histories for all 7 samples, from their fits to a lognormal as in Eq.~\ref{eq:intsfr}. 
The time $t_{1/2}$ is calculated from the lognormal fit and
corresponds to the time in the fit when galaxies drop to 1/2 of their
peak star formation rate.
 The legend also shows the fraction of galaxies with $t_{1/2}>30$ Gyr,
 i.e. those dropping 
to half of their peak star formation rate after 30 Gyr.   Although $t_{1/2}$ is related to the quenching time,
it is not exactly when the galaxy leaves the star forming main sequence,
as the latter also depends upon stellar mass and redshift.
The main samples are shown with a solid
line, the full samples are shown with a dashed line of similar color.
}
\label{fig:tpdist} 
\end{figure}

The distribution of $t_{\rm peak}$, is shown at top in
Fig.~\ref{fig:tpdist} for all 7 samples.  It is
weighted towards early times, 
especially in the $M^*$ and $M_h$ samples, which have the largest
fraction of massive and thus early forming galaxies.
Another characteristic time, as mentioned above, is when
 a galaxy drops to 1/2 of its peak star formation rate, $t_{1/2}$,
 shown at the
bottom of Fig.~\ref{fig:tpdist}.  Although related to quenching,
$t_{1/2}$
does not specify on its own when a galaxy leaves the star forming main
sequence, as the star forming main sequence changes with redshift and depends on the stellar
mass of the galaxy (see, \citet{Spe14}, for example, for different
estimates of where the star forming main sequence lies, depending upon
definitions of stellar mass and star formation rates).  

Galaxy by galaxy, on average, the full
samples have slightly earlier $t_{\rm peak}$ (0.12-0.25 Gyr) 
and larger $\sigma_t$ (0.74-1.35 Gyr). 
That is, the time evolution of the combined star formation rate of all the progenitor galaxies of a
final galaxy on average peaks earlier but decays more slowly than that for the
single main galaxy.  This effect has many contributing factors which
would be
interesting to better understand, including the smaller mass of the
galaxies which merge onto the main galaxy, their tendency to quench
when they fall into the main galaxy's halo, and the relation of the
merger rate to the star formation rate of the main galaxy.%

For these samples, the $t_{\rm peak}-\log \sigma_t$ correlation is about 80\% and the two
parameters obey a similar mean relation to that of Illustris, where $\sigma_t \sim 0.83 t_{\rm
  peak}^{3/2}$ \citep{Die17}.
In the cases here, the power law remains close
to 1.5 , but the prefactor varies by a factor
of two between samples with different mass distributions.\footnote{Fitting  $\log \sigma_t
  = \log t_{\rm
    peak} + C$ gave 
$\sigma_t = a
t_{\rm peak}^b$ where $a$ = (0.56,0.82,0.86,0.99,0.62,0.68,0.46) and $b=$
(1.54,1.44,1.44,1.43,1.59,1.62,1.57) for main, full $M_h$, main, full,
$M^*$,main, full {\tt ran}, and 
cen $M_h$ samples respectively.  }  
The full {\tt ran} sample, expected to have sampling closest to the
Illustris distribution, obeys
$\sigma_t \sim 0.68 t_{\rm peak}^{1.62}$.   \citet{Blu16} compared
both models to
observations and found 
that L-galaxies \citep{Hen15} quench too
quickly and Illustris galaxies not quickly enough, consistent with
Illustris having a larger $\sigma_t$
for a given $t_{\rm peak}$ as found here.

\begin{figure*}
\begin{center}
\resizebox{6.3in}{!}{\includegraphics{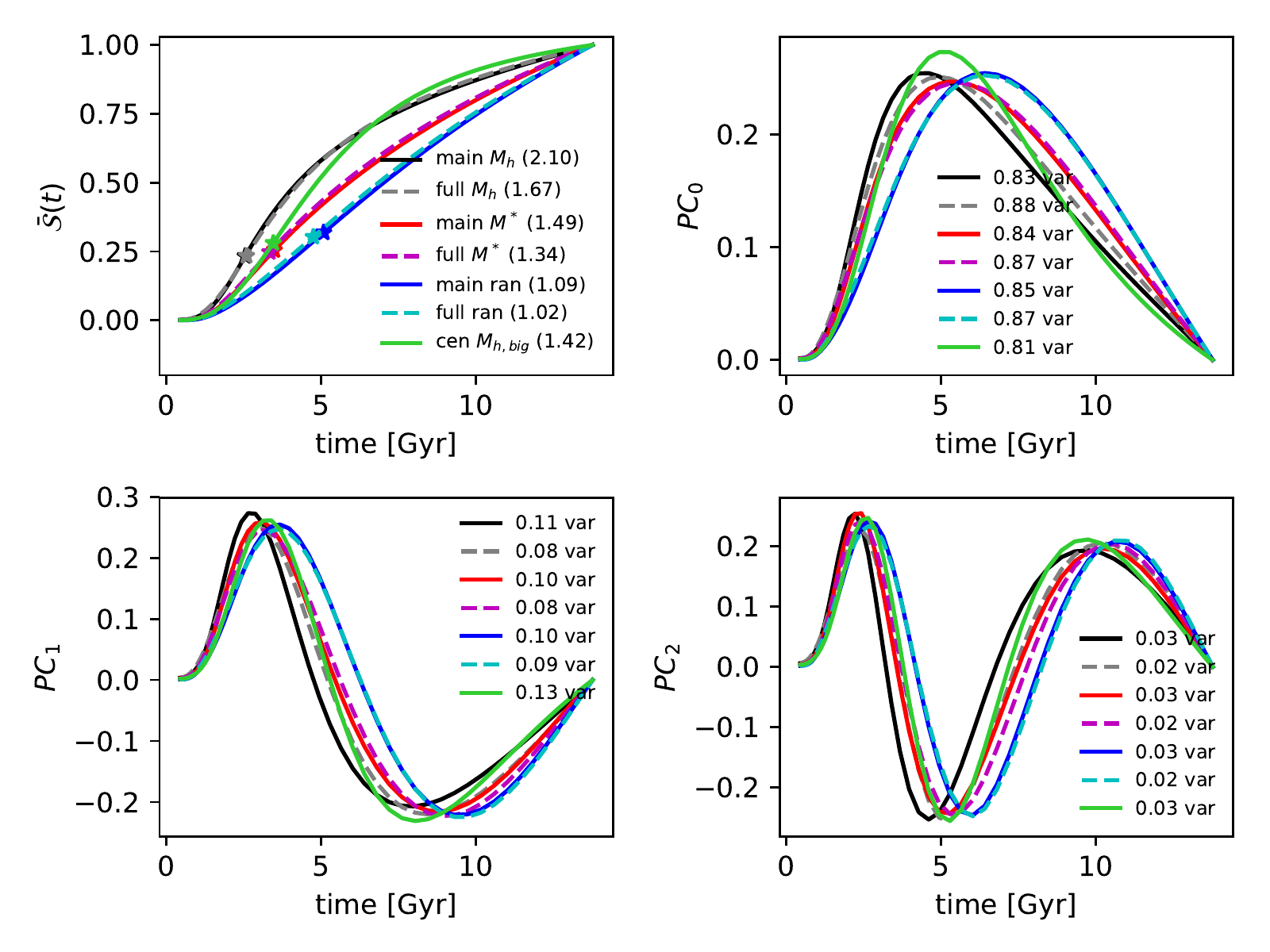}}
\end{center}
\caption{Basis PCA components for the expansion of galaxy histories (each
  normalized to 1 at final time).  Individual 
galaxy histories are approximated via  ${\cal S}(t) \approx {\cal \bar{S}}(t)$  + $a_0 PC_0(t) + a_1
PC_1(t) + a_2PC_2(t) $, 
Eq.~\ref{eq:pcaapprox}.  
Upper
  left: the average history $\bar{\cal S}(t)$ for all 7 samples,
  labelled with total
  variance for each.  Stars mark the
  lognormal fit $t_{\rm peak}$ for each average.  The earliest
  $t_{\rm peak}$ occurs for the samples with the largest number of massive galaxies.
The
leading three fluctuations around each average are $PC_0(t)$ at
 upper right, $PC_1(t)$ at lower left, and $PC_2(t)$ at lower right,
 labeled with the fractional contribution of each $PC_n$'s
 coefficients to the total variance.   
For each $n$, the $PC_n(t)$ show
ordering in their features similar to the $t_{\rm peak}$ in their
sample averages.
 The full and main averages essentially coincide for all samples.
The full and main $PC_n(t)$ all coincide for the {\tt ran} sample, but
differ in $PC_0(t), PC_1(t)$ for the $M_h$ sample, and
$PC_1(t),PC_2(t)$ in the $M^*$ sample.
}
\label{fig:leadingpcs} 
\end{figure*} 

\begin{figure}
\begin{center}
\resizebox{3.1in}{!}{\includegraphics{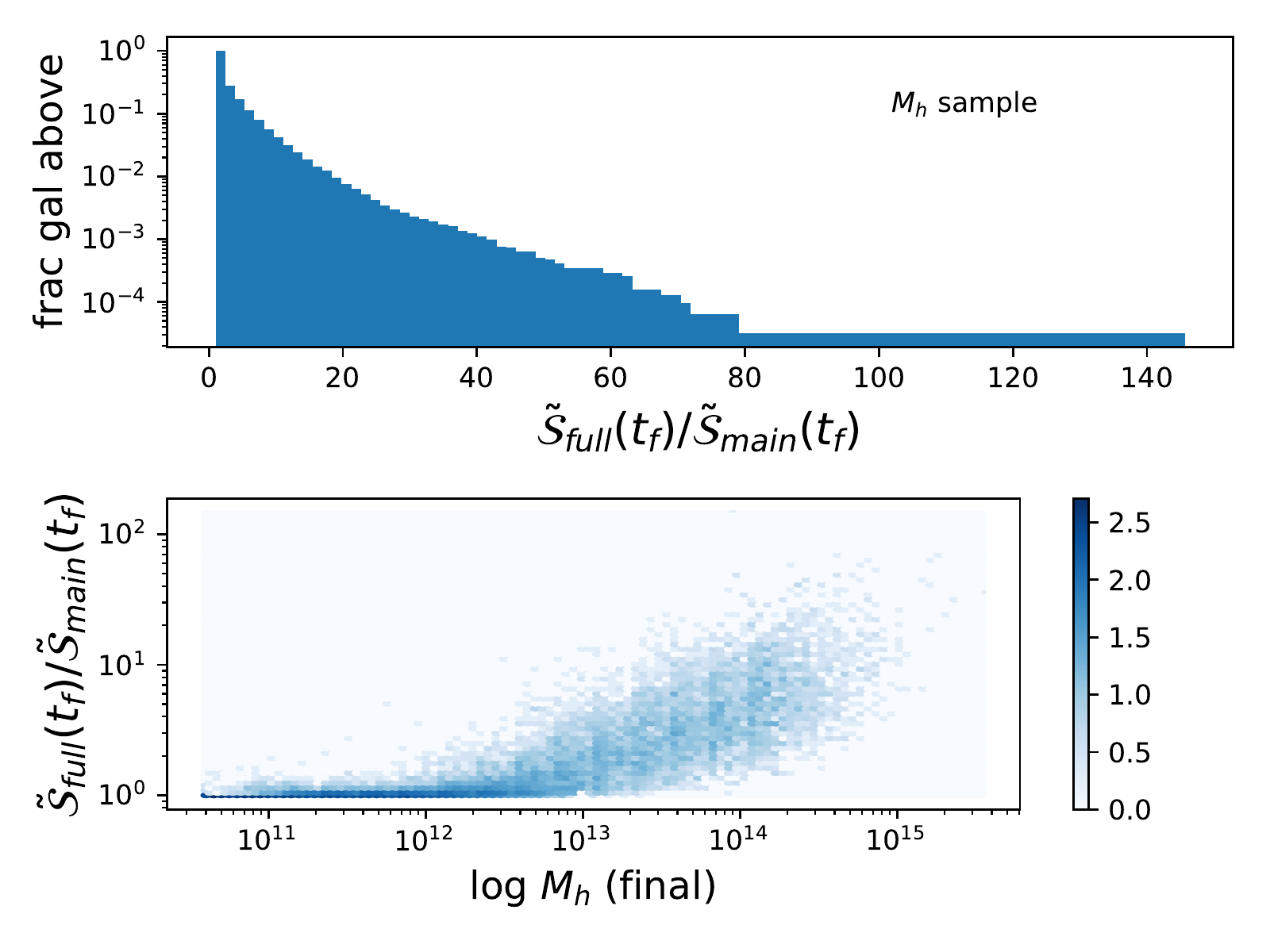}}
\caption{Top:
fraction of galaxies with given ratios of full to
main final integrated star formation rates, $\tilde{\cal
    S}_{\rm full}(t_f)/\tilde{\cal S}_{\rm main}(t_f)$.  After dividing by
  these normalizations, the full and main average integrated histories
 almost coincide, see upper left in  
  Fig.~\ref{fig:leadingpcs}. \newline Bottom: Final $M_h$ dependence of $\tilde{\cal
    S}_{\rm full}(t_f)/\tilde{\cal S}_{\rm main}(t_f)$, $\log_{10}$ number of
  galaxies in each pixel according to scale at right.
}
\label{fig:mainfullcomp} 
\end{center}
\end{figure}

\begin{figure}
\begin{center}
\resizebox{3.3in}{!}{\includegraphics{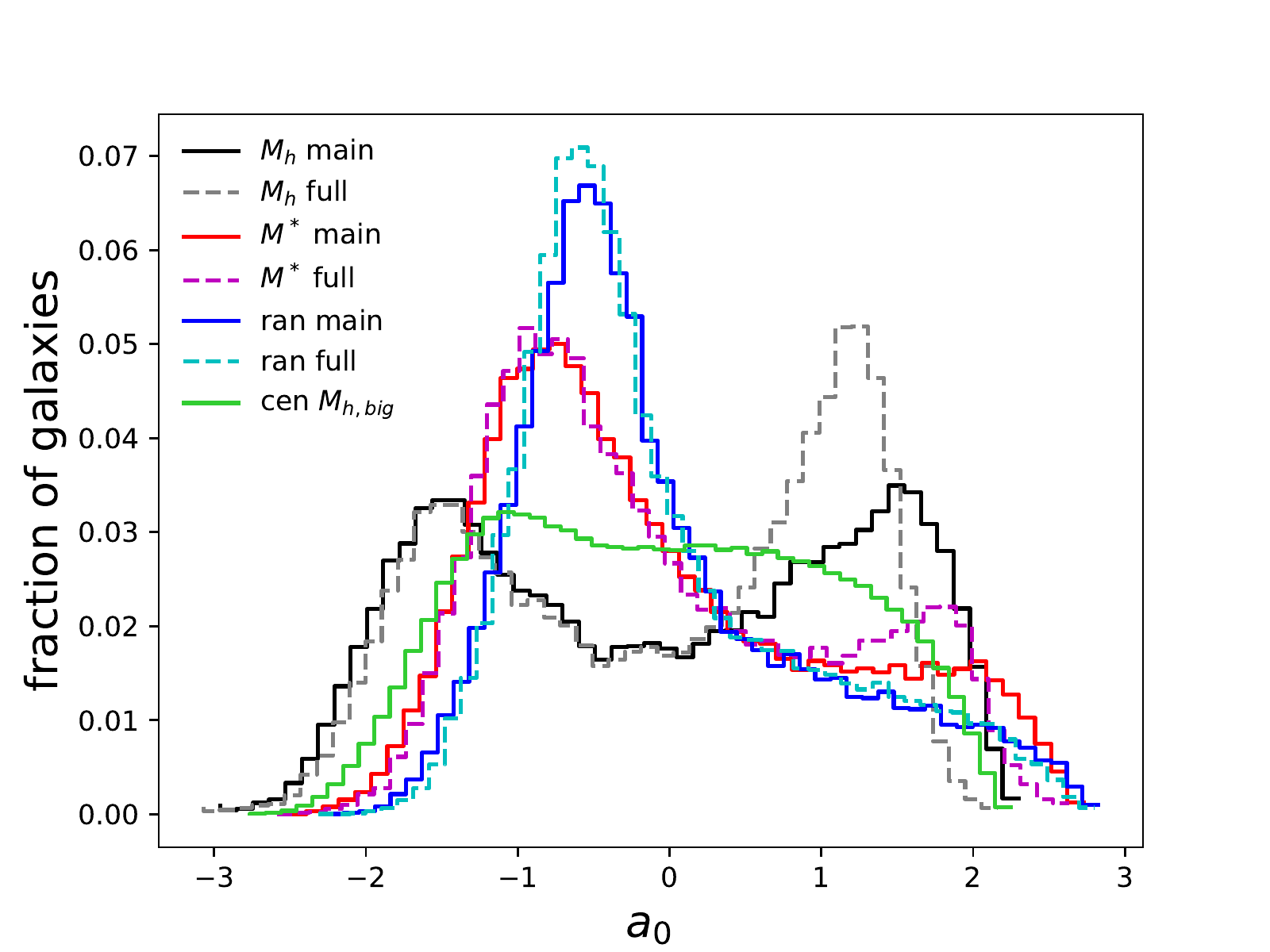}}
\end{center}
\caption{Distribution of $a_0$ for all 7 samples.  As $a_0$ is the
  coefficient of $PC_0(t)$ (shown in Fig.~\ref{fig:leadingpcs}, upper
  right),
positive $a_0$ increases the integrated star formation rate at early
times relative to the average, and negative $a_0$ decreases it,
corresponding to later star formation.
Full and main distributions
separate at positive $a_0$ for samples with many large final $M_h$
galaxies.  For these,
main histories tend to have a larger positive $a_0$, i.e., an earlier
rise in integrated star formation rate.  The variance in this
coefficient
captures (depending on
sample) from 81\% to 88 \% of the total variance around the average, as
noted in Fig.~\ref{fig:leadingpcs}.
}
\label{fig:a0dist} 
\end{figure}

\subsection{Principal Component Analysis}
Turning to principal analysis, the average histories $\bar{\cal
  S}(t)$ and leading three
fluctuations, $PC_0(t),PC_1(t),PC_2(t)$ are shown in Fig.~\ref{fig:leadingpcs} for all 7 samples.
The average history $\bar{\cal
  S}(t)$ is at upper
left.  The total variance around each $\bar{\cal S}(t)$ is listed in the legend
and a star marks the
lognormal fit $t_{\rm peak}({\bar{\cal S}(t)})$ for each.   
The other panels show the first 3
principal
components, and list their respective fractional contributions to the total
variance for each sample.  (Again, solid
lines are main histories, dashed are full histories.)  These first 3
fluctuations have 
$\geq 97 \%$ of the total scatter around the average.  This
  is a better approximation than that found by applying PCA
to the star formation rate histories themselves.   In the latter case,
again rescaling by $\tilde{\cal S}(t_f)$, all samples except cen
$M_{h,{\rm big}}$ require $>10 \; PC_n$ to capture 90\% or more of
  the variance around the average history.  (The cen $M_{h,{\rm big}}$
  sample requires 6 $PC_n$.)  The smaller fraction of
  variance in the first 3 fluctuations around the instantaneous star
  formation rate makes the PCA approximation, 
  Eq.~\ref{eq:pcaapprox}, much less useful.  

Comparing samples, as the number of lower $M^*$ galaxies (which tend to be
star forming) increases, there is a trend towards later sample average
$t_{\rm peak}$ 
and correspondingly later times for the peaks
of the principal components.  This is in line with the tendency of
lower $M^*$ galaxies to quench at later times.
The average histories of each of the subsamples seem
independent of whether the full or main histories are used.
This is in spite of very different full to main normalizations, a
comparison of 
$\tilde{\cal S}_{\rm full}(t_f)$ and $\tilde{\cal S}_{\rm main}(t_f)$
is in
Fig.~\ref{fig:mainfullcomp} for the $M_h$ sample.  
The bottom
panel shows their ratio as a function of final $M_h$ (the trend
with
final $M^*$ was weaker).  Higher $M_h$ halos have larger ratios of full
to main ${\tilde{\cal S}}(t_f)$, that is, they have more star formation in
their full history which was not ``in situ'', i.e., not in the main star
formation rate history.  

Most of the variance around the average history is
captured in the coefficient of the  leading fluctuation
$PC_0(t)$,  $a_0$.  Fig.~\ref{fig:a0dist} shows the distribution of $a_0$ for all 7 samples.
From
Fig.~\ref{fig:leadingpcs}, it can be seen that adding $PC_0(t)$ to the
average history with 
a positive coefficient $a_0$ will cause the integrated star formation rate to rise earlier
than the average history, and a negative $a_0$ will cause a later rise
in the integrated star formation rate history.
Although the full and main
average histories (and $PC_0(t)$, except for the $M_h$ sample)
closely overlap,
the positive $a_0$ distributions strongly different between the full and main
histories for the 
$M_h,M^*$ samples, which have a large number of high mass halos.  (The
full cen $M_{h,{\rm big}}$ sample was not downloaded, as mentioned earlier.)

\begin{figure}
\begin{center}
\resizebox{3.7in}{!}{\includegraphics{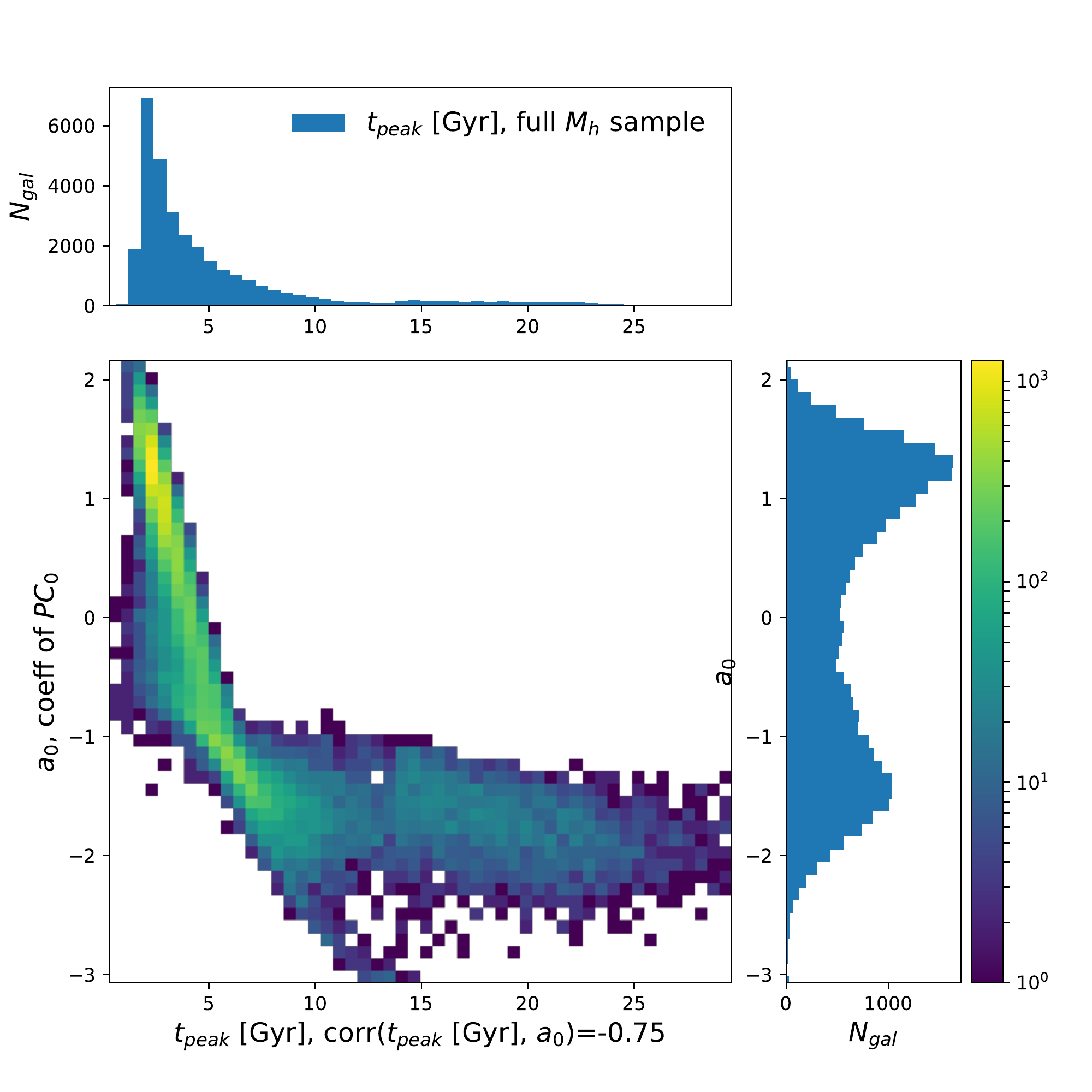}}
\end{center}
\caption{Comparison of lognormal and PCA approximations via their leading parameters
  $t_{\rm peak}$ and $a_0$. 
Top: the $t_{\rm peak}$ distribution
 for the full
  $M_h$ sample. Far right: the $a_0$ distribution.  Bottom left:
  the logarithm of the number of galaxies sharing each pair of
  values.  Although a correlation is visible, the relation between
  $t_{\rm peak}$ 
  and $a_0$ changes noticeably for low $a_0$ and
  large $t_{\rm peak} > 7$ Gyr, i.e. for galaxies which have star
  formation at later times. 
Even with this flat tail, there is a high $a_0, t_{\rm
  peak}$ correlation of galaxies shown (the highest $t_{\rm peak}>30$ Gyr
objects are dropped, about 0.1\% of this sample).  
These trends in the joint distribution of are seen in all samples;
correlations range from
-0.70 to -0.75, again using galaxies with $t_{\rm peak}<30$ Gyr (up to
0.4\% of galaxies in the {\tt ran} sample).  Similar correlations are found with $t_{1/2}$.
}
\label{fig:tppc0hex} 
\end{figure}

\subsection{Comparison of lognormal and PCA approximations}
\subsubsection{Relation of leading parameters}
The two parameterizations are related.  In particular, the PCA leading
contribution,
$a_0$, is correlated with the lognormal fit parameter $t_{\rm
  peak}$.    Their relation is shown for the full $M_h$
sample in Fig.~\ref{fig:tppc0hex}.   Roughly, a late $t_{\rm
  peak}$ corresponds to a negative $a_0$, meaning the rise in the
integrated star formation rate occurs at a later time.    The correlation
is similarly strong for $a_0$ with $\ln \sigma_t$, expected given
the mean relation for $\sigma_t(t_{\rm peak})$, and with
$t_{1/2}$, the  time when star formation rate in the fit falls
to half of its maximum.
Relations for the other 6 samples are comparable in shape and size.\footnote{Restricting to
  galaxies with good fits, e.g. with $D<0.05$, changes the correlations
  slightly, but not the shape of the plot.}
The relation between the two parameters visibly
changes for larger $t_{\rm peak}\gtrsim 5$ Gyr, presumably because the
shape of $PC_0(t)$ is not flexible enough to approximate star
formation rates peaking at late times, see below.
In addition, the integrated histories for galaxies 
with later $t_{\rm peak}$ tend to be very close
to each other (more elaboration in \S\ref{sec:split} below).

In spite of the many correlations,
the fits have key differences.
In particular, $t_{\rm peak}$ for each galaxy is
independent of the full galaxy sample used, defined solely in terms of
fitting to a predetermined lognormal
shape.  In contrast, $a_0$ depends
upon the full sample (which determines both $PC_0(t)$ and the average)
but has no prior assumptions about the shape of the histories.
Also, for the $t_{\rm
  peak}$ parameterization, for different galaxies the
peak moves in position and changes in width
(the integral of the height is fixed). In the PCA approximation,
varying the PCA coefficients can alter the
sign and amplitude of each of the fluctuations $PC_n(t)$ around the fixed average, but not their
shape.  The $t_{\rm peak}$
parameterization enforces that the integrated star formation rate is
monotonic, while the approximation using the first 3 $PC_n(t)$ does not
require this (as mentioned earlier, its derivative can thus lead to negative
instantaneous star formation rates, these unphysical star
formation rates are set to zero).

\subsubsection{Approximating histories}
\label{sec:approx}
\begin{figure}
\begin{center}
\resizebox{3.3in}{!}{\includegraphics{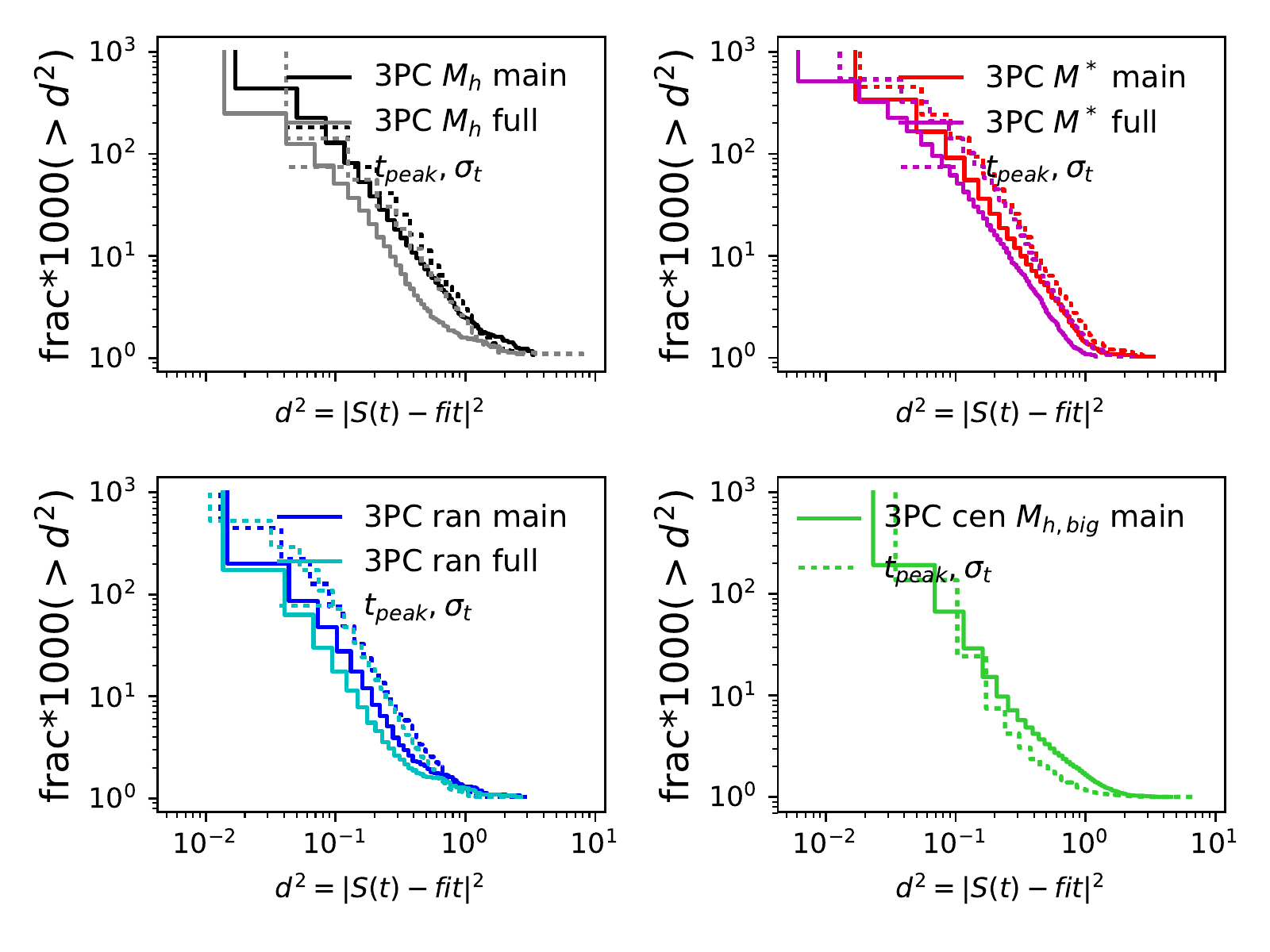}}
\resizebox{3.3in}{!}{\includegraphics{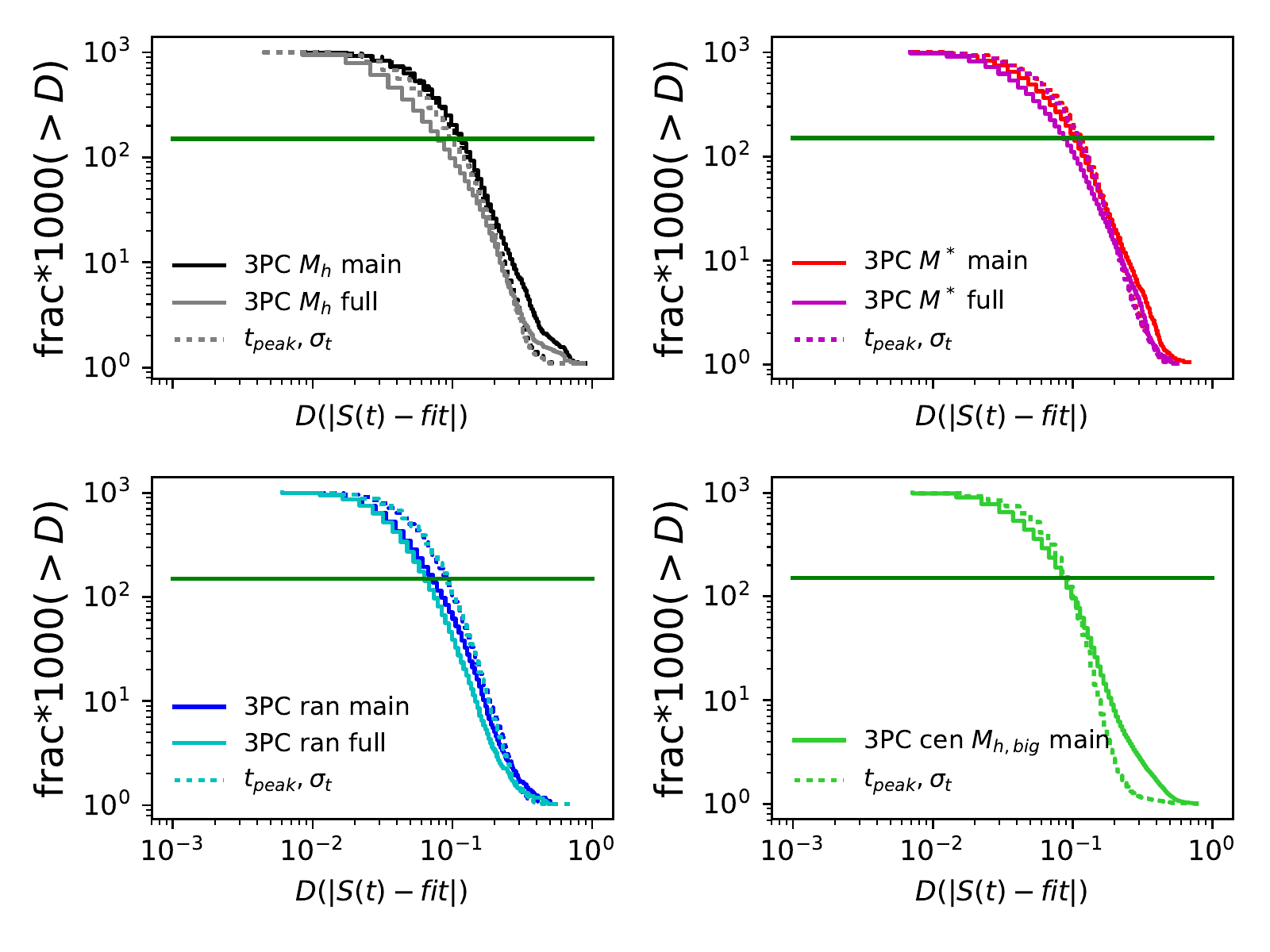}}
\end{center}
\caption{Accuracy of approximating normalized integrated
  histories ${\cal S}(t)$ by first three PCA
  coefficients $a_0,a_1,a_2$ (solid
  line) or by 2 parameters of the
  lognormal fit, e.g.,$t_{\rm peak},\sigma_t$ (dotted line). 
Top: the cumulative fraction of galaxies with a square of the separation between the approximate
and simulated ${\cal S}(t)$ above a given value, times 1000.  Bottom:
1000 times the cumulative number of galaxies above a separation
measure $D$, defined in Eq.~\ref{eq:goodfitcrit},
along with a line marking 15\% of the galaxies. In \citet{Die17}, for
Illustris, the 15\% line crosses the distribution at a lower $D$, $D=0.05$.  In both, the
best fits are for the {\tt ran} samples.
}
\label{fig:dsep}
\end{figure}

Two measures of the goodness of fit to ${\cal S}(t)$, the squared ``distance''
$d^2 =|fit-{\cal S}(t)|^2$ and the goodness of fit criterion $D$ in
Eq.~\ref{eq:goodfitcrit}, roughly the maximum spacing between the
history and the fit, are shown in
Fig.~\ref{fig:dsep}.   Solid lines are the PCA
approximation, dependent upon $a_0,a_1,a_2$, and dashed lines are the lognormal
fit, dependent upon $t_{\rm peak}$ and $\sigma_t$.   (As ${\cal S}(t)$ is
used here,
the parameters $\tilde{\cal S}(t_f)$ and $A$ drop out.  Scaling out
these factors is automatic in $D$, and for $d^2$ it prevents the high
mass galaxies from swamping the signals as well as making intercomparisons
more difficult. ) 
The two methods give roughly the same quality of fit by these
measures, sample by sample.  The {\tt ran} samples have the best fits.  Relative to the other samples, the {\tt ran} samples also have more satellites (noted to have
worse lognormal fits in \citet{Die17}),  but fewer
high mass halos.\footnote{The distance $d^2$ 
is minimized to calculate $t_{\rm peak}, \sigma_t$.  The
parameter $D$ was introduced in part to undo the
cumulative effects of using the integrated star formation rate rather
than the star formation rate itself.  (However, one can also take the
integrated star formation rate as the quantity of choice for
considering the history, and then use $d^2$ alone.) }  The $d^2$ for
the two approximations are correlated, good or bad fits tend to occur
together.
Many of the $D>0.05$ fits can be seen by eye to be due to rejuvenating
histories, where two bouts of star formation occur, separated by a
period of quiescence.
\citet{Die17}
found 15 \% of the Illustris galaxies had $D>0.05$, the 15\% line
horizontal crosses each 
sample's distribution in Fig.~\ref{fig:dsep} at larger values, that is, the fits are worse for
these samples than for Illustris.\footnote{Aside from many of the samples having different
  compositions relative to Illustris, the number of steps in the
  histories may have also contributed to the difference in goodness of
  fit, as the Illustris simulation has about twice as many outputs
  over the same period covered by the Millennium outputs, with equal
  spacing in time rather than in scale factor.}

\begin{figure}
\begin{center}
\resizebox{3.3in}{!}{\includegraphics{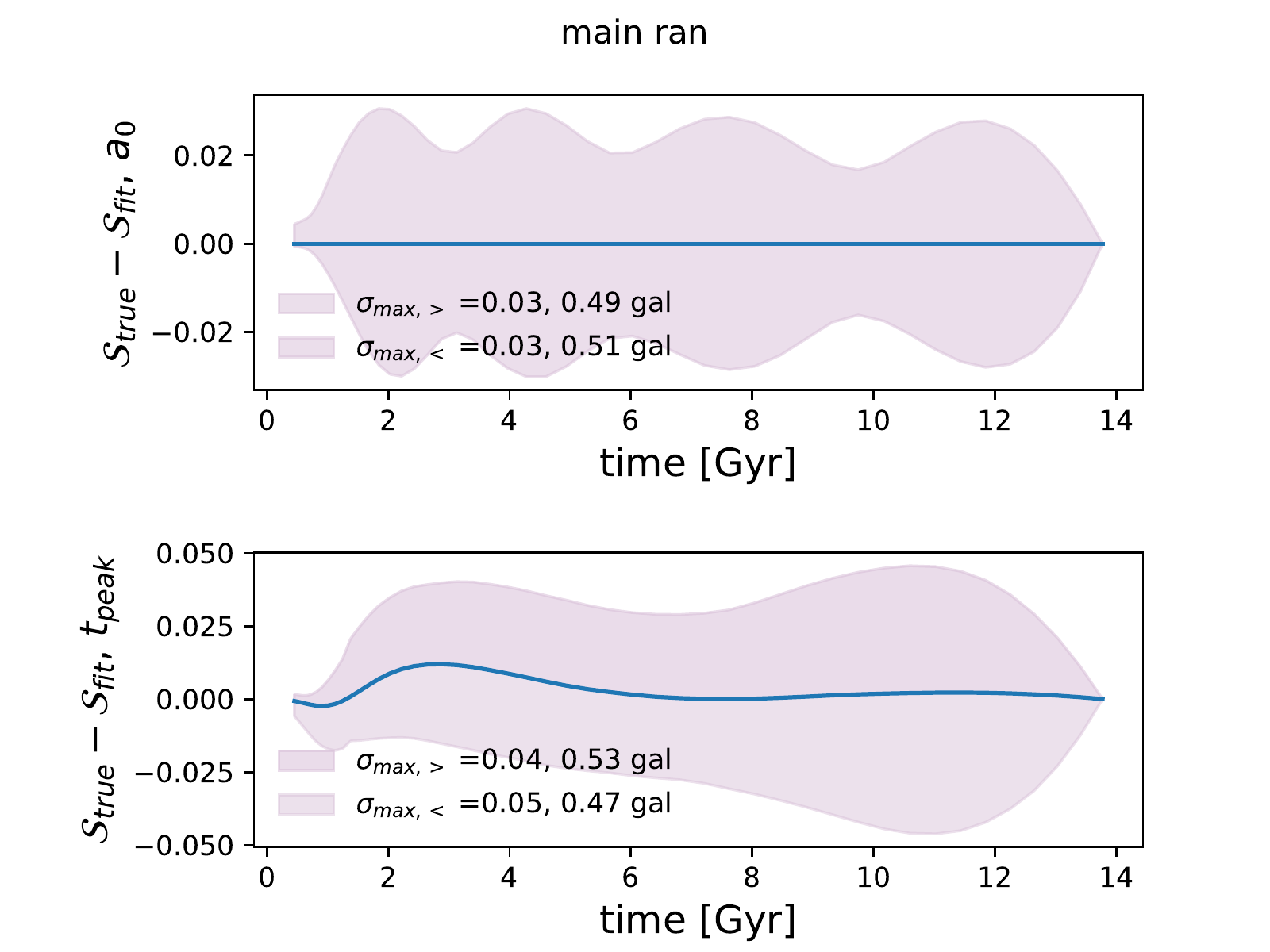}}
\end{center}
\caption{Average difference of the approximations and the true (simulation) normalized histories $\langle {\cal S}_{\rm true}(t) -{\cal
  S}_{\rm fit}(t)\rangle$ and scatter, as a function of history time step.
The top panel corresponds to the
PCA approximation, and the bottom panel to the lognormal approximation, in
the main {\tt ran} sample.
The shaded regions are one standard deviation, calculated separately for galaxies with
${\cal S}_{\rm true}(t) -{\cal S}_{\rm fit}(t) $ above or below 
the average.
The total
fractions of galaxies above and below the average, over all 48 time
steps, are given by the numbers shown, as well the maximum
standard deviation (``var max'')  in each direction.  See text for more discussion.
}
\label{fig:sfrsigdist}
\end{figure} 
To get a sense of how the histories deviate from their fits in more detail,
the average of
 ${\cal S}_{\rm true}(t)-{\cal S}_{\rm fit}(t)$ is shown for the main {\tt ran} sample in
Fig.~\ref{fig:sfrsigdist}.  The  full {\tt ran} sample is similar.   
This average is zero for the PCA
approximation by construction, but slightly nonzero for the lognormal
fit.
The shaded regions are the
standard deviations (calculated for top and bottom
separately) for each time step.  These
are up to 5\% of the final value
(which is 1) for this sample.  The PCA approximation error
is the
sum of the neglected principal components in Eq.~\ref{eq:pcdecomp},
its standard deviation
generally has an oscillatory envelope, and the envelope is $\leq 0.05$ across samples,
compared to
$\leq 0.07$ for the lognormal fit (and in every sample the deviation for
PCA was $\leq$ that for the lognormal - 0.02).

\subsubsection{Approximating final time star formation
  rate}
\label{sec:finaltime}
\begin{figure}
\begin{center}
\resizebox{3.3in}{!}{\includegraphics{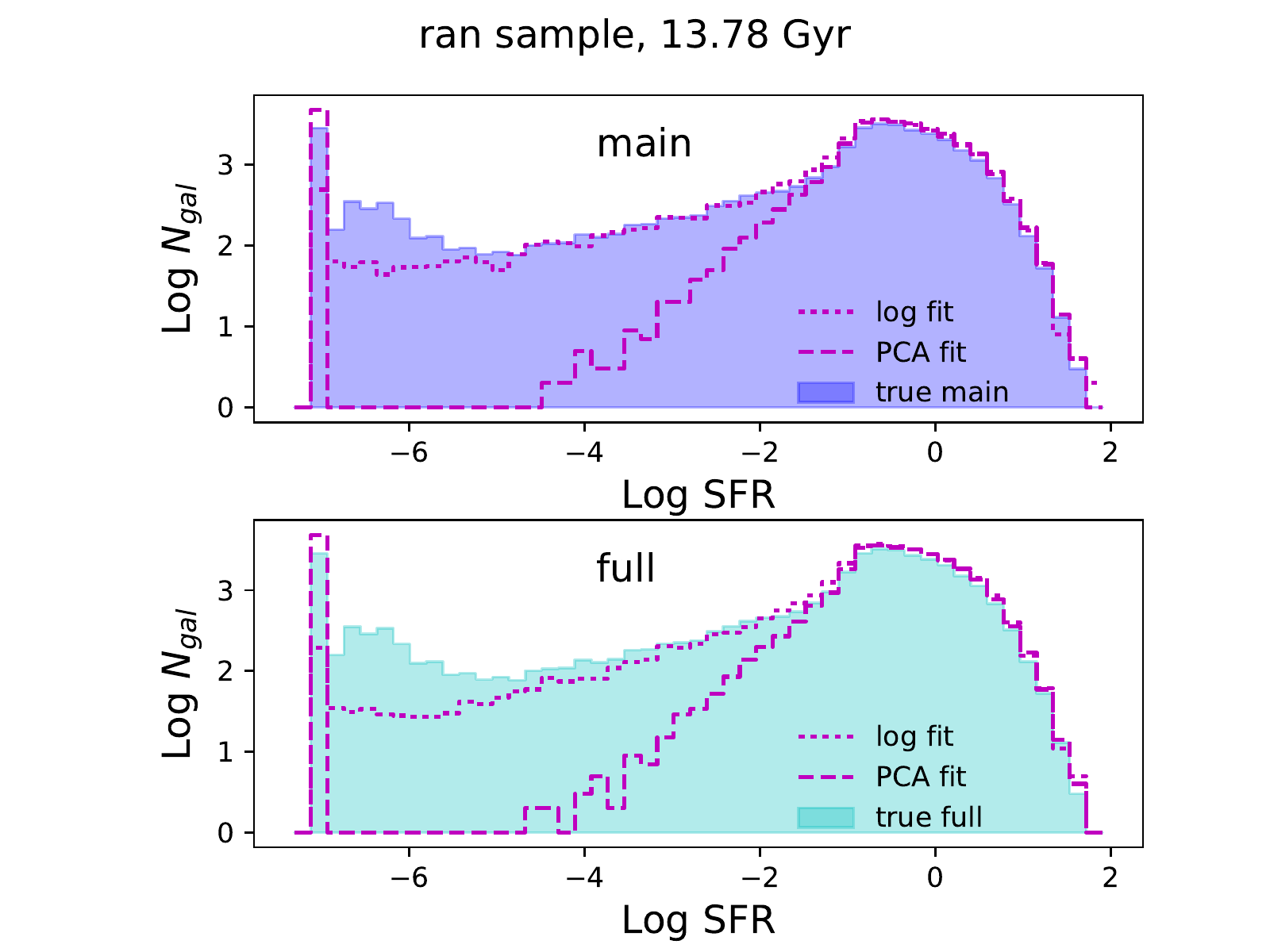}}
\end{center}
\caption{Final time SFR distribution in the main (top) and
full (bottom) {\tt ran} samples.  The true (simulation) values are given by the shaded
histogram.  Galaxies with SFR $<10^{-7}M_\odot yr^{-1}$ are assigned SFR
$10^{-7}M_\odot yr^{-1}$.  Dots show the lognormal fit and dashed show the
PCA fit.  Starting at the highest star formation
rates, the number of galaxies rises as the star formation rate
decreases, and then flattens out, eventually decreasing more, and  ending with a peak at the minimum
star formation rate.  
The other samples, with more high mass
halos, have worse fits at higher star formation rates.  For these, again
starting with the highest star formation rates, the number of galaxies
tends to continue to increase beyond (below) star formation rates
where the simulation number of galaxies 
flattens, giving an excess.
}
\label{fig:sfrhist}
\end{figure} 
\begin{figure*}
\begin{center}
\resizebox{6.2in}{!}{\includegraphics{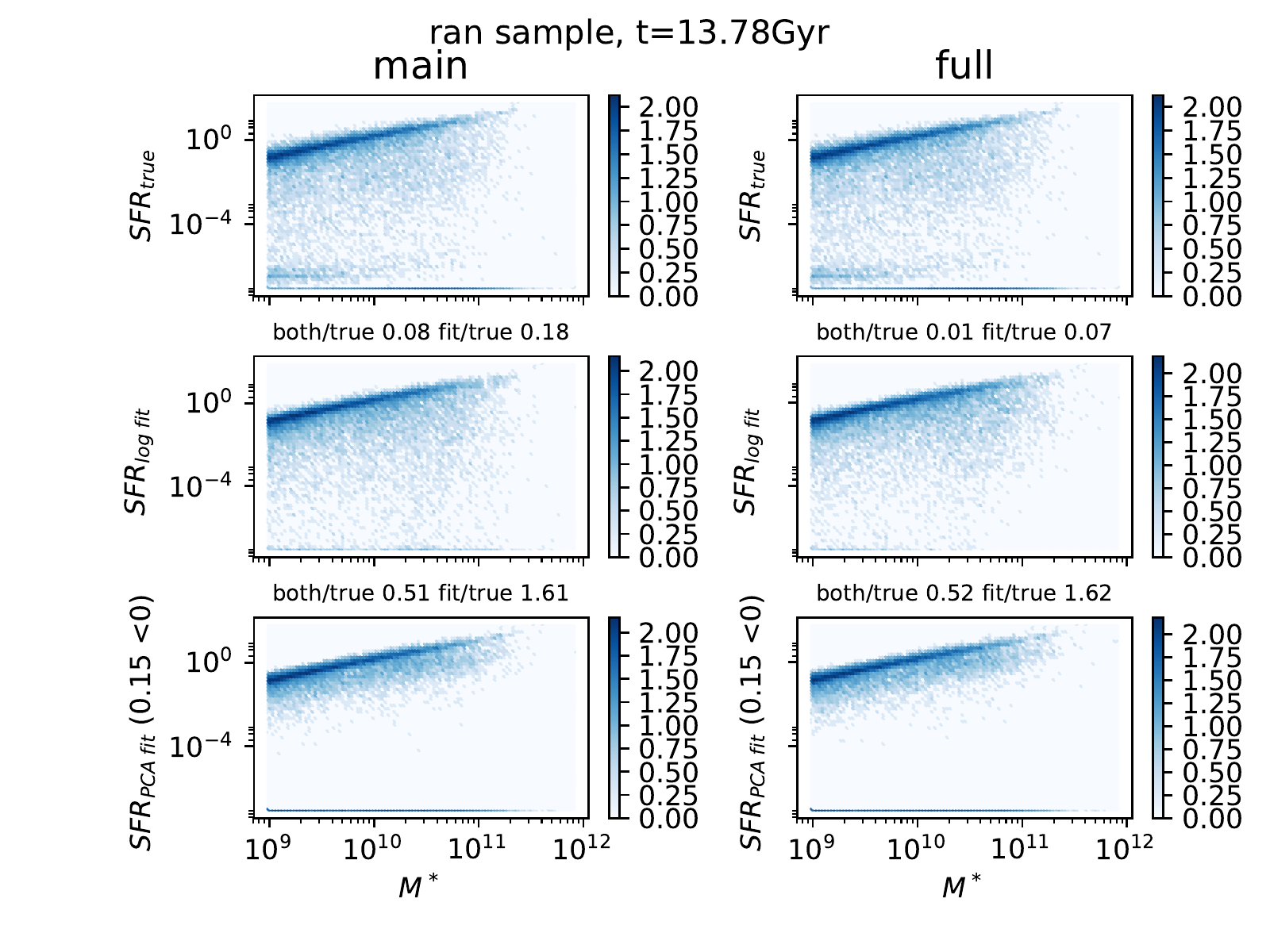}}
\end{center}
\caption{Stellar mass (in units of $M_\odot$) vs. star formation rate
  (in units of $M_\odot$ yr${}^{-1}$) at final time
  in the simulation (top panel, same at left and right), as compared
  to the
  lognormal (middle) and PCA (bottom) approximations. 
  Main histories are at left, to the full histories at right.
The star forming main sequence is visible at top
 ($\log N_{\rm
    gal}$ per pixel shown in the bars at right). 
Galaxies with $SFR< 10^{-7} M_\odot yr^{-1}$ are assigned the minimum 
$SFR=10^{-7} M_\odot yr^{-1}$, including the
15\% of PCA fit galaxies with $SFR<0$ in this sample.  
The ``green valley'' between the minimum $SFR$ and the star forming
main sequence has fewer galaxies in the PCA fit, compared to the
lognormal fit or simulation.  ``both'', ``true'', and ``fit''  refer to the number of
galaxies with minimum SFR common to the simulation and fit, in the
simulation, or in the fit, respectively, with ratios as shown.
Relative to the simulation, in the {\tt ran} sample more galaxies are quiescent in
the PCA fit (fit/true $>1$) and fewer are quiescent in the lognormal
fit (fit/true $\ll$ 1); these numbers vary with sample.
}
\label{fig:mstarsfr}
\end{figure*}

One can also step back and 
compare the fits to the histories to the simulation at a fixed time, for
instance at $z=0$.
The final star formation rate distribution is shown
in Fig.~\ref{fig:sfrhist} for the {\tt ran} sample.   
The shaded region is the final time star formation rate distribution
of the simulation,\footnote{The PCA construction gives the prediction for ${\cal
    S}(t)$; SFR$(t_f)$ is approximated as $(\tilde{\cal S}(t_f)-\tilde{\cal
    S}(t_j))/(t_f-t_j)$ where $t_f-t_j$=0.36 Gyr.  This is also a 
  possible approximation for SFR$(t_j)$, 
 which has a similar distribution, although there is scatter between
 the two.}  
 identical for the main (top) and full (bottom)
samples.  Any rates $< 10^{-7} M_\odot yr^{-1}$, including possible
negative ones from the PCA construction, are set to $10^{-7} M_\odot
yr^{-1}$.  
The lognormal fit has many more galaxies in the green valley,
closer to the shape of the star formation rate distribution in the
simulation.  In contrast, there are more galaxies with the
minimum star formation rate in the PCA fits; their number then
drops precipitously
in the green valley.  Adding more principal components can increase the number of galaxies lying in the green valley,
but even using 38 components, i.e. including up to $a_{37}PC_{37}$, did
not reach the approximate agreement at final time in the green valley found by using
the lognormal fits.\footnote{I thank M. Sparre for suggesting this
  test.  At the final time, successive principal components
 to the star formation roughly oscillate.  Lying
in the green valley requires a close but not exact cancellation
between these successive terms, which might explain why so many terms are
required.  This green valley gap also appears if one uses the PCA fit to the
instantaneous star formation rate, dividing first by the final stellar
mass.}
In the other samples, with more high mass galaxies, the agreement
between the fit at higher star formation rates is worse.   An excess
appears for these other samples at the higher star formation rates,
which persists to lower star formation rates for the lognormal fit.

There is slightly different information in the instantaneous stellar
mass-star formation rate diagram, with the star formation rates again
calculated 
from the fits to the integrated histories.
This relation is shown in Fig.~\ref{fig:mstarsfr}.
The top two 
panels are the simulation, which is identical on the left and right,
again because this is the final time.
Below are the final time star formation rates based upon fits to the
the main (left)
 and full (right) integrated
star formation rate histories.  The middle panel is the lognormal fit, the bottom
panel the fit from PCA. 
(The fraction of galaxies with negative rates in the PCA fit
is listed on the y-axis for the lower 2 panels.)
In the simulation and both fits,
a star forming main sequence is evident, but in the PCA fit, the absence of
galaxies in the ``green valley'' between star forming and
quiescent is again noticeable.  The numbers of galaxies with the minimum star
formation rate
$\leq 10^{-7} M_\odot yr^{-1}$ are compared in the simulation and the
fits.  Those which are
common to both the fit and the simulation (``both''), and those
present in the fit (``fit'')  are divided by the number in
the simulation (``true'').  The
lognormal fit lacks some of these quiescent galaxies, while
the PCA fit has too many, relative to the simulation.  For other
samples,``fit/true''$<1$ for all of the lognormal fits, and for the
$M_h$ and cen $M_{h,{\rm big}}$ PCA fits.

To summarize, the galaxy histories were approximated with two
fits. The lognormal description assigns each 
history a peak, which can move in time, and a width of lognormal
shape, while the PCA approximation treats all histories in a sample as
the sum of the same
average history plus perturbations with fixed position and shape,
derived from the sample, with the perturbation coefficients changing
for different galaxies.
The PCA approximation, which normalizes the integrated star formation
rate histories before expanding them, has similar averages and basis fluctuations for
the full and main histories.  The
PCA normalization factors, i.e., the final
integrated star formation rates, differ the most between full and main
histories for galaxies with higher final $M_h$.  In the lognormal fit,
the peak time is slightly earlier for the full samples, and the width
slightly larger.  The
lognormal and PCA approximations have correlated leading parameters and
give similar ``distances'' (as shown in Fig.~\ref{fig:dsep})  from the
simulated histories, using two estimates of goodness of fit.

One use of these fits is to compare their parameters to final time galaxy
properties and histories of other galaxy properties, 
explored next.

\section{Galaxy star formation rate fits compared to other galaxy properties}
\label{sec:otherprops}
With the parameterizations based upon a lognormal fit or PCA in hand,
their relation to
other galaxy properties can be explored, such as the observable final
time $M^*$ and
$SFR$, the in principle observable final time $M_h$, and
properties of main histories for halo mass and stellar mass.
Recall that the full (rather than main) histories of galaxy
halos and other dark matter properties are combined 
with the full semi-analytic model to create
the detailed star formation rate histories in the first place.
Both correlations and machine learning can be used to analyze these
relations.
\newpage
\begin{figure} %
\begin{center}
\resizebox{3.3in}{!}{\includegraphics{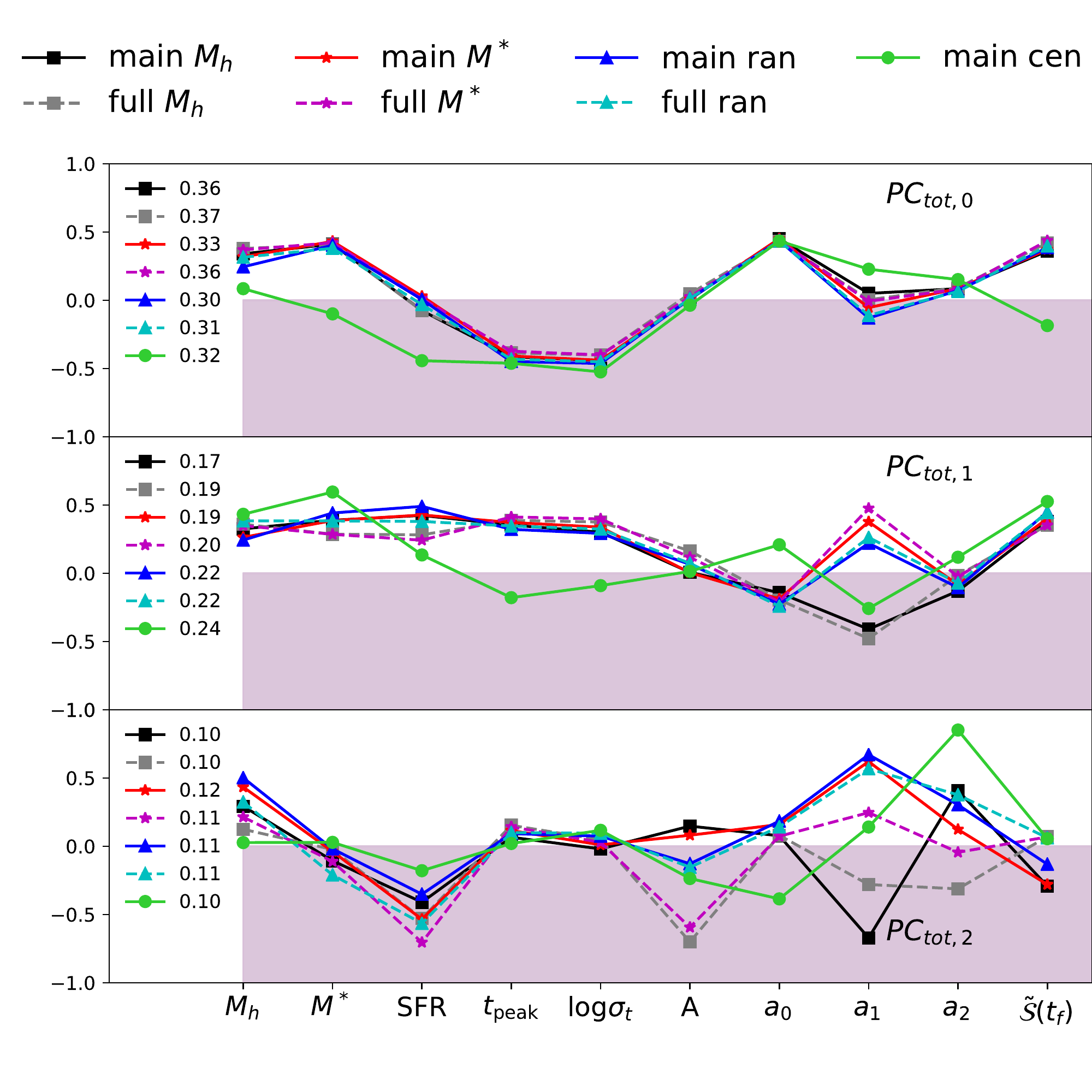}}
\end{center}
\caption{First 3 principal components for joint changes in final
  $M_h$, final $M^*$, final SFR, lognormal fit parameters
  $t_{\rm peak}$, $\log \sigma_t$, $A$, and PCA fit parameters $a_0$, $a_1$,$a_2$,
  $\tilde{\cal S}(t_f)$.  These
  three vectors are the combinations of variations which dominate the normalized scatter
  of the 10$\times$ 10 correlation matrix of each sample.  The
  fraction of the scatter in each $PC_{{\rm tot},n}(t)$ is shown at left, for
  each sample.
  The shading shows where zero correlation lies.  The leading contribution to
  the scatters, $PC_{{\rm tot},0}(t)$, shows that  final $M^*$, final $M_h$ and $a_0$ tend
  to fluctuate opposite to $t_{\rm peak}, \sigma_t$.  The subleading
  contributions show  $a_1$ as related to final $M_h$, final  $M^*$, and, in the
  next leading combination, having sample dependent relations.  The
  different samples have different fractions of high final $M_h$ or
  $M^*$ galaxies, Fig.~\ref{fig:massdists}.  }%
\label{fig:pcaeverything} 
\end{figure}

\subsection{Correlations}

Because there are correlations between
galaxy histories and galaxy final properties, for example, more
massive galaxy halos tend to have galaxies which quenched earlier, 
correlations are expected between $a_0$, $t_{\rm peak}$ and final time
 $M_h,
M^*$ and $SFR$.   The halo mass
$M_h$ refers to the host (sub)halo of a galaxy (``mvir'' in the
Millennium simulation, i.e., its $M_{200c}$ mass when it was last a
central galaxy).

Quantities which are expected to be related to each other include
final $M_h$, final $M^*$, final SFR, and parameters from both fits: 
  $t_{\rm peak}$, $\log \sigma_t$, $A$, $a_0, a_1, a_2$ and
  $\tilde{\cal S}(t_f)$.  Although the
  $a_i$ are uncorrelated with each other by construction, the rest of the parameters
  tend to be correlated with each other (e.g. the $\sigma_t(t_{\rm
    peak})$ relation, or the $M^*(M_h)$ relation in the final time observables).
The list of cross correlations between all 10
  components, for 7 samples, is unwieldy.  Here, to get some idea of
  how variations are related, the combinations of the correlations
  between all of these quantities which dominate the (normalized)
  scatter are found, i.e., PCA, but for the correlation matrix.  Correlations are used because of the wide 
  ranges of the different quantities.
 Instead of the first 3
combinations of variations comprising $\geq$95\% of the scatter, such as for the
integrated star formation rate histories,
here the first 3 combinations, shown in Fig.~\ref{fig:pcaeverything} capture 
  $\sim $2/3 of the (normalized) scatter.  However, the leading
  combinations do
  capture a large amount of scatter and can be used to look at trends. 
The fraction of (normalized) variance captured in each $PC_{{\rm
  tot},n}(t)$ is listed to the left for $PC_{{\rm tot},0}(t)$ (top), $PC_{{\rm
  tot},1}(t)$ (middle), and $PC_{{\rm tot},2}(t)$ (bottom) .  
For example, 
$PC_{{\rm tot},0}(t)$, with $\sim 1/3$ of the normalized scatter, has the parameters
$t_{\rm peak},\sigma_t$ change in the same direction and by
similar amounts
(as expected due to their mean relation), and in opposition to
$a_0$, with the expected relations to halo mass and stellar mass also
visible.  That is, high $M^*$ or high $M_h$ is associated with low
$t_{\rm peak}$, which is the
familiar relation of high stellar mass galaxies forming
stars earlier.  Differences between samples can
be seen, which might in particular indicate some mass dependence (the $M_h$, $M^*$
samples have more high $M_h$ or $M^*$ galaxies, see Fig.~\ref{fig:massdists}).  

There are also, of course, correlations of the integrated star
formation rate parameters 
with the main $M_h$ or $M^*$ histories, as all these histories are related.  There are
several ways to compare the $M_h$ histories, $M_h(t)$, to the star formation rate
history parameterizations.  (The notation $M_h(t)$ refers to the main
halo history, analogous to the main star formation rate history,
i.e. the red line in Fig.~\ref{fig:histtypes}.)
For the lognormal fit,
\citet{Die17} compared $t_{\rm peak},\sigma_t$
from the Illustris simulation with
parameters for halo history fits of
\citet{Wec02,Tas04,McBFakMa09}, using only the times the galaxies were
in halos, rather than as satellites in subhalos. 
They found trends\footnote{using $z_{\rm wechsler}= \ln 2/\alpha$ for
a fit to halo histories of $M_h(z)= M_0 e^{-\alpha z}$\citep{Wec02}}
for $t_{\rm peak}$ with halo formation redshift $z_{\rm wechsler}$ for
$t_{\rm peak}$, in 3 different stellar mass bins.  (They noted their
scaling $\sigma_t \sim 0.83 t_{\rm peak}^{3/2}$ did not seem to arise
naturally from the analytic \citep{Dek13} mass accretion rate for a
halo.)\footnote{ \citet{Die17} also
measured correlations between the lognormal fit parameters and other galaxy
quantities, including final $M^*$ (which can be traded for
another parameter
in the fit), maximum halo mass, $z=3$
environment, halo age (using 2 measures), black hole mass, and size.}
For the samples here, the correlation of $t_{\rm peak}$ with half mass redshift was
highest for the samples with the latest average $t_{\rm peak}$, $\sim$
-50\% 
for {\tt ran}, dropping to magnitude $<10\%$ for the $M_h$ samples.
Correlations were similar, with opposite sign, for $a_0$ and half
mass redshift. 
Using PCA for $M_h(t)$, normalized to end at 1 (unlike the
integrated star formation rate history, $M_h(t)$ does not have to be
monotonic), correlations between $a_{0,h}$ and $a_0$ and $t_{\rm
  peak}$ were small (below 20\%) for all samples except the {\tt ran}
samples, where they were
$\sim \pm$30\%.\footnote{
Halo histories $M_h(t)$ were analyzed via PCA in
\citet{WonTay12}, and (sub) halo main histories were
compared to stellar mass histories in
\citet{CohvdV15}. \citet{WonTay12} found that the
largest principal component for halo histories was most closely
correlated with concentration. Instead of dividing by
the final halo mass, \citet{WonTay12} set the mean of each history to zero and the
variance to 1 and then did PCA, i.e., on correlations.}

A close relation is expected between the integrated star formation
rate and the main stellar mass history, as $M^*(t)$
is the sum of stellar mass formed
within the galaxy (the ``main'' integrated star formation rate) plus
contributions from mergers, stripping by and of other galaxies, and
ageing 
(instantaneously applied in the semi-analytic models).\footnote{The
full stellar mass histories, not considered here, would include another degree of
computational complexity and should be extremely close to the integrated full star
formation rate history.}  Correlations between
$a_0$ and its stellar mass history counterpart $a_{0,*}$ are $\sim
76-97$\%, with the random sample having the largest
correlation (for samples with both main and full integrated star formation rate,
both are similarly correlated with
$a_{0,*}$).
The fluctuation $PC_0$ in the integrated star formation rate is
associated with 
more of the scatter than its counterpart in the stellar mass histories.
\footnote{For stellar mass history PCA, \citet{CohvdV15} found that galaxies sharing
approximately the same final stellar mass ($z=0$) were 
well characterized, $\gtrsim$ 90\% of variance, by their average values plus their first
3 $PC_n(t)$ fluctuations.}

\subsection{Halo mass at $t_{\rm peak}$: $M_{h,{\rm peak}}$}
\begin{figure} %
\begin{center}
\resizebox{3.3in}{!}{\includegraphics{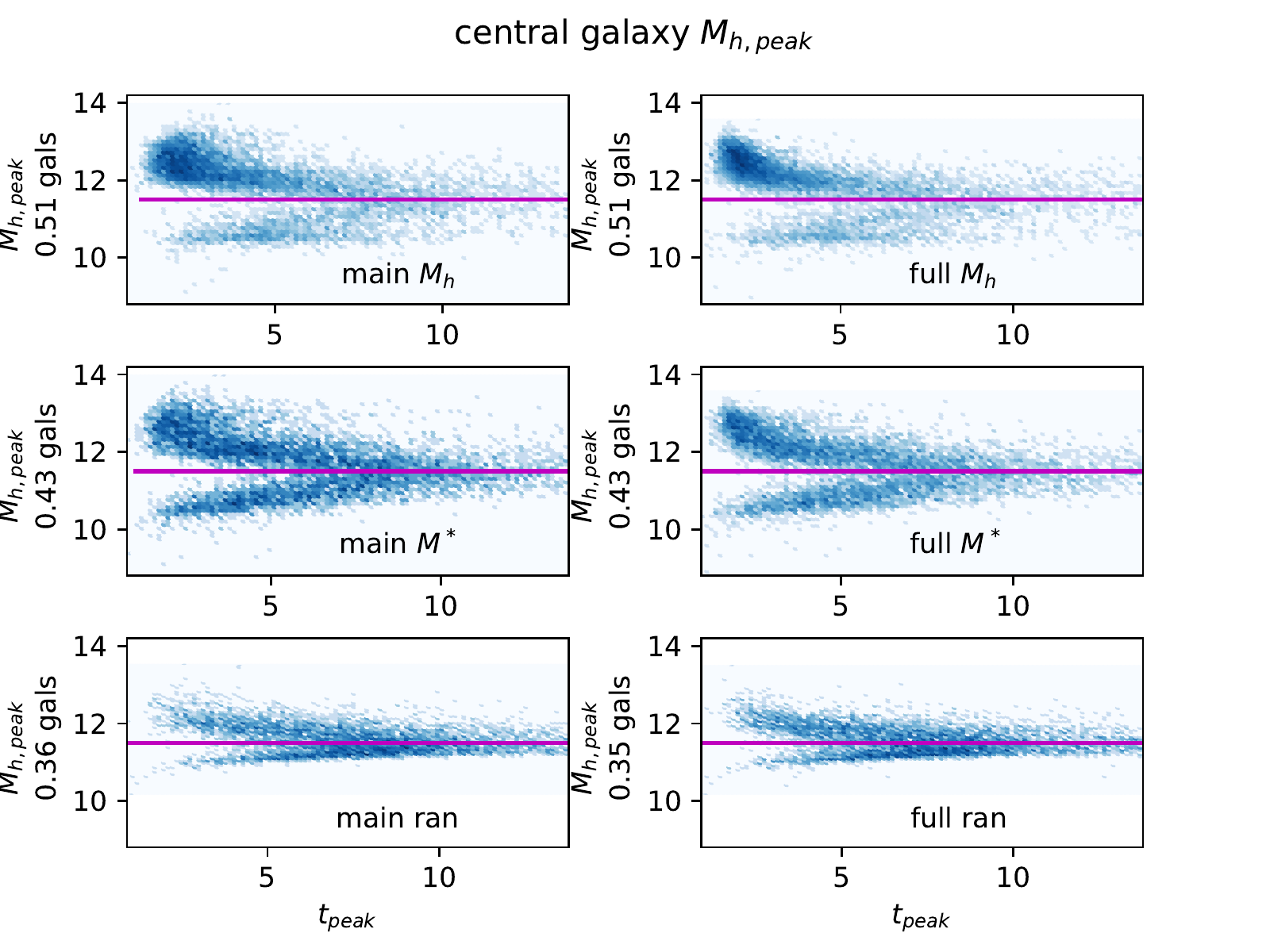}}
\resizebox{3.3in}{!}{\includegraphics{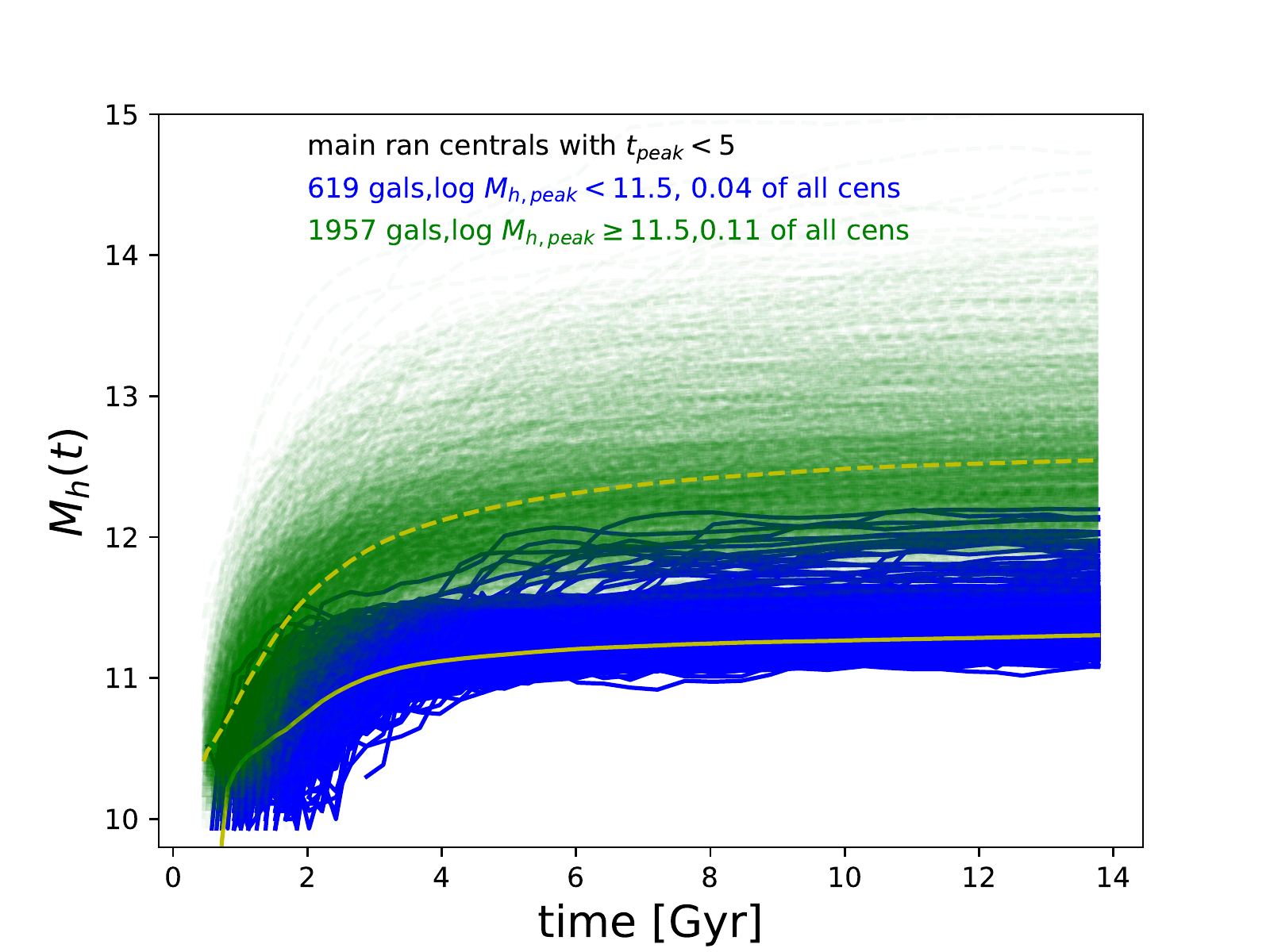}}
\end{center}
\caption{Top:  central galaxy main halo mass $M_h$ at peak time $t_{\rm peak}$, $M_{h,{\rm
      peak}}$, as a function of
  $t_{\rm peak}$.  The fraction of galaxies in each sample which are
  central at all times and
  have 
  $t_{\rm peak}$ in the past is noted at left for each panel.  Samples
  are as listed
  for $M_h,M^*$ and {\tt ran}, top to bottom, with
  main integrated star formation rate histories at left,
   full integrated star formation rate histories at right.  The magenta
  horizontal line at $\log M_{h,{\rm peak}}/M_\odot = 11.5$
divides these galaxies into two groups.    
For the majority of central galaxies, which are above the
horizontal magneta line, $M_{h,{\rm peak}}$
seems to decrease with increasing $t_{\rm peak}$. 
Bottom:  In blue, the halo
  histories for the 619 central galaxies (of 17278 total central
  galaxies)  below the dividing line shown in panels above,
  i.e. with  $M_{h,{\rm peak}} < 10^{11.5} M_\odot$, 
  and with $t_{\rm peak}<5$ Gyr. 
Unlike most other halo
  histories for galaxies with $t_{\rm peak}<5$Gyr, shown as green lines, the low
  $M_{h,{\rm peak}}$ galaxy halo masses
  seem to stall at low values.  The averages of the two samples,
  including only nonzero histories at each step, are
  shown by the solid yellow and dashed yellow lines.}
\label{fig:mhtp} 
\end{figure}

In the lognormal fit,
another way of comparing the halo history to $t_{\rm peak}$ is to
consider $M_h(t_{\rm peak}) \equiv M_{h, {\rm
    peak}}$.  This characteristic mass at peak star formation rate is shown in
in Fig.~\ref{fig:mhtp} for all central galaxies in the $M_h$, $M^*$
and {\tt ran} samples.  This mass is only available for galaxies with
$t_{\rm peak}$ in the past, and satellites are excluded because their
$t_{\rm peak}$ is expected to also depend on their time of infall into
a larger halo.  Only galaxies which have been central at
all times are counted as central.

In the top figures, for the $M_h,M^*$ and {\tt ran} samples, main (left) and
full (right), 
a bimodal feature is evident.  This is clearest at
low $t_{\rm peak}<5$ Gyr, and is highlighted by the separating line at $M_h =
10^{11.5}M_\odot$.  (The cen $M_{h,{\rm big}}$ sample by construction has no galaxies
  below $10^{12}M_\odot$ at final times, and so is not shown.)  
Galaxies with low $M_{h,{\rm peak}}$ are a small fraction.  Those with
with $M_{h,{\rm peak}}<10^{11.5}M_\odot$ and $t_{\rm peak}<5$ Gyr
comprise (main and full) 7\%,
11\%, and 4\% respectively of the
central galaxies for the $M_h$, $M^*$ and {\tt ran} samples. 
About half of these low $M_{h,{\rm tpeak}}$ galaxies have relatively poor fits
(with $D>0.1$; distributions of $D$ for the full samples are shown in
Fig.~\ref{fig:dsep}).

 These galaxies not only quench at a lower halo
mass, but often have their halo masses remaining low afterwards.
This can be seen in the bottom of Fig.~\ref{fig:mhtp}.\footnote{Not all galaxies with low $M_{h,{\rm
      peak}}$ ``stall''.
 In the main ($M_h, M^*$, {\tt ran})
samples, a very small fraction (2\%, $<$1\%, 4\%) of these stalled galaxies
surpass $M_h = 10^{11.75}M_\odot$, 
in the full samples, these fractions are
(3\%, 1\%, 8\% ) respectively.}
For the main {\tt ran} sample, these low $M_{h,{\rm peak}}$ galaxy 
halo histories over time are shown as blue lines.
 The other central galaxy histories with $t_{\rm peak}<5$ Gyr are
 shown as the 
green shaded lines, and tend to reach much higher
halo masses over time.  
It would be interesting to
  find out more about this sample of galaxies, and whether this split
  in $M_{h,{\rm peak}}$ arises in other models or can be tested
  observationally.  
In simplified models based on halo histories, 
stalling of halo growth or
  hitting a specific host halo mass are often used as criteria to determine when
  star formation quenches (e.g., \citet{HeaWat13}). However, the
  reason 
  for stalling is not clear; it would be interesting to pursue this further.  The
  trend of lower $M_{h,{\rm peak}}$ with increasing $t_{\rm
    peak}$, for galaxies which have not ``stalled'' seems to reflect the known trend of downsizing.\footnote{I
    thank B. Diemer for pointing this out.}


\subsection{Machine learning}
\label{sec:ml}
\begin{figure*} 
\begin{center}
\resizebox{3.3in}{!}{\includegraphics{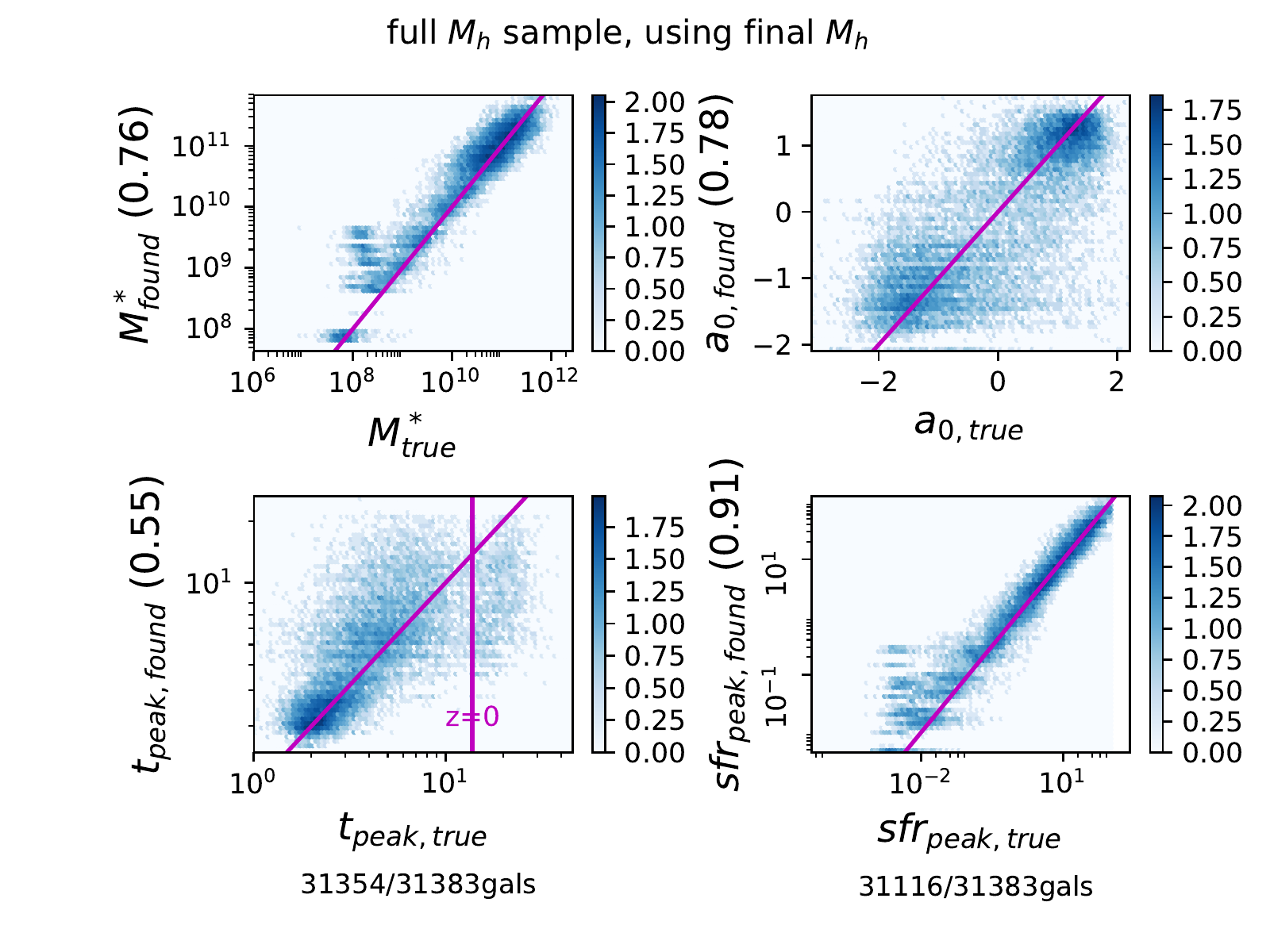}}
\resizebox{3.3in}{!}{\includegraphics{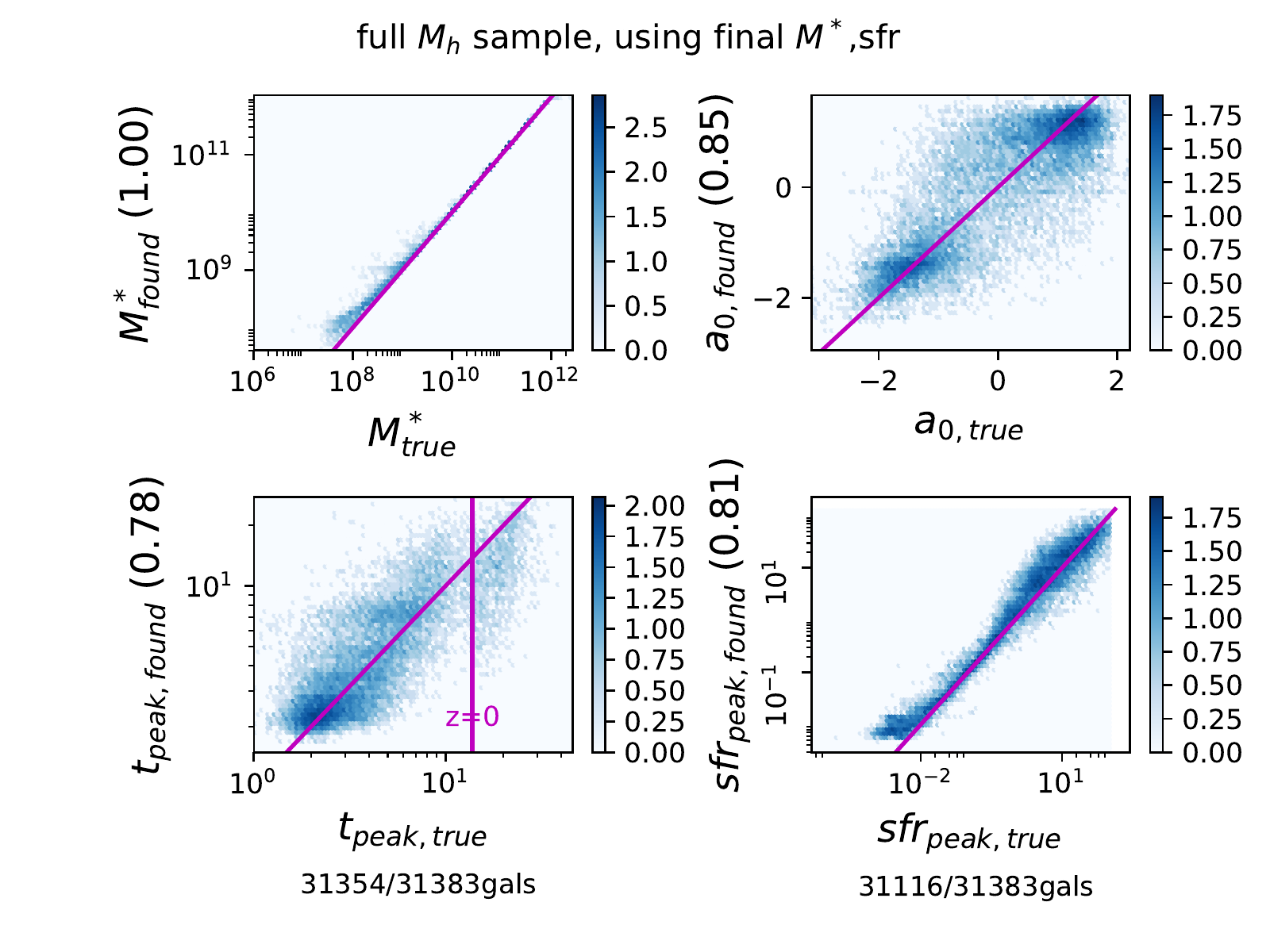}}
\resizebox{3.3in}{!}{\includegraphics{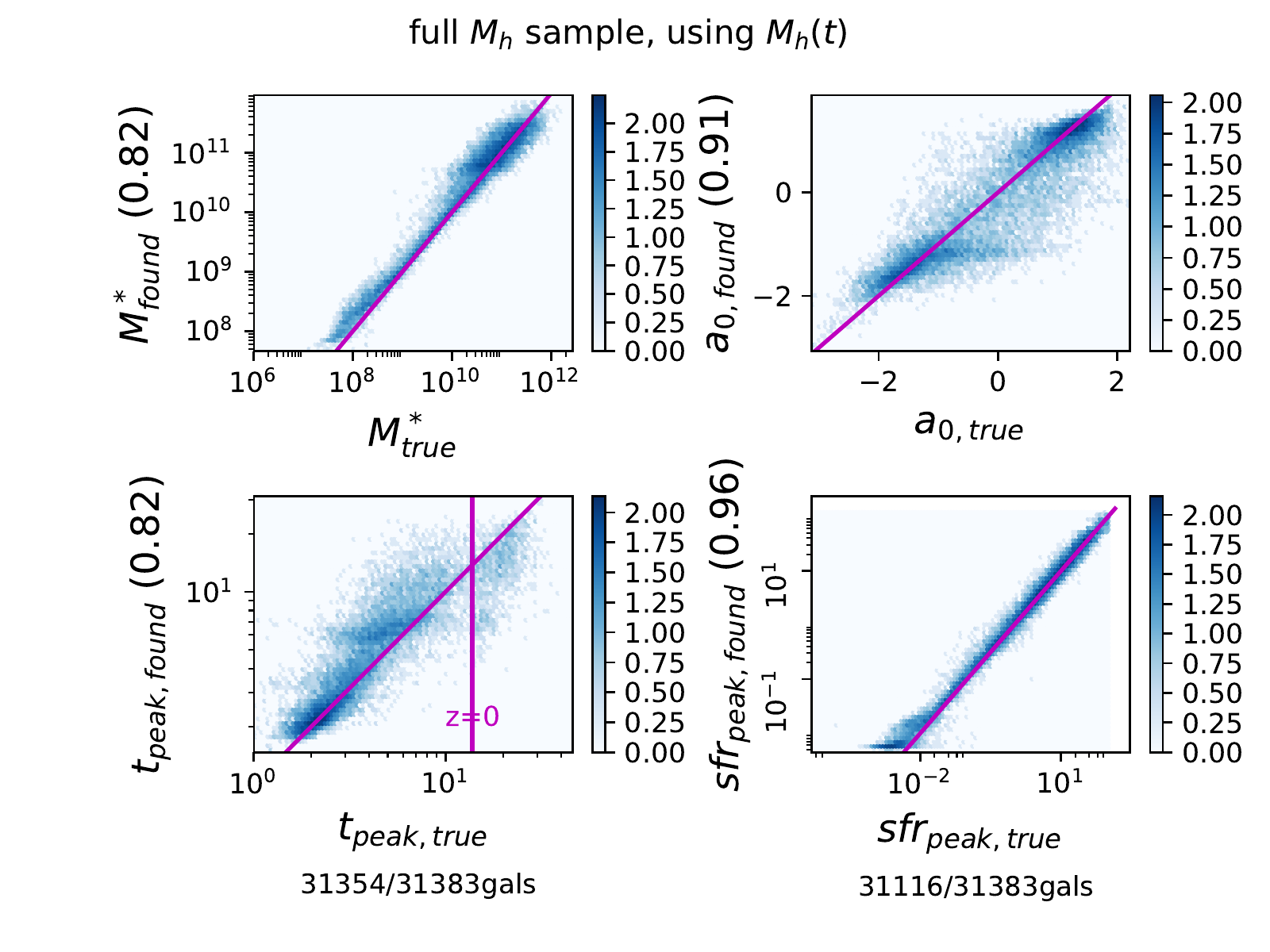}}
\resizebox{3.3in}{!}{\includegraphics{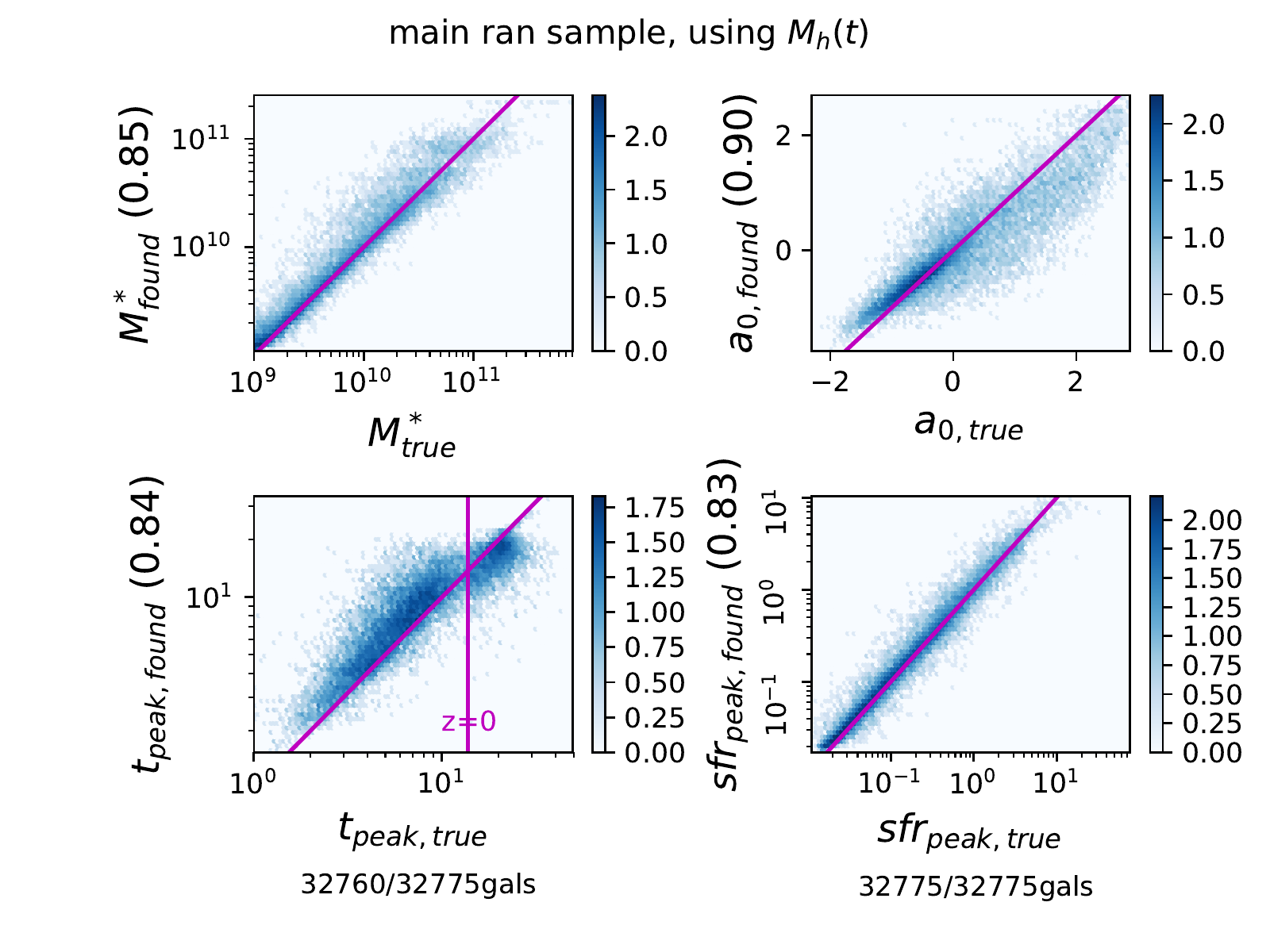}}
\end{center}
\caption{Found and true galaxy properties as
  listed for the full $M_h$
  sample (top and lower left) and the main {\tt ran} sample (lower
  right), using machine learning.  Correlations between found and true properties are listed
  in parentheses.  Training set galaxies were 1/10 of the sample, trimmed by
  requiring a good $t_{\rm peak}$ fit, and are included
  in the plot and correlation.  Different inputs were explored to
  obtain final galaxy properties: final
 $M_h$ (upper left), final $SFR,M^*$ (upper right), and the main halo history
 $M_h(t)$ for the lower two panels. 
  The color scale shows the
  log of the number of galaxies in each pixel.  Galaxies with
  $t_{\rm peak,true}>100 Gyr$  and $SFR_{\rm peak,true}>100$ 
were excluded for comparisons between found and true
$t_{\rm peak}$ and 
  $SFR_{\rm peak}$ respectively, fractions of galaxies used are given.
  The fit in Eq.\ref{eq:logfit} was used to
  calculate the ``true'' $SFR_{\rm peak,true}$.
}
\label{fig:truefound} 
\end{figure*}
\begin{figure*} 
\begin{center}
\resizebox{3.3in}{!}{\includegraphics{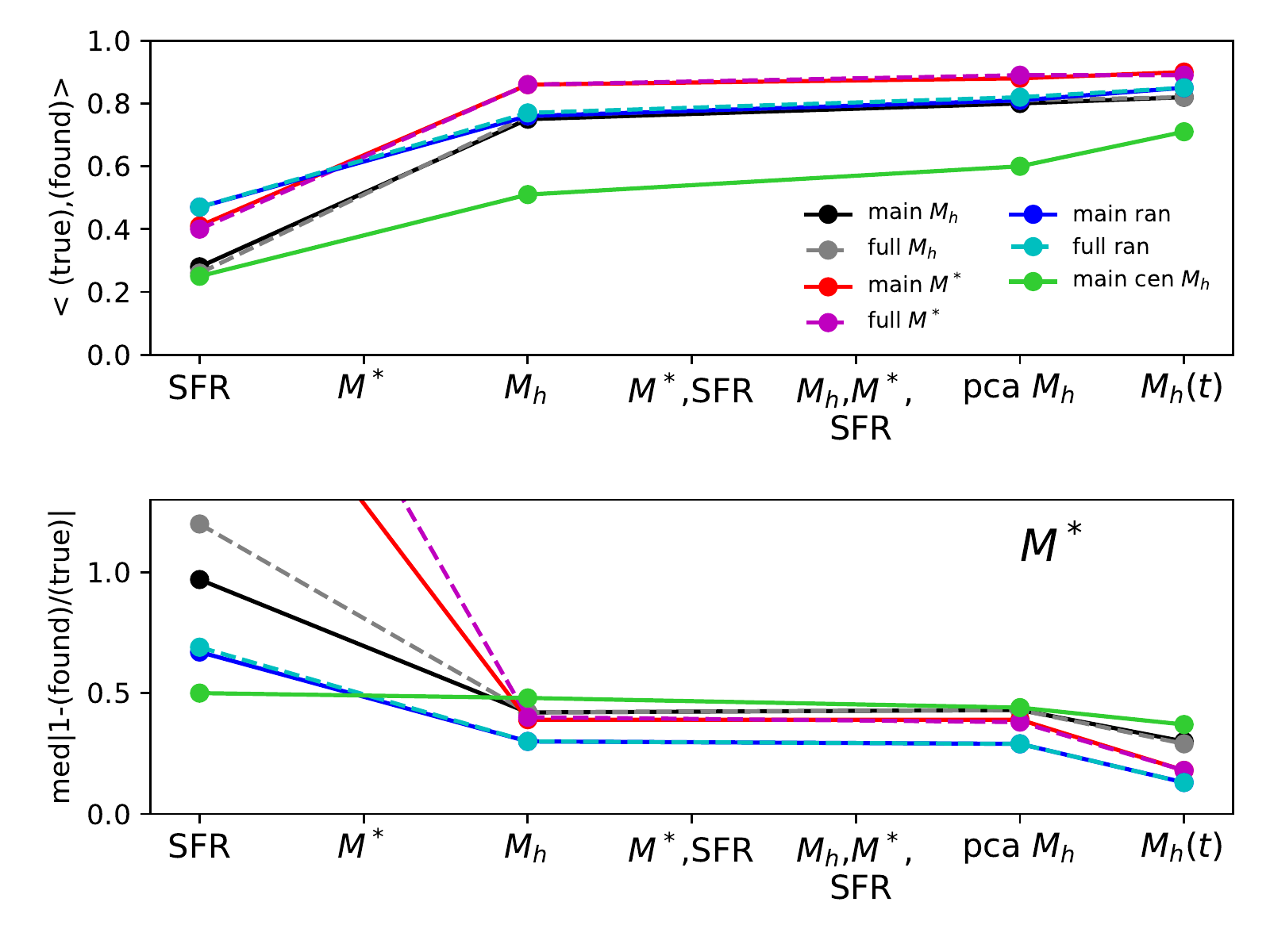}}
\resizebox{3.3in}{!}{\includegraphics{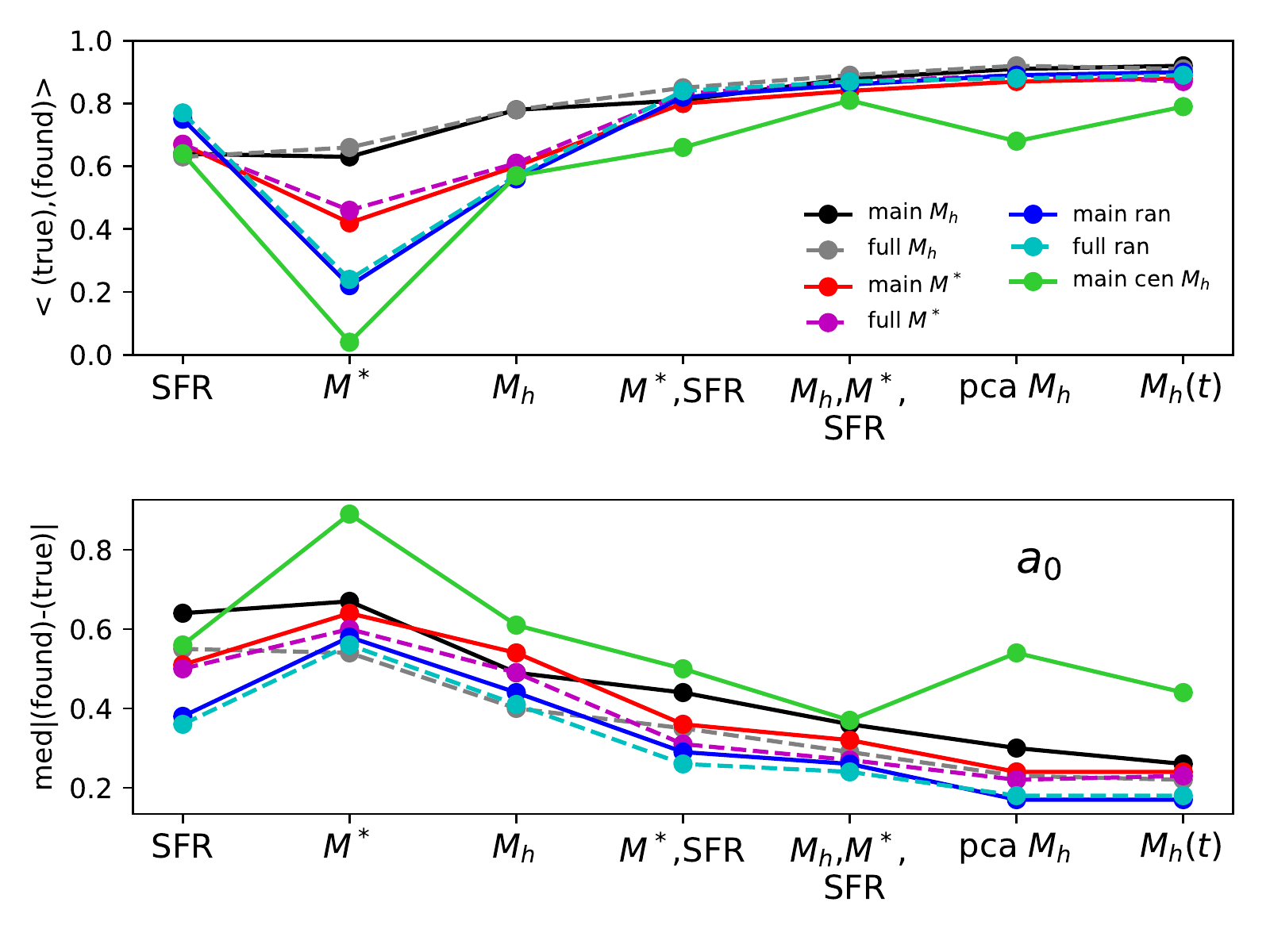}}
\resizebox{3.3in}{!}{\includegraphics{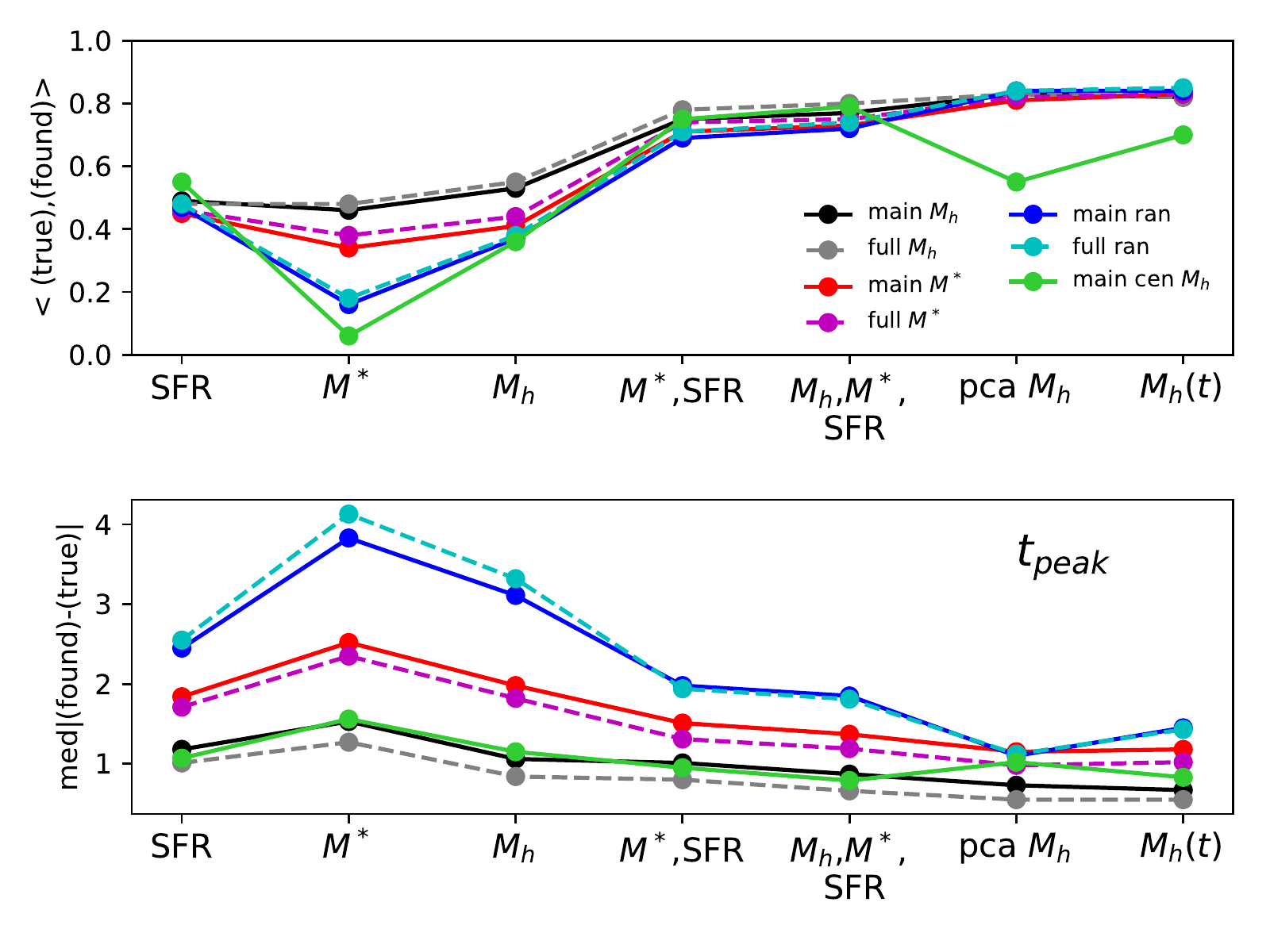}}
\resizebox{3.3in}{!}{\includegraphics{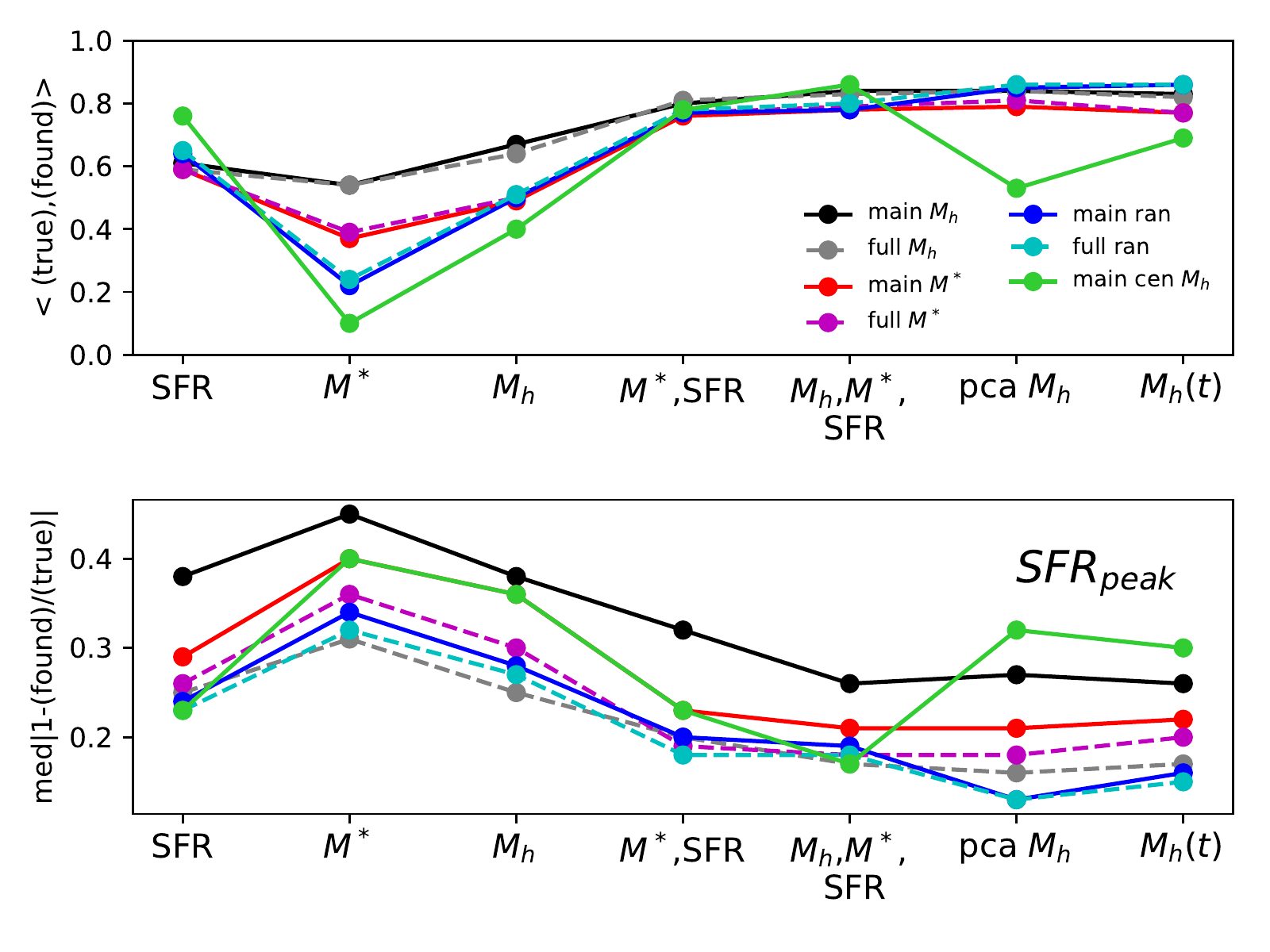}}
\end{center}
\caption{Results for machine learning of 
galaxy properties:  final $M^*$ (upper left), $a_0$ (upper right), $t_{\rm peak}$
  (lower left) and $SFR_{\rm peak}$ (lower right),
for each of the 7 samples, with inputs as listed on $x$-axis ($M^*,
SFR,M_h$ refer to values at final time).
The correlations between true and found
  values are in each top panel of the 4 sets of panels, the medians of
  $|\frac{\rm true-found}{\rm true}|$ or the difference $|{\rm found- true}|$, as
  indicated, are in the bottom panels.  Results for full and main
  histories are similar in many, but not all, cases.
}
\label{fig:predresults} 
\end{figure*}
One can go beyond correlations and try to predict $t_{\rm
  peak}$, $a_0$, and more, following the
machine learning approach of \citet{KamTurBru16a,KamTurBru16b}.
If machine learning is successful in using smaller numbers of galaxy
properties to reproduce 
properties of the full models, then it can be used to get these properties instead of, for
instance, the full semi-analytical models.  In
addition, the success of obtaining galaxy final and history properties based upon a
smaller set of inputs can
help guide the choice of properties to include in simplified models.

 The details of the
methods of \citet{KamTurBru16a,KamTurBru16b}, in particular, python notebooks, are available publicly at
https://github.com/ProfessorBrunner/ml-sims .  (See also
\citet{Xu13,Nta15} for some other applications of machine learning to
galaxy formation in particular.\footnote{\citet{AgaDavBas17,Nad17} also
  appeared as this work was being written up.} )  \citet{KamTurBru16a} used main
galaxy halo histories $M_h(t)$ and a few other halo properties as inputs, predicting several final time observables. Again, 
the
semi-analytic models which produce the star formation rate histories 
 use the full, not main, halo history, plus additional
dark matter simulation halo information (see, e.g. \citet{Fu13} for a
recent summary).

Here, machine learning is applied to predict
  $t_{\rm peak}$, $\log \sigma_t$, $SFR_{\rm peak}$,
$a_0$,$a_1$,$a_2$, the final integrated star formation rate,
$\tilde{\cal S}(t_f)$, and final $M^*$ for all 7 samples.\footnote{Results 
for $A$, although just a combination of $SFR_{\rm peak}$,$t_{\rm peak}$ and
$\sigma_t$ via equation Eq.~\ref{eq:logfit}, were much worse that these
other quantities; $A$ was thus calculated from $SFR_{\rm peak}, t_{\rm
  peak}, \sigma_t$.}
The method \citet{KamTurBru16a} found most promising for $M^*$ and
several other properties, 
extremely randomized trees \citep{Bre84,GeuErnWeh06}, also 
gave the strongest correlations
between the predicted and true values of $a_0$ and $t_{\rm peak}$,
although RandomForestRegressor was very close, again, similar to what
they found.\footnote{
Small parameter variations from the \citet{KamTurBru16a} choices did not 
improve the true-found correlations. 
\citet{AgaDavBas17}
have more 
 comparisons and comparison methods.} 

For all but the cen $M_{h,{\rm big}}$ sample,
the initial training set was a random selection of 10\% 
of the galaxies, subsequently trimmed to keep only those with a 
good lognormal fit for $t_{\rm peak}$ ($D < 0.06$, defined in
Eq.~\ref{eq:goodfitcrit}).  
For the much larger cen
$M_{h,{\rm big}}$ data set, \footnote{The cen $M_{h,{\rm big}}$ sample is
  analogous to that of \citet{KamTurBru16a}, more
discussion in the appendix, \S\ref{sec:moreml}.} 10,000 random galaxies were chosen (due to
limited computing power), and requiring $D<0.06$ left $\sim 7000$ galaxies,
closer to 3\% of the sample total.  

Although main halo histories $M_h(t)$ are a key part of the
\citet{KamTurBru16a} training set, is it also interesting to understand
how well fewer or other inputs recover parameters.  This helps to
clarify which inputs contain the most
predictive power.  Inputs considered are:
\begin{itemize}
\item final time $SFR$ only
\item final time $M^*$ only
\item final time $M_h$ only
\item final time $SFR$ and $M^*$ together (both observable)
\item  final time $SFR, M^*, M_h$ together
\item first 3 PCA components for
 $M_h(t)$ (again, $M_h(t)$ histories normalized to 1 at final time)
\item  main halo mass histories $M_h(t)$ (not normalized)
\end{itemize}

For all combinations of inputs listed above, correlations between predicted and
true values, and median differences or ratios were measured. 
A few distributions of true versus predicted values are
shown in Fig.~\ref{fig:truefound}.\footnote{Changing the random subsample
scattered the results around by 3\% or so, except for the
$SFR_{\rm peak}$ predictions, which sometimes would fluctuate to a
very small number, e.g.,$\sim$24\%, presumably due to outliers.
The feature around the current age of
the universe in the $t_{\rm peak}$ fit is due to the change in the lognormal fitting routine for
the histories at
that point.  As it is hard to extrapolate beyond the present day, peak
times beyond today were downweighted, using the same method as
\citet{Die17}.  I thank B. Diemer for explaining in detail how he did
his fits.}
The training data for these measurements
  are final $M_h$ at upper left, final $SFR, M^*$ at upper right, and the
  main halo histories, $M_h(t)$, for different samples at lower left
  and right.

Summary statistics using all combinations of the inputs to predict
$M^*$, $a_0$, $t_{\rm peak}$ and $SFR_{\rm peak}$ are
in Fig.~\ref{fig:predresults}. The (fewer than 10\% of the)
galaxies in the training set are included in the plots and the
correlations.  The best results
 came from using the whole (main) halo history, closest to the
galaxy information used by \citet{KamTurBru16a}.  However, many of the
variants starting with
smaller numbers of inputs exhibited significant success, in
particular using the three leading principal components for the halo 
history.

Some expected trends are visible.
For instance, the success of using final $M_h$ to predict final $M^*$ is
presumably due to the stellar mass-halo mass relation.  The larger
median separations between true and found $t_{\rm peak}$ for the
random sample, followed by the $M^*$ sample, are likely related to
their higher fractions of low $M^*$ and thus large $t_{\rm
  peak}$ galaxies, especially the harder-to-estimate future $t_{\rm peak}$ values.
Although the correlations between true and found were similar for the
full and main histories, the difference in the median values of the
fit parameters sometimes varied more between main and full histories
than between samples, for example, for $SFR_{\rm peak}$.

The poorer results for the cen $M_{h,{\rm big}}$ sample seems to be not due to
the training 
of the fits,
but from the galaxy distribution in the cen
$M_{h,{\rm big}}$ sample itself.
Using the full $M_h$ sample, e.g., to train a network to predict
parameters for the
cen $M_{h,{\rm big}}$ sample, gave predictions similarly bad to those found
by training the network on the 
cen $M_{h,{\rm big}}$ sample.\footnote{
 Beyond the halo virial mass, \citet{KamTurBru16a} also
trained on 
the halo number of particles, maximum
velocity and velocity dispersion, as well as, for the final time, the
halo half mass radius, virial velocity, virial radius, and $r_{\rm
  crit,200}$.  Virial velocity and maximum velocity histories did not strongly improve 
the cen $M_{h,{\rm big}}$ sample correlations between true and found.}
More generally, overall correlations between true and found values are strongly dependent
on the makeup of the sample.  For instance, if only a small range of final halo
masses is considered, the correlations between true and found values
of $t_{\rm peak}$ decreases, because much of the strength of the
correlation between true and found values of $t_{\rm peak}$ is due to
machine learning using the final $M_h$ dependence of $t_{\rm peak}$
(see Fig.~\ref{fig:truefound}).  

The mismatch between true and found values of the parameters
translates into worse approximations for the fits to the original
histories and
for the final time stellar mass to star
formation rate relation, and a different shape of the scatter around the
histories.  Results and some comparisons to the
earlier direct fits (shown earlier in Fig.~\ref{fig:dsep},
Fig.~\ref{fig:mstarsfr}, and Fig.~\ref{fig:sfrsigdist}) are in
the appendix, \S\ref{sec:moreml}.  In particular, the number of
galaxies assigned the
lowest star formation rates ($\leq 10^{-7} M_\odot yr^{-1}$) via
machine learning never reaches 1/3 of those in the simulation, and in
the {\tt ran} sample is $\leq$ 1\% of the simulation number for the lognormal fit.

To summarize, many of the galaxy properties at final time and
their main halo histories $M_h(t)$ are strongly correlated with the star formation rate
history parameters.  Machine learning can find fairly good fits to the
peak time $t_{\rm peak}$ or leading PCA fluctuation coefficient $a_0$ 
by using the leading
3 principal components of the halo history, or by using the main halo
history $M_h(t)$.  However, although these parameters and final stellar masses
are fairly well
approximated, the approximations to the true simulation integrated histories and
final time values are
noticeably worse, and the machine learning
determined instantaneous star formation rates
at final times have significantly fewer quiescent galaxies in the {\tt
  ran}
sample (doing slightly better in the samples with more high mass galaxies).

\section{Bimodality and beyond}
\label{sec:split}

Galaxies are often classified as star forming and quiescent 
(separated at
 SSFR  $=10^{-12} yr^{-1}$ for the samples here, from considering the
SSFR distribution in the simulation outputs).  This division
can help identify common properties and correlations within
the set of star
forming or quiescent galaxies,\footnote{But the description using the
  first three PCA components assumes
  one average history and captures most of the scatter, and using an average history also works in some
  descriptions of galaxy evolution more generally \citep{BWCz8}.  See also
  \citet{Eal18,Kel14}. }  
and guide the search for mechanisms which cause transitions between
these two categories.  Since
both $t_{\rm peak}$ and $a_0$ give one parameter characterizations for
galaxy histories, they can also be used to group galaxies, into
subfamilies that share similar integrated star formation rate histories.

Whether a galaxy is star forming or quiescent is, not surprisingly,
related to its star formation rate history, and thus to $t_{\rm
  peak}$ and $a_0$, with quiescence tending to imply low $t_{\rm
  peak}$ (or $t_{1/2}$), and high $a_0$, that is, early star
formation.
However, although related, these separations of galaxy histories are all distinct.
The number of
galaxies with high $t_{\rm peak}$ and high SSFR matches that of
galaxies with low $a_0$,
and number of galaxies with low SSFR and high $a_0$ matches that of
galaxies with low $t_{\rm
  peak}$, but for other pairings of SSFR, $t_{\rm
  peak}$, and $a_0$, the number of galaxies in subfamilies cut on one quantity differs from 
that in a subfamily found by a cut in another quantity. 
\begin{figure} 
\begin{center}
\resizebox{3.3in}{!}{\includegraphics{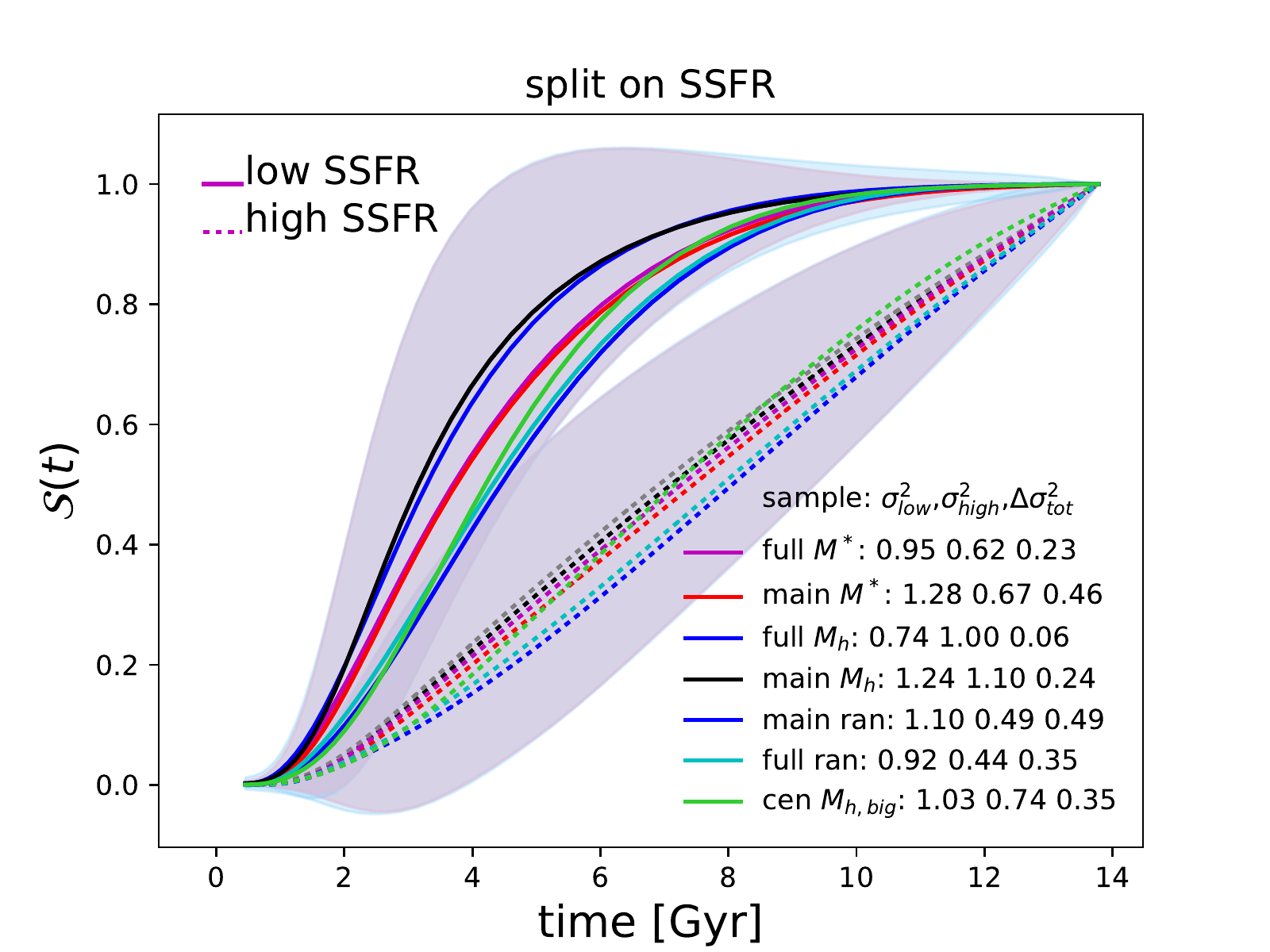}}
\resizebox{3.3in}{!}{\includegraphics{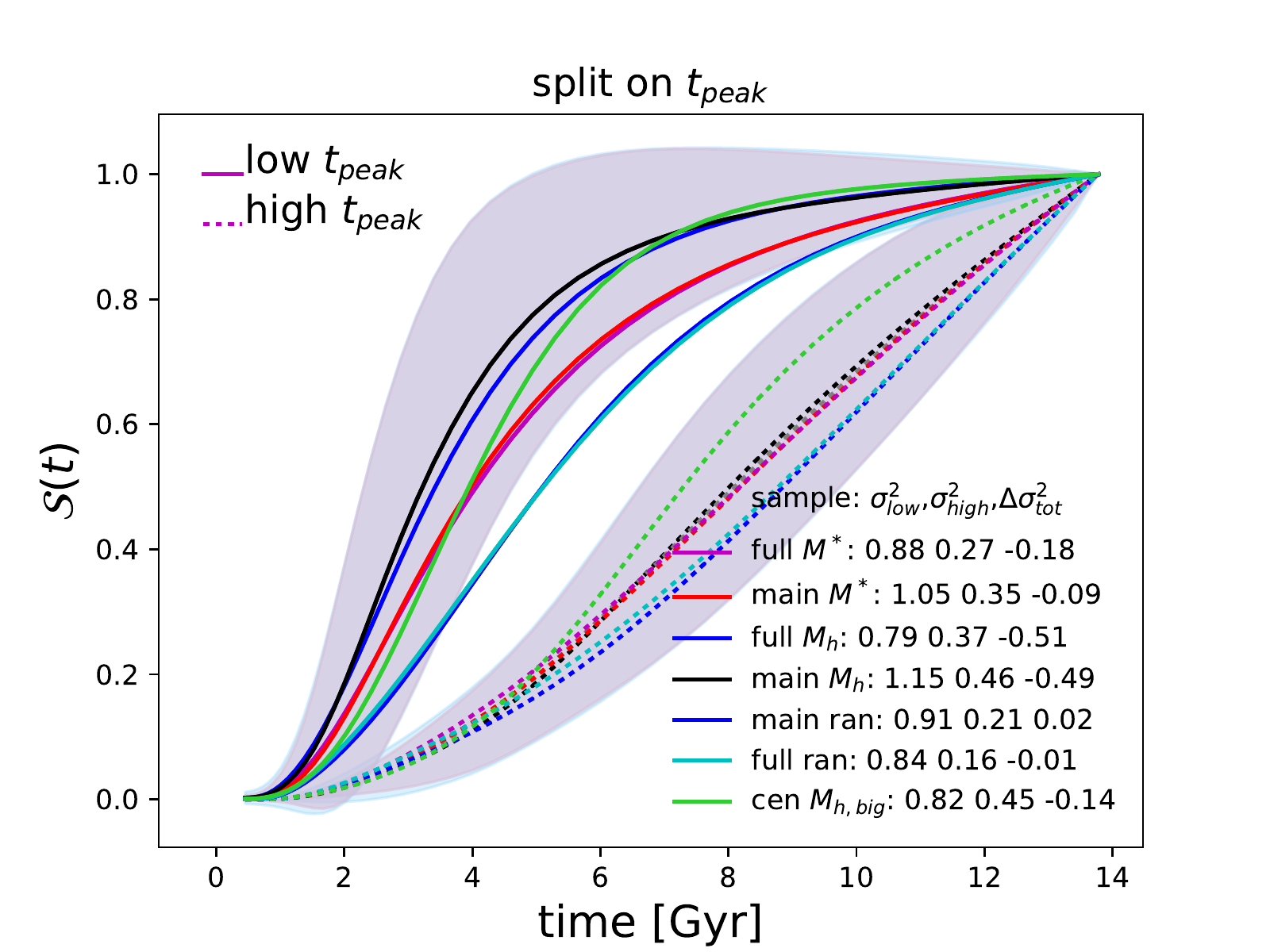}}
\resizebox{3.3in}{!}{\includegraphics{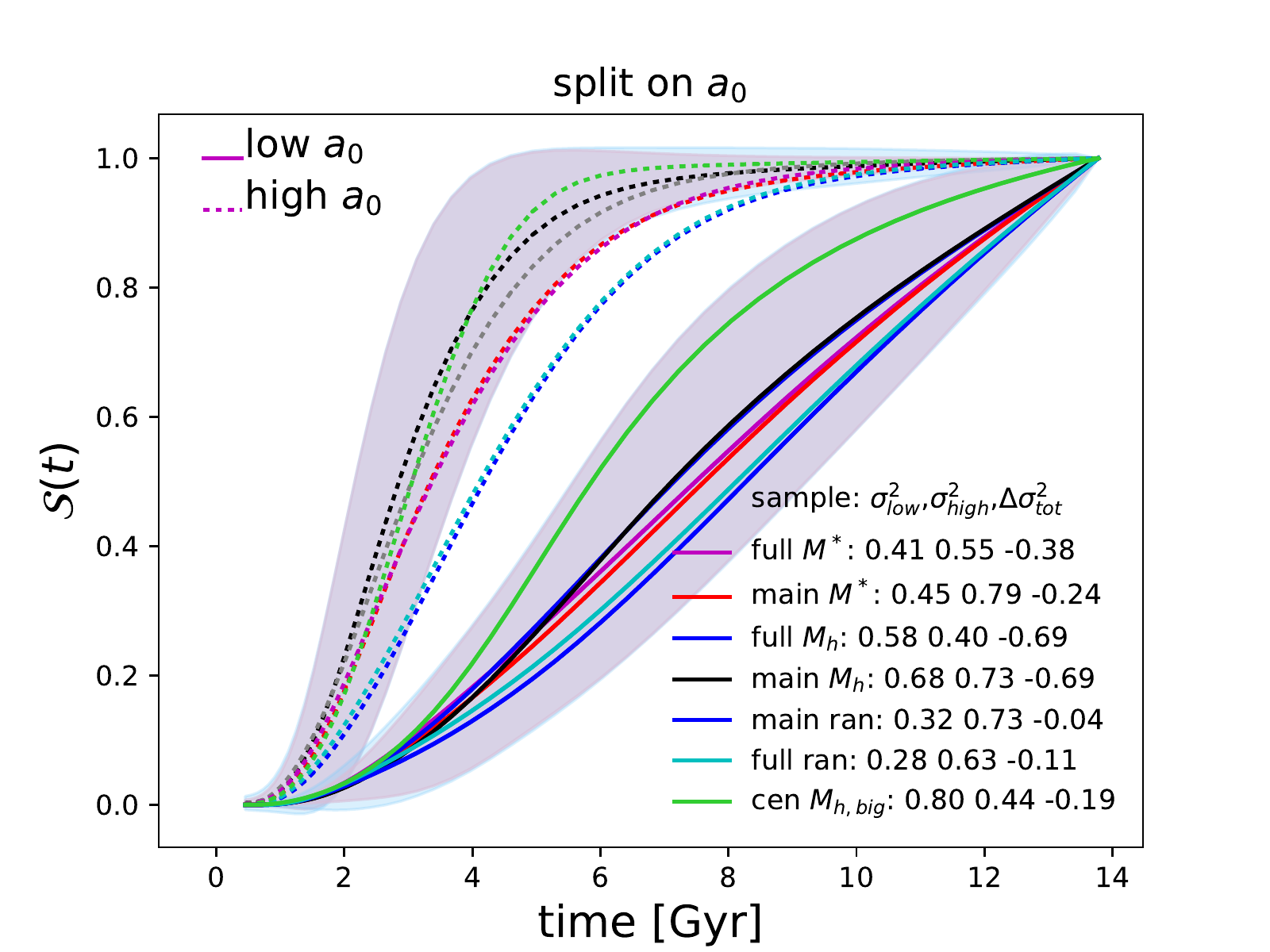}}
\end{center}
\caption{Average integrated star formation rate histories for all 4
  samples, split into high and low specific
   star formation rate (top, $SSFR \lessgtr 10^{-12} yr^{-1}$),
 $t_{\rm peak}$ (middle) and
 $a_0$
  (bottom).  Adding a parameter, i.e. splitting into subfamilies, should
  reduce the scatter ($\Delta \sigma_{\rm tot}$ negative), however the SSFR
  subfamilies have $\Delta \sigma_{\rm tot}>0$.
Thistle shading shows the contribution of the first 3
principal components to the variance, blue (only visible at the edges)
the full variance, for the main $M_h$ subfamilies only; see text for details of
separating into subfamilies.
}
\label{fig:splitthree} 
\end{figure}

Although all three quantities can be used to separate galaxy samples,
the integrated star formation rate
histories of quiescent and star forming galaxies do
not separate as well as those with high and low $t_{\rm peak}$ or
$a_0$.
In particular, there is larger scatter around the average values for the quiescent and star forming histories, as
shown in Fig.~\ref{fig:splitthree}.
In each panel, each sample's galaxies are split into high and
low $SSFR$ (top panel), $t_{\rm peak}$ (middle panel) and $a_0$
(lowest panel). Averages for the high and low
  subfamilies are shown by lines as
indicated. 
To get a sense of the scatter around the two subfamilies, that for the main $M_h$
subfamily is shown in each figure.   The subfamily variances around 
the two averages are shown in blue, and
superposed  (purple, which almost coincides) are the shapes of the fluctuations due
to the first 3 subfamily principal components
(within each subfamily).  The variance due to the first three
principal components at time $t_i$ is $ \sum_{n=0}^{n=2}
<a_n^2>PC_n(t_i)^2$. 
This gives a visual estimate of the overlap and shows that the two
subfamilies again have a few parameters capturing a large amount of
scatter around their respective averages. 

There are also two
quantitative ways of classifying a separation into subfamilies as successful.
The first is the change
from the total (original) variance to that around the two samples,
\begin{equation}
\Delta \sigma^2_{\rm tot} = \sum_{{\rm samples} \; i} \sigma^2_{i,{\rm final}} -
\sigma^2_{\rm initial}  < 0\; .
\label{eq:deltavar}
\end{equation}
When $\Delta \sigma^2_{\rm tot}<0$, the separation into subfamilies
reduces the total scatter.  

 The second quantity is the distance between the two
subfamily averages, relative to the
overlaps of their populations (roughly estimated by the variance
around each average).  If this ratio is less than one, it suggests that
the two populations do not overlap significantly,
\begin{equation}
\frac{\sigma_0^2 +\sigma_1^2}
{|\bar{\cal S}_0(t)-\bar{\cal S}_1(t)|^2 }<1 \; .
\label{eq:scattereq}
\end{equation}
Here, $\sigma_i^2$ is the variance around each subfamily, with
respective 
average $\bar{\cal S}_i(t)$.  

In the legends in Fig.~\ref{fig:splitthree}, $\Delta \sigma^2_{\rm tot}$ for
each separation is given for each galaxy sample and separations.
Although the SSFR separation is fixed, the separations for $t_{\rm peak}$ and $a_0$
are chosen by scanning through values to minimize $\Delta
\sigma^2_{\rm tot}$ and the overlap,
Eq.~\ref{eq:scattereq}. \footnote{The separations are at
$t_{\rm peak}=$ 5 Gyr for all but the {\tt ran} sample ($t_{\rm peak}=6.4
Gyr$), and at
 $a_0=(0,0.4,0.5,0.8)$ for $\log M_h$, $\log M^*$, {\tt ran}, cen
$M_{h,{\rm big}}$ respectively.  Scanning through different splittings, $\Delta
  \sigma^2_{\rm tot}$ is negative, with slow variation, 
for a wide range of $a_0$ splits, while there is a clear minimum for
splitting on a specific $t_{\rm peak}$.  The minimum position changes with each galaxy
sample.}  
For the SSFR split, $\Delta \sigma^2_{\rm tot}$ and
the overlap between the regions which lie in the scatter of both
average paths are larger (this is true for all 7 samples).  Subfamilies of galaxies
sharing high or low $t_{\rm peak}$ or high or low $a_0$ have
more distinct integrated star formation rate histories.

\subsection{Subsets of galaxy histories} 
As splitting on specific star formation rate does not separate galaxy
histories into distinct families as well as using $t_{\rm peak}$ or
$a_0$, and the distribution of these latter two parameters (Fig.~\ref{fig:tpdist}) is not
necessarily bimodal, it seems possible to group galaxies
into more than two subfamilies, with each subfamily sharing similar
integrated star formation rate histories.  One motivation for this is
to compare properties of galaxies lying in different subfamilies,
besides the parameters used to sort into subfamilies.
This might be useful in identifying shared trends in subfamilies or
general physical causes of certain properties.  For instance, if
massive galaxies are present in several different subfamilies, one might ask
what properties caused their different integrated star formation rates,
in spite of their sharing the same final halo mass?
These subfamily classifications can serve as starting points
for such lines of investigation.  

Here a first step  is taken in exploring
separations into many subfamilies.
Whether subfamilies are well separated can again be decided by comparing whether
the final sum of scatters around each subfamily is smaller relative to the that of the full sample around
its average ($\Delta \sigma^2_{\rm tot}<0$, Eq.~\ref{eq:deltavar}) and whether adjacent subfamilies are sufficiently separated,
\begin{equation}
\frac{ \sigma_i^2
+\sigma_{i+1}^2}
{|\bar{\cal S}_i(t)-\bar{\cal S}_{i+1}(t)|^2} < 1 \; .
\label{eq:scattersep}
\end{equation}
The split is now into many ($i=1,\dots N$) subfamilies, each with 
individual averages $\bar{\cal S}_i (t)$.

\begin{figure*} %
\begin{center}
\resizebox{3.3in}{!}{\includegraphics{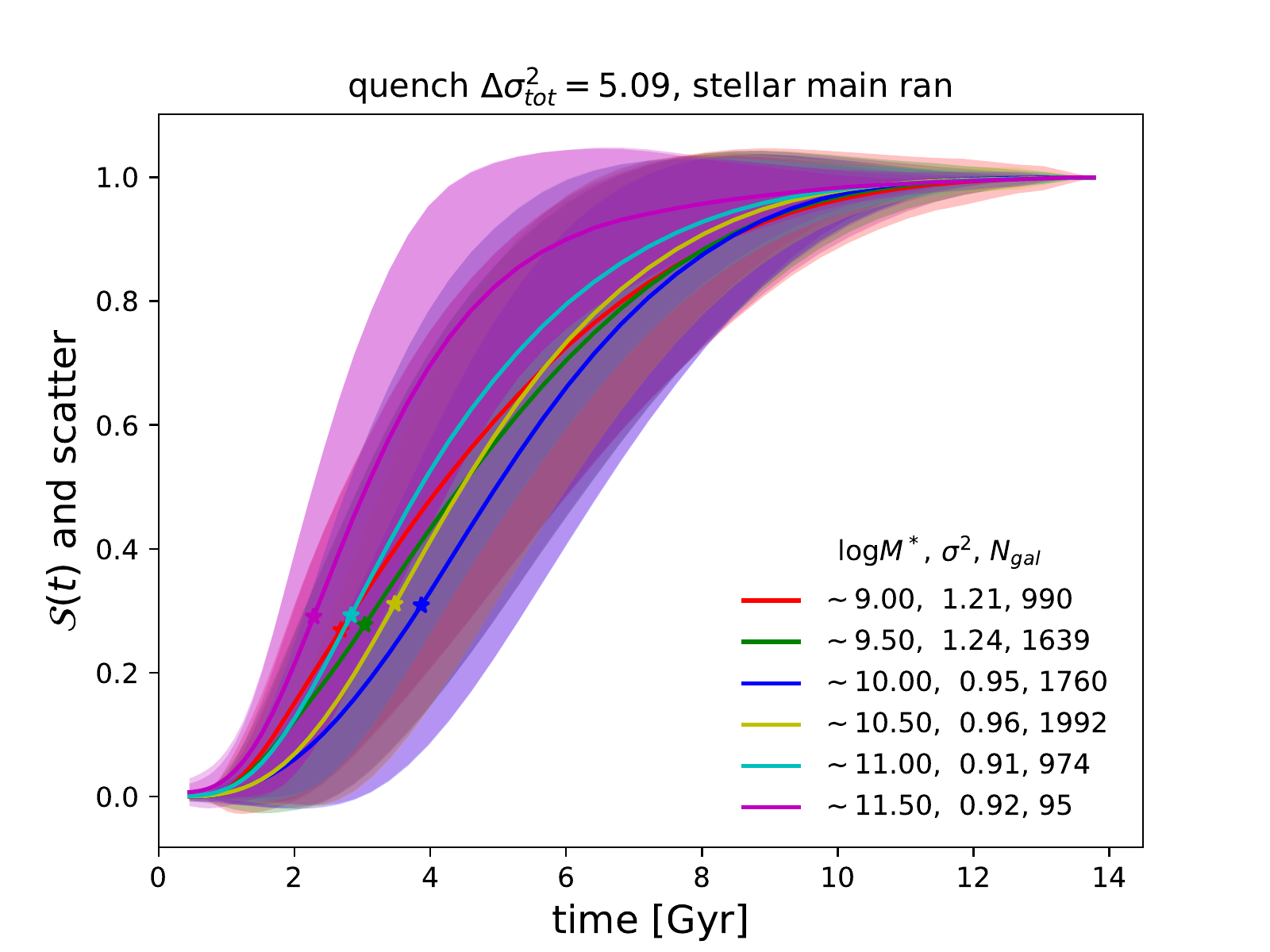}}
\resizebox{3.3in}{!}{\includegraphics{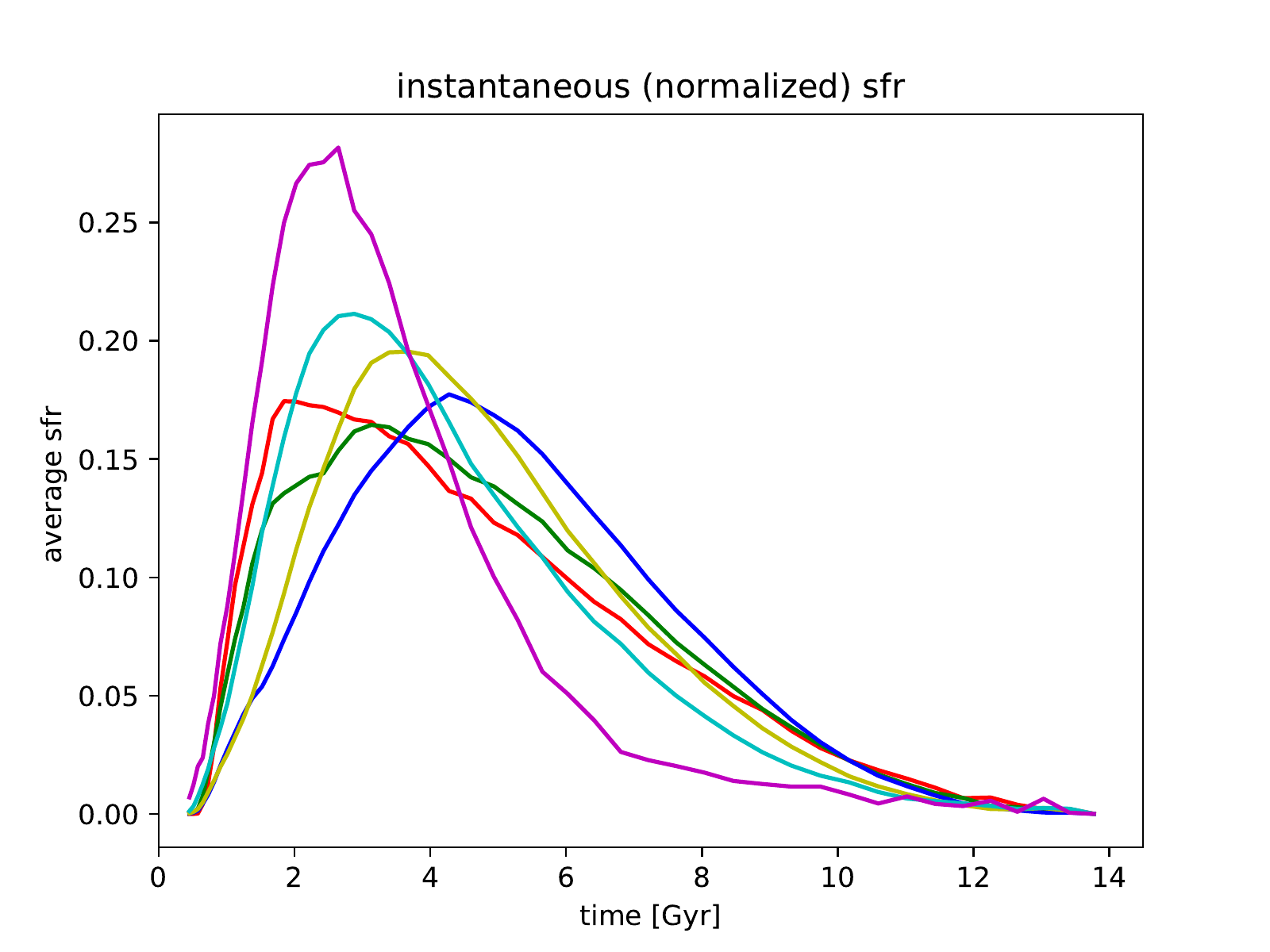}}
\resizebox{3.3in}{!}{\includegraphics{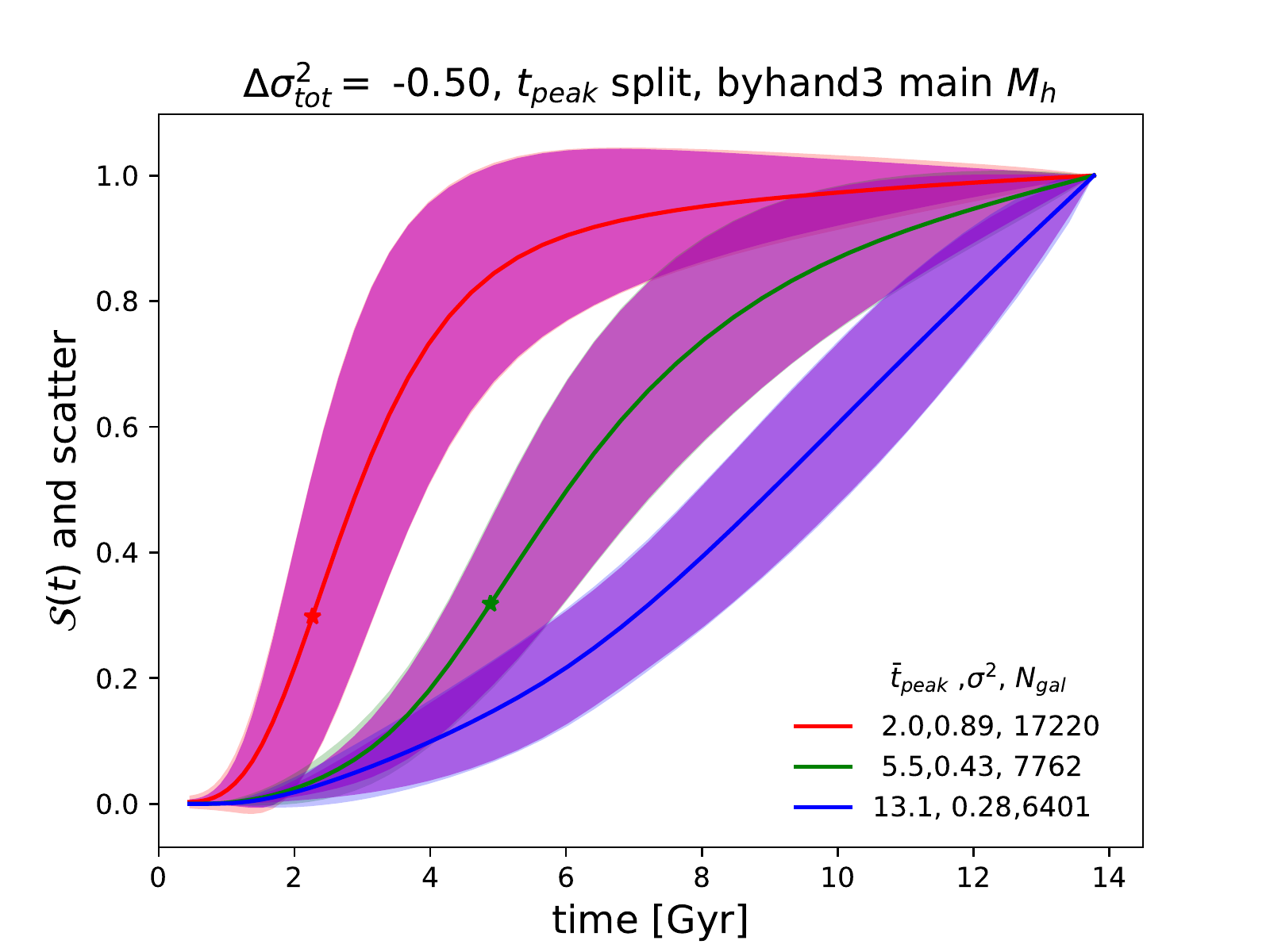}}
\resizebox{3.3in}{!}{\includegraphics{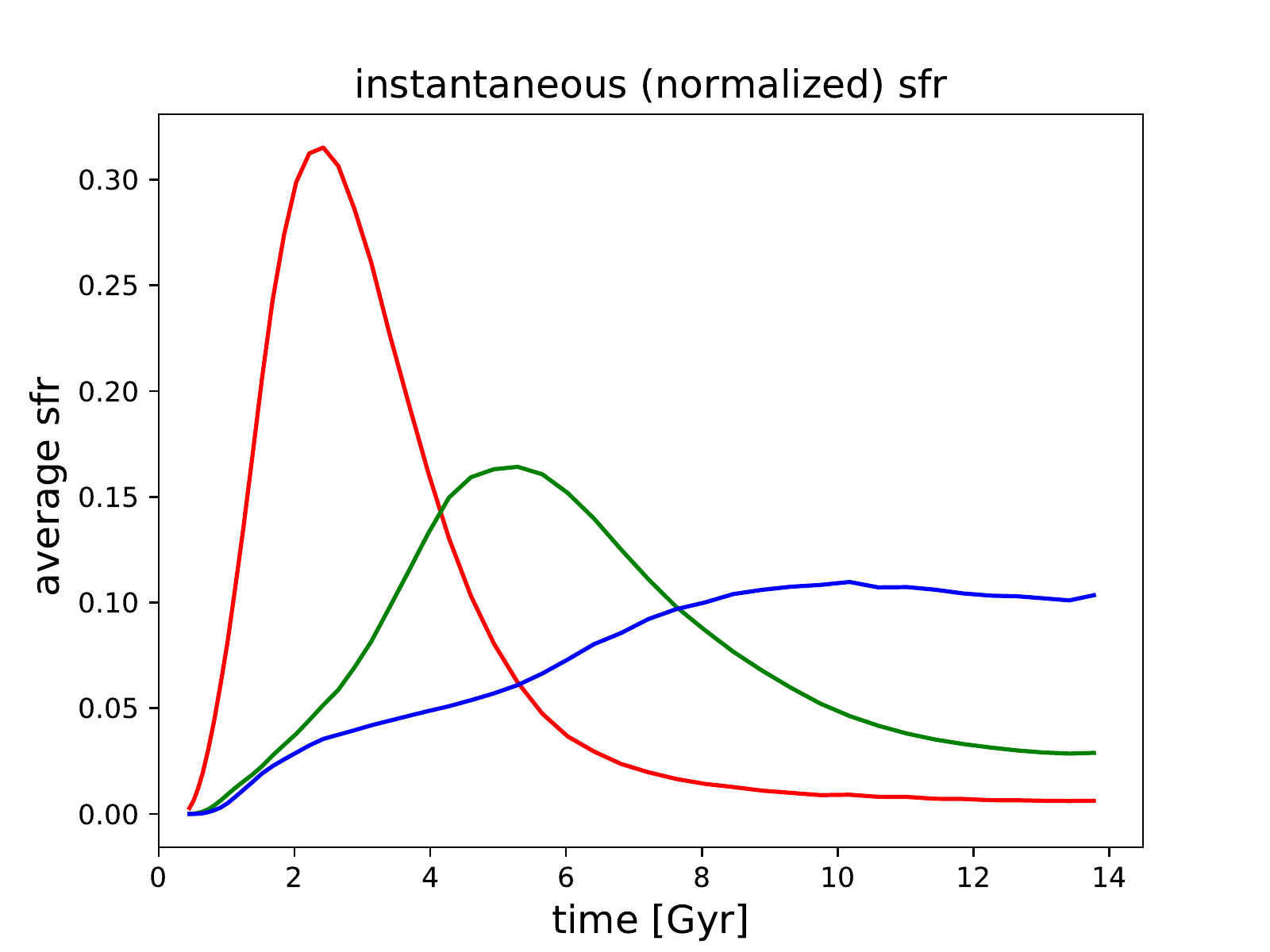}}
\resizebox{3.3in}{!}{\includegraphics{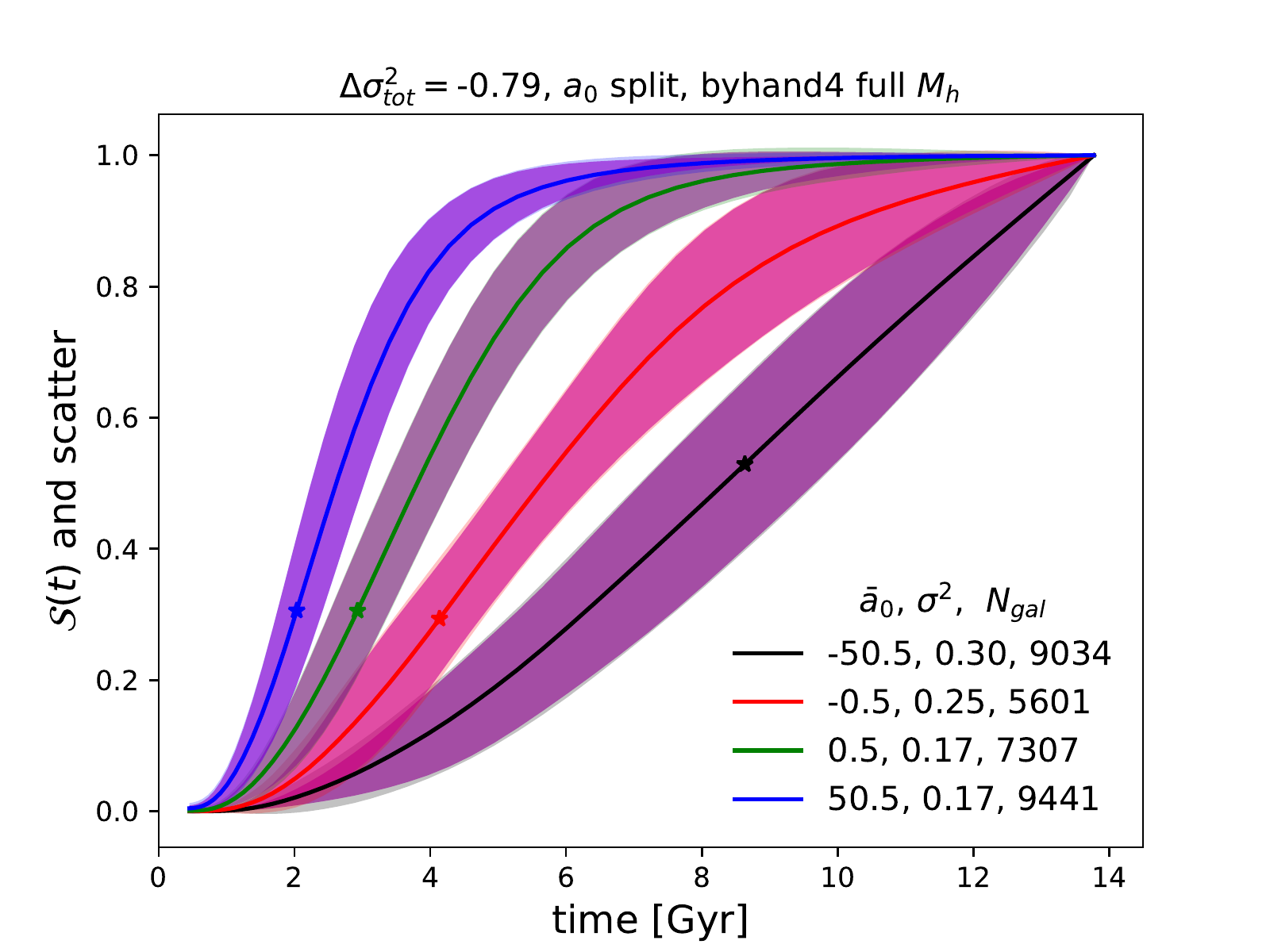}}
\resizebox{3.3in}{!}{\includegraphics{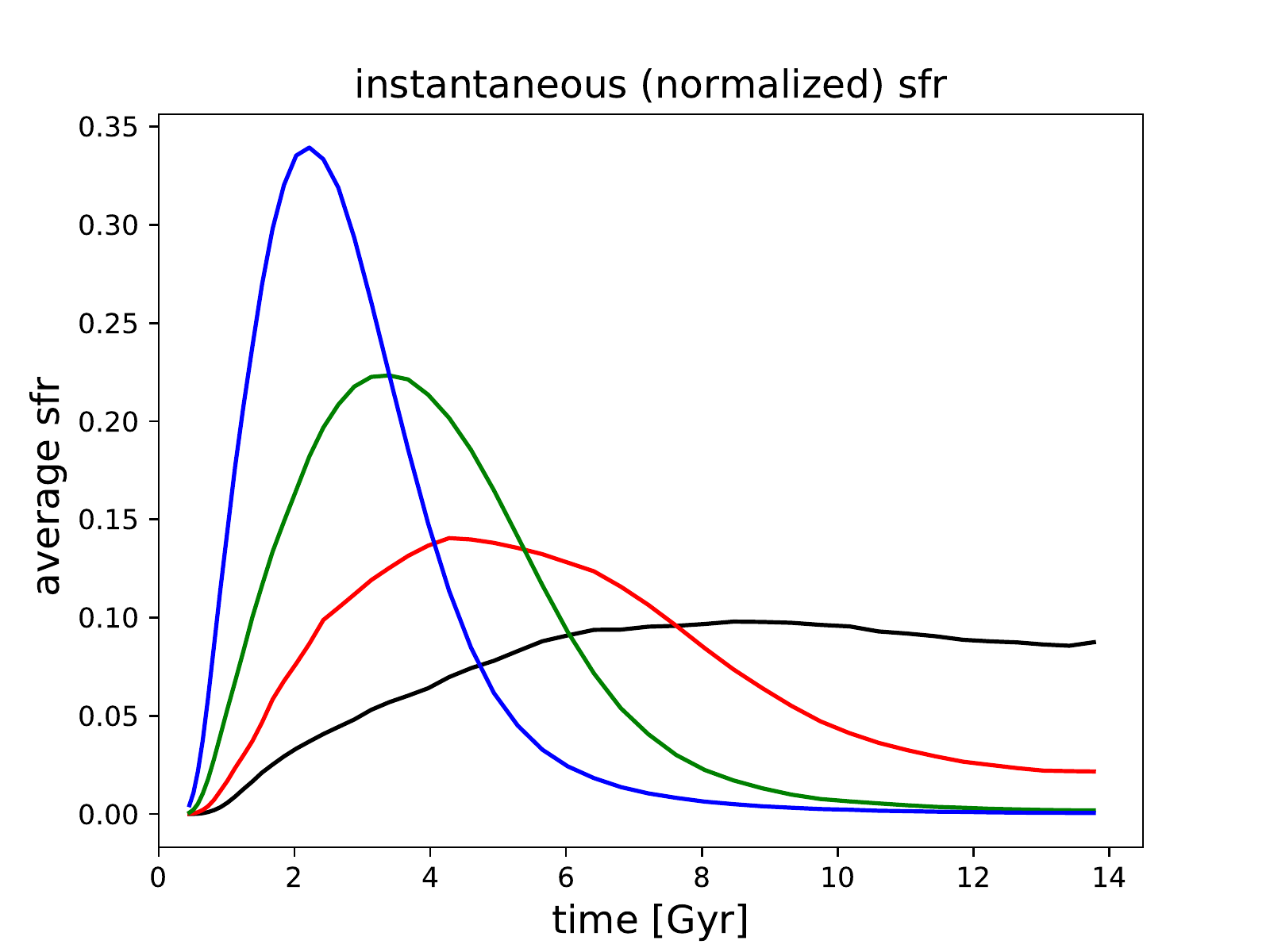}}
\end{center}
\caption{Three examples of splits of integrated histories into families.
Top:  quenched galaxies in the main {\tt ran} sample (SSFR $< 10^{-12} yr^{-1}$), split by final stellar mass
  as in \citet{Pac16}.  Middle:
  the main $M_h$
  sample split by $t_{\rm peak}$.  Bottom: the full $M_h$ sample split
  by
  $a_0$.  In each, $\Delta \sigma^2_{\rm tot}$, Eq.~\ref{eq:deltavar}, is shown
  at top.  At left, lines are the average in each subfamily, stars are
  the lognormal fit $t_{\rm peak}$ to these averages.
 The dark shaded region is the variance around each average history due to the
  first 3 principal components in each subfamily, as in
  Fig.~\ref{fig:splitthree}.  The light shaded region
  (indistinguishable in most places, the first 3 principal components
  dominate the scatter) is the corresponding full
  variance.  At right are the averages of the
  instantaneous star formation rates in each subfamily (normalized by each galaxy's 
 final integrated star formation rate). 
For the stellar mass separated
  sample, the different subfamilies overlap significantly;
  final stellar mass is not that closely
  correlated with galaxy integrated star formation rate history.
   Using $t_{\rm peak}$ and $a_0$ gives better separations into subfamilies.
}
\label{fig:samp3} 
\end{figure*}

The wide range of histories shared by galaxies with the same final
stellar mass was noted by \citet{Pac16}, who
separated quiescent and star forming galaxies and then stacked star
formation rate histories within
these categories based upon stellar mass.  The averages $\bar{\cal S}_i(t)$ and variances
around them for 
six stellar mass families 
(quiescent galaxies only, i.e. SSFR $\leq 10^{-12} \; yr^{-1}$) are shown in
in Fig.~\ref{fig:samp3}, top, for 
the main {\tt ran} sample.  The other samples are similar.  
In addition to the large scatter overlaps between average
histories for the subfamilies, i.e., Eq.~\ref{eq:scattersep} does not hold,
the sum of scatters around the individual $\bar{\cal S}_i(t)$
is much larger than original scatter around the single average
history ($\Delta \sigma^2_{\rm tot}$ is listed above the panel).

In comparison, $a_0$ and $t_{\rm peak}$ can separate integrated histories more
cleanly.  Two examples are shown in Fig.~\ref{fig:samp3}.  The middle
example splits all galaxies based upon $t_{\rm peak}$, the lower
example, using $a_0$.\footnote{It is not clear whether the
  inflection point in the integrated histories of galaxies with high $t_{\rm
    peak}$ is due to bad fits (inclusion of some galaxies with
a lower $t_{\rm peak}$) or a physical property of such galaxies.}

An 
exhaustive study of possible
separations into categories is beyond the scope of this note.
An assortment of subfamily splits were tried.  Their
$\Delta \sigma^2_{\rm tot}$ values are compared in appendix
\S\ref{sec:moresplit}, and those tried which reduced the total scatter also
had subfamilies separated enough according to Eq.~\ref{eq:scattersep}.
Some general features
were noted, for instance, when using $t_{\rm peak}$ to determine subfamilies,
galaxies with $t_{\rm peak}>$7 Gyr had integrated histories which seemed too close
together to lie in different subfamilies.  One other way of separating
integrated histories was also considered, suggested by \citet{Pac16}, a quenching
time $t_q$\footnote{A rough way to define $t_q$, used here, is as the last
time that the interpolated specific star formation
rate is above the cutoff for quiescence, $10^{-12} yr^{-1}$.  
To get the SSFR, the (smoother) lognormal fit to the galaxy's star formation rate
history is divided by the stellar masses at each time, and interpolated to all
times between the peak time, $t_{\rm peak}$ from the fit, and the
current age of the universe. }.   In the few examples explored, $t_q$
did not seem to work as well as $t_{\rm peak}$ or $a_0$, for instance 
in terms of $\Delta \sigma^2_{\rm tot}$, however, again, an exhaustive
comparison was not made, and the definition could also be refined.

Once a sample is split according to $a_0$ or $t_{\rm peak}$, galaxies
which share similar integrated star formation rate histories can be compared in
terms of other properties, such as final time $M^*$ or $M_h$, for
instance, to look for reasons that a common final time property is associated
with different subfamilies, when that occurs.
Distributions of several properties for
the galaxies in the 3 or 4 subfamilies of Fig.~\ref{fig:samp3}
are compared side by side in the appendix, \S\ref{sec:moresplit}, as examples.

In summary, as might be expected, splitting integrated star formation rates of 
galaxies according to whether they are
star forming or quiescent doesn't separate their histories as well as
splitting based upon their lognormal fit $t_{\rm peak}$ 
or $a_0$.   For a bimodal split, using $a_0$ to sort each galaxy
reduced the total scatter 
more generically than using $t_{\rm peak}$, however, for splits into
several subfamilies, both $t_{\rm peak}$ and $a_0$ can be seen to reduce
the full scatter and give what seem to be reasonably separated
histories.  A few other general trends seemed to occur.  
For instance, all galaxy integrated star formation rate histories
with a fitted $t_{\rm peak} \gtrsim 7$ Gyr 
tended to have large overlap with each other.  And again, as $a_0$ dominates the
scatter around the average history, it is not surprising that
subfamilies split via its value are less likely to overlap than those
split via final time properties.
These separations may be useful as starting points for comparing
galaxies which share one property but not another in a simulation (for example,
$t_{\rm peak}$ but not final $M^*$, or final $M_h$ but not $a_0$).

\section{Summary, Discussion and future directions}
\label{sec:discussion} 
In this note, two different methods for parameterizing integrated star formation
rate histories were considered: a fixed lognormal form (following
\citet{Gla13,Die17}) and a PCA approximation, treating all histories as the 
ensemble average plus a combination of the leading three fluctuations
(principal components)
around it.
The lognormal parameterization treats the star formation
rate history as having a peak at a certain time, plus a width with a
fixed shape, while PCA views all histories as
fluctuations around one average history (independent of whether the
galaxy is quiescent or not), plus fluctuations of fixed shape and 
coefficients varying
in size and sign.  The PCA approximation, using the first 3 principal components, has one more parameter than
the lognormal fit, and is more closely tied to the properties of the
ensemble of galaxies it describes, as the principal components and the
average history around which they fluctuate are both determined using
the galaxy sample itself.  

These fits were explored with data from the
\citet{Hen15} model built upon the Millennium simulation \citet{Spr05,Lem06}.
 To illustrate how to compare samples (or more generally models), four sets of
 simulated galaxy histories were created:
one approximately uniform in $\log M_h$,
 one approximately uniform in $\log M^*$, one randomly
selected, and including all final time central galaxies in
massive halos.

The samples of galaxies were characterized by their 
lognormal fit parameters (especially $t_{\rm peak}$) and by their average
histories,  PCA fluctuations, and distributions of the fluctuation
coefficients (especially $a_0$).
For the PCA approximation, the shapes of the averages and fluctuations
were similar across
different samples, with most variations between samples easily
interpreted as due to changes in the number of galaxies with high
final halo
mass (expected to quench earlier).  The first 3 PCA components
captured a large fraction of the scatter around the average history
for every sample.  
The lognormal and PCA fits have correlated leading parameters, especially for galaxies with
an early $t_{\rm peak}$ and high $a_0$.  The lognormal fit parameter $t_{1/2}$, when
a galaxy dropped to half of its peak star formation rate, was also strongly correlated with $a_0$.

Star formation rates of both the main (following one galaxy through time) and full
(including all the galaxies which eventually merge to form the the final
galaxy) integrated star formation rate histories were considered for 3 of the 4 samples. 
The full histories have an earlier $t_{\rm peak}$ in the lognormal fit,
and smaller variance around the average history in PCA.  The full and
main histories differed more strongly 
for samples with larger numbers of high final mass galaxies, 
Different samples (or models) of galaxies can be compared via parameters of the lognormal fit  ($t_{\rm
  peak},\sigma_t,A$, and 
the average relation $\sigma_t \approx a t_{\rm peak}^b$)\footnote{As
  in \citet{Die17}} and by the 
PCA fit parameters ($a_0,a_1,a_2, \tilde{\cal S}(t_f)$),
the average history of the sample and its PCA basis fluctuations and
the variance in the fluctuations.

Two fixed time properties were studied in more detail.  The
instantaneous star formation rates in the simulations were compared to
that given by the fits.  The lognormal fit worked
better in tracing the distribution of the instantaneous star formation rate at the
final redshift (in particular, it was difficult to get the correct
number of green valley galaxies in the PCA approximation, even when
many principal components were included).   
It is possible that even with the visible differences
from the true (i.e. simulation) values that the fits can provide 
useful approximate star formation rates, depending on what the rates
are used for; the average deviation between
the fit and simulated values, over time steps, is zero by construction
for the PCA approximation, and small for the lognormal fit.  

Secondly, the peak star formation rate halo mass $M_{h,{\rm peak}}$
is bimodal as a function of $t_{\rm peak}$.
(It is not seen in the high final mass cen $M_{h,{\rm big}}$ sample
which excludes low mass halos by construction.)  Downsizing is also
evident on the dominant (higher $M_{h,{\rm peak}}$) branch.  It would be interesting
to understand what is happening with the lower $M_{h,{\rm peak}}$ galaxies.  Perhaps
environmental effects are starving their growth, for example.  It
would also be interesting to
understand if this feature appears in other galaxy formation models
and in nature.

The parameterizations for both fits were correlated with final time
properties ($M^*,M_h$ and SFR), and with properties of the galaxy main halo
histories. 
Machine learning, following \citet{KamTurBru16a}, 
was used to estimate the PCA and lognormal approximation parameters,
using a range of inputs,
including just the final halo mass and the main halo history $M_h(t)$.
(The galaxy histories are the product of a detailed and complex
semi-analytic model, following all halo and subhalo contributions and many physical
properties of a galaxy throughout time, and so are automatically
related to the full, rather than main, halo histories.)  The final
halo mass could already give a significant correlation between the
true and found values of several fit parameters.  The first 3
principal components of $M_h(t)$ almost worked as well as $M_h(t)$
itself in predicting fit parameters, and the true and found values of
several quantities
were highly correlated.  However, the machine learning
predicted final time star
formation rates were even further from their simulation (true) values than 
the original fits.
Machine learning shows that a relation can be found, but does not
detail the relation, aside from providing importances.  For these
galaxy halo histories, it seemed that halo masses at a wide range of
times in the history were important for predicting final values.
It suggests promise for linking halo main histories directly to star
formation rate histories through these parameterizations, perhaps in a
simplified galaxy formation model.\footnote{
In addition, the first principal component of halo history (with a
slightly different definition) has been
associated with concentration \citep{WonTay12}, which has also been
suggested as a key parameter controlling the scatter in galaxy quenching, for instance by S. Faber in her Berkeley Astronomy Colloquium of fall 2017.}

Using the
leading parameter of either approximation, $a_0$ or $t_{\rm peak}$, better
separates galaxies into subfamilies with similar histories than using
whether a galaxy is quenched or star forming at final times.  However, once a
continuous parameter is used to separate histories, there is no
obvious reason to only split galaxies into two groups. Separations into
more families of galaxy histories were explored.  Many were found which
both reduced overall scatter and had subfamilies separated further
than
the variances around each subfamily average. 
These might be useful to compare galaxies with similar histories but
different final properties or vice versa, to help identify which
changes create these different populations within a single galaxy formation
model, or to compare between models.

All of these calculations were done within the context of the
\citet{Hen15}, or L-galaxies, semi-analytic model built upon the Millennium
simulation.
The lognormal fit was applied to the Illustris simulation 
in the \citet{Die17} paper inspiring this work.  Illustris incorporates
different physics, has a different number of time steps, and better
(using the measure $D$ in Eq.~\ref{eq:goodfitcrit}) lognormal fits to its
histories than the fits to the histories here.  
Both of these simulations have some disagreement with
observations, as noted earlier, for instance, one comparison has found that L-galaxies quenches too
quickly and Illustris not enough \citep{Blu16}, which was seen, for
example, in comparison of their $\sigma_t(t_{\rm peak})$ relations.
It would be interesting to compare these approximations between other
simulations and models.

Not only can galaxy formation models be changed, other definitions of peak time may
also be more effective at either separating the integrated star
formation rate histories into families, or
matching the instantaneous star formation rates.  For instance, there are
different fits such as the double
power law fit of \citet{BWCz8} studied in \citet{Die17}, which has less
scatter in many cases, but also many singular fits at least when tried
for the galaxy histories studied here, see also \citet{Gla13,Car17,Mar18}
for examples of other
studies of a variety of functional forms.  (\citet{Car17} 
identify several distinct star formation history shapes, depending
upon the particular galaxy.) 
Two other obvious possibilities for special times, even within the lognormal
fit definition, 
are $t_{1/2}$ and the quenching time.
The quenching time also requires the stellar mass, time, and choosing a definition
and width of the
star forming main sequence, so it was not pursued in detail here.

In summary, these two ways of viewing galaxy integrated star formation
rate histories provide examples of how to distill some of the huge variation and complexity
of galaxy formation into
a few characteristics of galaxy histories and their populations.  These
characteristics often have simple meanings and
 can be compared between models, and ideally, eventually to physical mechanisms.  In the examples here, the variations
 between these characteristics revealed the
 different underlying mass distributions in the subsamples.
Comparing the average histories, fluctuations\footnote{Care must be
  taken to rescale the variance around the average when two models
  have different numbers of time steps.}, distribution of
parameters (and their relations to each other, e.g., here for the lognormal fit), separations into subfamilies and other properties across
simulations may allow identifying 
properties charaterizing the full samples, and thus the models which created
them.  These properties may not be evident in the detailed
prescriptions for the individual galaxies, but instead emerge in the
samples, and perhaps in observations, as a whole. 

\section*{Acknowledgements}

I thank P. Behroozi and C. Pacifici for the inspiring discussions to work on
this, to B. Diemer, G. Lemson, Y. Feng for help in various steps of obtaining and analyzing the
data, and many others for conversations, comments and questions, including S. Alam, M. van
Daalen, N. Dalal, R. Dave, and A. Kravtsov.  I also thank other participants at the
Santa Barbara Galaxy-Halo connection workshop in June 2017, 
members of the Royal Observatory of Edinburgh, and participants at the
Berkeley Center for Cosmological physics workshop and Nordita workshop
in July 2017 for the opportunity to talk on this work and for their
feedback and suggestions, and the Royal Observatory and Nordita for
their kind hospitality.  
B. Diemer and C. Pacifici generously provided many helpful
suggestions and criticisms of an earlier draft. 
I am especially grateful to M. White for
innumerable discussions and suggestions, as well as encouragement.
I also thank the referee for many helpful suggestions and questions.
The Millennium Simulation databases used in this paper and the web application providing online access to them were constructed as part of the activities of the German Astrophysical Virtual Observatory (GAVO).

\appendix
\section{The cen $M_{h,{\rm big}}$ sample}
\label{sec:morekam}
The cen $M_{h,{\rm big}}$ sample is based upon
the sample used by  \citet{KamTurBru16a}, who took
central galaxies which had $M_h\leq 10^{12} M_\odot$
at final times, and found that using main halo histories plus some other
histories and parameters as inputs for machine learning gave a $\sim$ 88\%
correlation between predicted and true $M^*$.  For their machine learning,
they use information beyond $M_h(t)$ as input for the
  learning algorithm, including circular velocity $V_{\rm max}(t)$
and velocity dispersion $V_{\rm disp}(t)$.  

But there are also differences with their $M_h(t)$ sample from cen
$M_{h,{\rm big}}$. 
In particular, their outputs are traced back to $z=5.7$, while
  here the starting time is $z\sim 10$ (they had 45 outputs compared
  to the 48 here).   They also used a
different, earlier, semi-analytic model \citep{Guo10} from the
Millennium database.
 For machine learning, it seems the most important difference is that 
their sample was trimmed in two ways.  It was trimmed
  explicitly reject outliers.   It was also trimmed implicitly through the SQL
  query to reject the $\sim$7\%  of the halos meeting the final halo
  mass and central galaxy criteria which were not present at
  all time steps.  That is, they used
``INNER JOIN'' rather than ``FULL OUTER
  JOIN'' used here. 
  This may have
  resulted in a sample which was not only better behaved, but easier
  to model via machine learning.   

\section{Machine learning fits}
\label{sec:moreml}
Machine learning, discussed in \S\ref{sec:ml} in the text, was applied
to all 7 samples, using $M_h(t)$ to predict the parameters $t_{\rm
  peak}$, $\sigma_t$, $A$, $a_0$, $a_1$, $a_2$, $\tilde{\cal S}(t_f)$ and
the final time $M^*$.
Below, the resulting fits to the integrated star formation rates are compared to the simulation outputs, in direct
parallel to the comparison of the direct fits with the simulation
outputs in \S\ref{sec:approx} and \S\ref{sec:finaltime}.

\subsection{Goodness of approximations using ML parameters}
\begin{figure}
\begin{center}
\resizebox{3.3in}{!}{\includegraphics{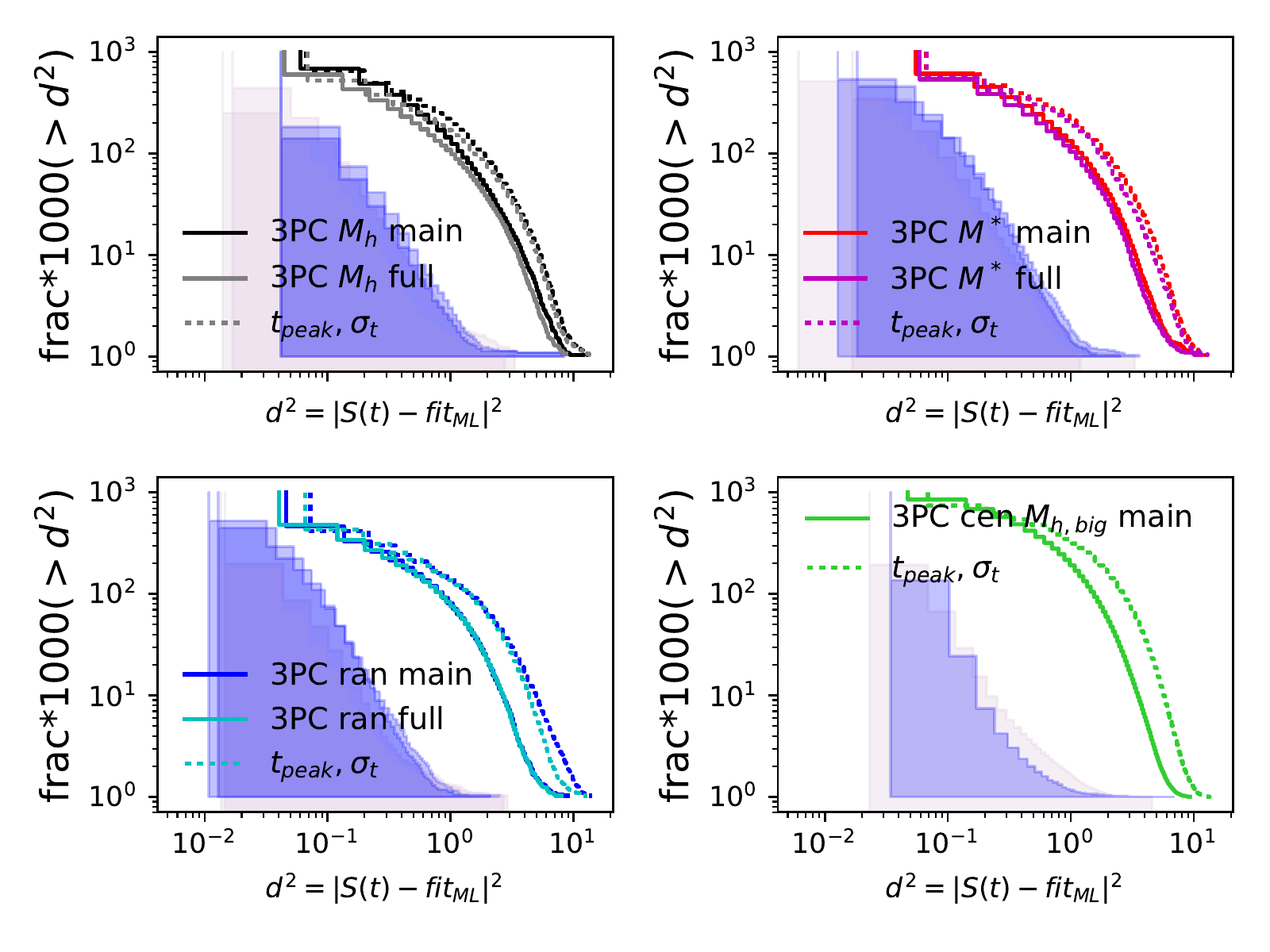}}
\resizebox{3.3in}{!}{\includegraphics{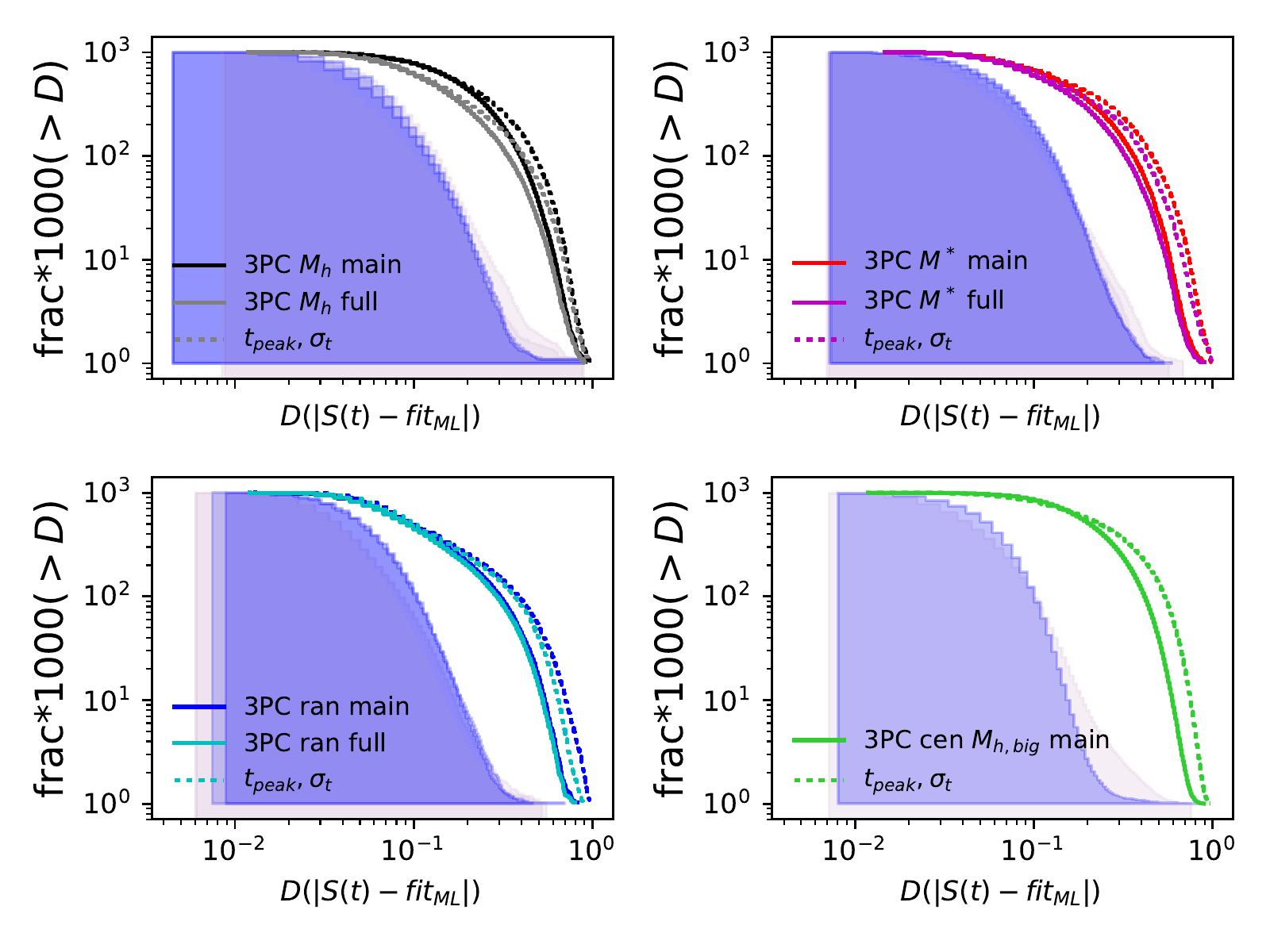}}
\end{center}
\caption{Separation between simulated integrated star formation rate
  histories ${\cal S}(t)$ and their 
approximations using machine learning to find $a_0,a_1,a_2$ (solid lines) or $t_{\rm peak},\sigma_t$ (dashed lines).
The top four panels give $d^2=|{\cal
    S}(t)-fit_{ML}|^2$, while the bottom panels give $D$
  (Eq.~\ref{eq:goodfitcrit}), as in Fig.~\ref{fig:dsep}.
The shaded regions are the corresponding distributions for the direct fits to the simulation histories,
from  Fig~\ref{fig:dsep}.
}
\label{fig:mlcomp}
\end{figure}

\begin{figure}
\begin{center}
\resizebox{3.3in}{!}{\includegraphics{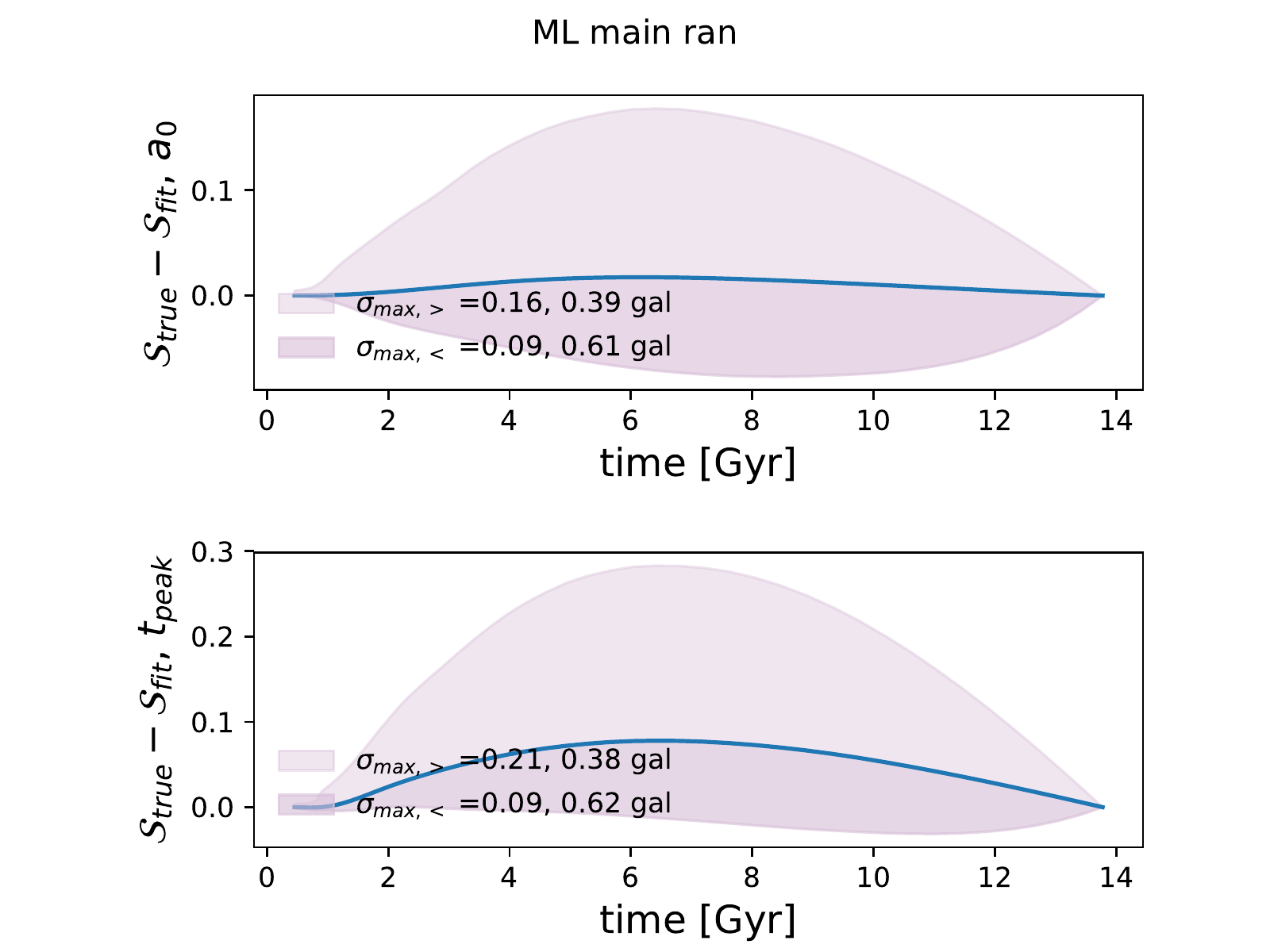}}
\end{center}
\caption{Average difference  $\langle {\cal S}_{\rm true}(t) -{\cal
  S}_{\rm fit}(t)\rangle$, solid line, and
its scatter, shaded, for the machine learning
PCA fit (top) or lognormal
  fit (bottom), in the main {\tt ran} sample. The simulation gives ${\cal
    S}_{\rm true}(t)$.
This can
  be compared to the average and scatter from the direct fits shown in
  Fig.~\ref{fig:sfrsigdist} in the main text.  
Again, the shaded regions are one standard deviation, calculated separately for galaxies with
${\cal S}_{\rm true}(t) -{\cal S}_{\rm fit}(t) $ above or below 
the average.
The scatter is much larger in the machine learning fits, and
has changed shape.  Relative to the direct fits, the standard
deviations have increased from $\leq 0.05$ to 0.15-0.20 (PCA), and from
$\leq 0.07$ to
 0.19-0.24 (lognormal).  Note 
the fit on average tends to overestimate rather than underestimate
${\cal S}_{\rm true}$,
unlike the earlier fit.
}
\label{fig:sfrsigdistml}
\end{figure} 
 Just as for the original
fits in \ref{sec:approx}, the machine learning fits can be compared to the simulated
histories using  
$d^2$ and $D$.  These are shown in
Fig.~\ref{fig:mlcomp}, with the original distributions
from Fig.~\ref{fig:dsep} shown as
 shaded regions for comparison.\footnote{The average histories and principal
 components for the PCA approach are assumed to be fixed for the
 sample to their true value, rather than calculated from the training set alone.}
  The
shading gives an estimate of how much the fits degrade when machine
learning is used.
For the direct fits to the simulations, the squared separation between
the simulated and direct fits, i.e., 
$d_{t_{\rm peak}}^2$ and
$d_{\rm PCA}^2$, were 22\%-
53\% correlated. The correlations of scatter from the true values,
between the two kinds of approximations, increased to 80\%-92\%
for the machine learning 
fits based upon $M_h(t)$.  
That is, large or small separations between the actual history and their
machine learning reconstructions tended to occur together for both
kinds of approximations, perhaps indicating something about which $M_h(t)$
were harder/easier to associate with the correct fit parameters.

A slightly more detailed characterization of the difference between
the approximations from machine learning and the simulated ${\cal
  S}(t)$, similar to Fig.~\ref{fig:sfrsigdist}, is shown in Fig.~\ref{fig:sfrsigdistml}.
This again shows,  for the {\tt ran} sample,  the average deviation between
true and fit ${\cal S}(t)$ at each time step,
and standard deviations from it (again calculated separately for above and
below).  The average deviations when using machine learning fits rather than
direct fits are much larger, the standard deviations go up by a factor
of $\sim 3$ and 
have a different shape. 
Note 
the machine learning fit on average has a larger bias, and a larger
tendency to overestimate rather than underestimate
${\cal S}_{\rm true}$.

 This shape is similar for all samples and for
both the lognormal and PCA fits.  The average deviation of
the fit and the size of the scatter around this average deviation are
also much bigger for the machine learning fits.

\begin{figure}
\begin{center}
\resizebox{3.3in}{!}{\includegraphics{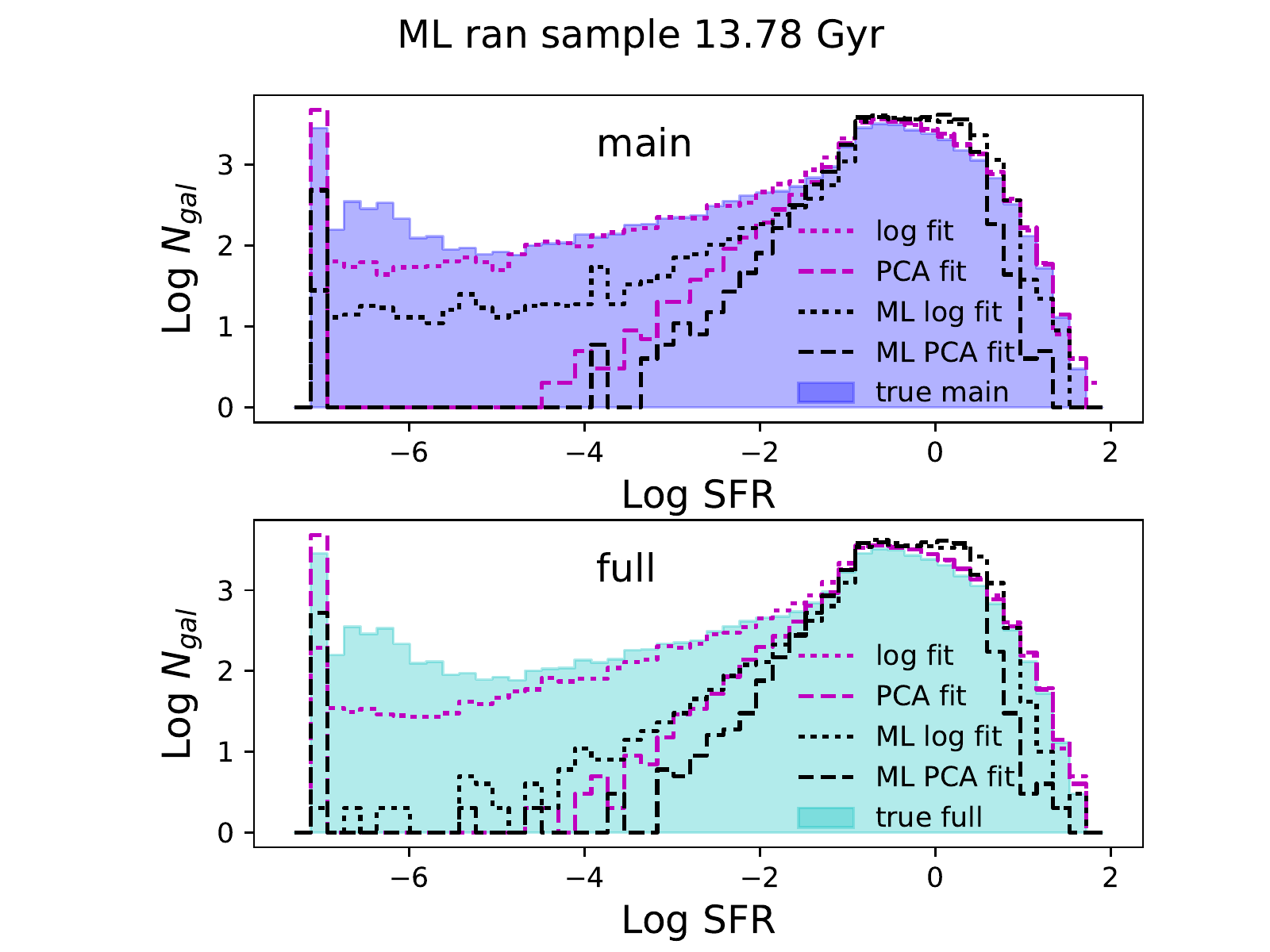}}
\end{center}
\caption{Final time star formation rates in the {\tt ran} simulation (top
  panel uses main integrated star formation rate history fits, bottom
  panel uses full integrated star formation rate history fits).  The filled in region shows the
  simulation (true) final star
formation rate distributions. The star formation rates for the  direct
fits to the histories,
shown earlier in magenta in Fig.~\ref{fig:sfrhist}, are shown
again here.  The machine learning
fits are in black.  Dotted lines correspond to the lognormal fit and dashed
lines to the PCA fit. 
Again, galaxies with negative star formation rates in the PCA fit are
set to $10^{-7} M_\odot yr^{-1}$  along with any other galaxies with star formation
rates $< 10^{-7} M_\odot yr^{-1}$.  The machine learning fits tend to give a
narrower peak at high star formation rate.
}
\label{fig:sfrhistml}
\end{figure}

\begin{figure*}
\begin{center}
\resizebox{6.6in}{!}{\includegraphics{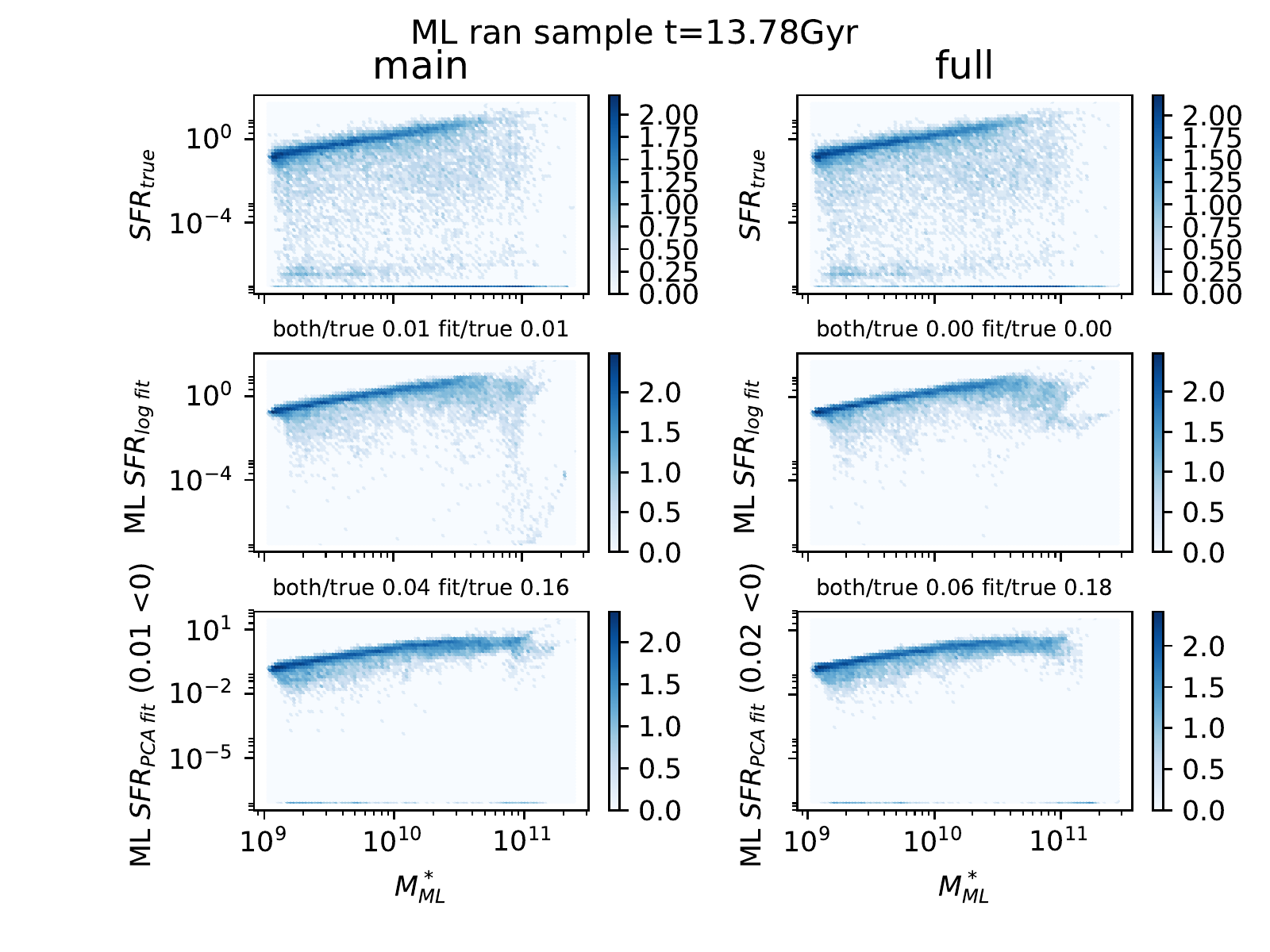}}
\end{center}
\caption{Stellar mass versus star formation rate at
  redshift zero from machine learning for $M^*$ and the lognormal
  fit (middle panels) and the PCA fit (bottom panels), compared
  to the actual distribution in the simulation (top panels).  This is
  the same comparison, for the same {\tt ran} sample, as shown in
  Fig.~\ref{fig:mstarsfr}, however in that case, the fits were direct,
  rather than via machine learning.
At left,
the fit to the main star formation rate history is used, at right, the
fit to the full star formation history is used. 
Again, galaxies with negative star formation rates in the PCA fit are
counted and then
set to $10^{-7} M_\odot yr^{-1}$  along with any other galaxies with star formation
rates $< 10^{-7} M_\odot yr^{-1}$.  Other quantities are defined as in
Fig.~\ref{fig:mstarsfr}.  See text for more discussion.
}
\label{fig:mstarsfrml}
\end{figure*}

\begin{figure} 
\begin{center}
\resizebox{3.3in}{!}{\includegraphics{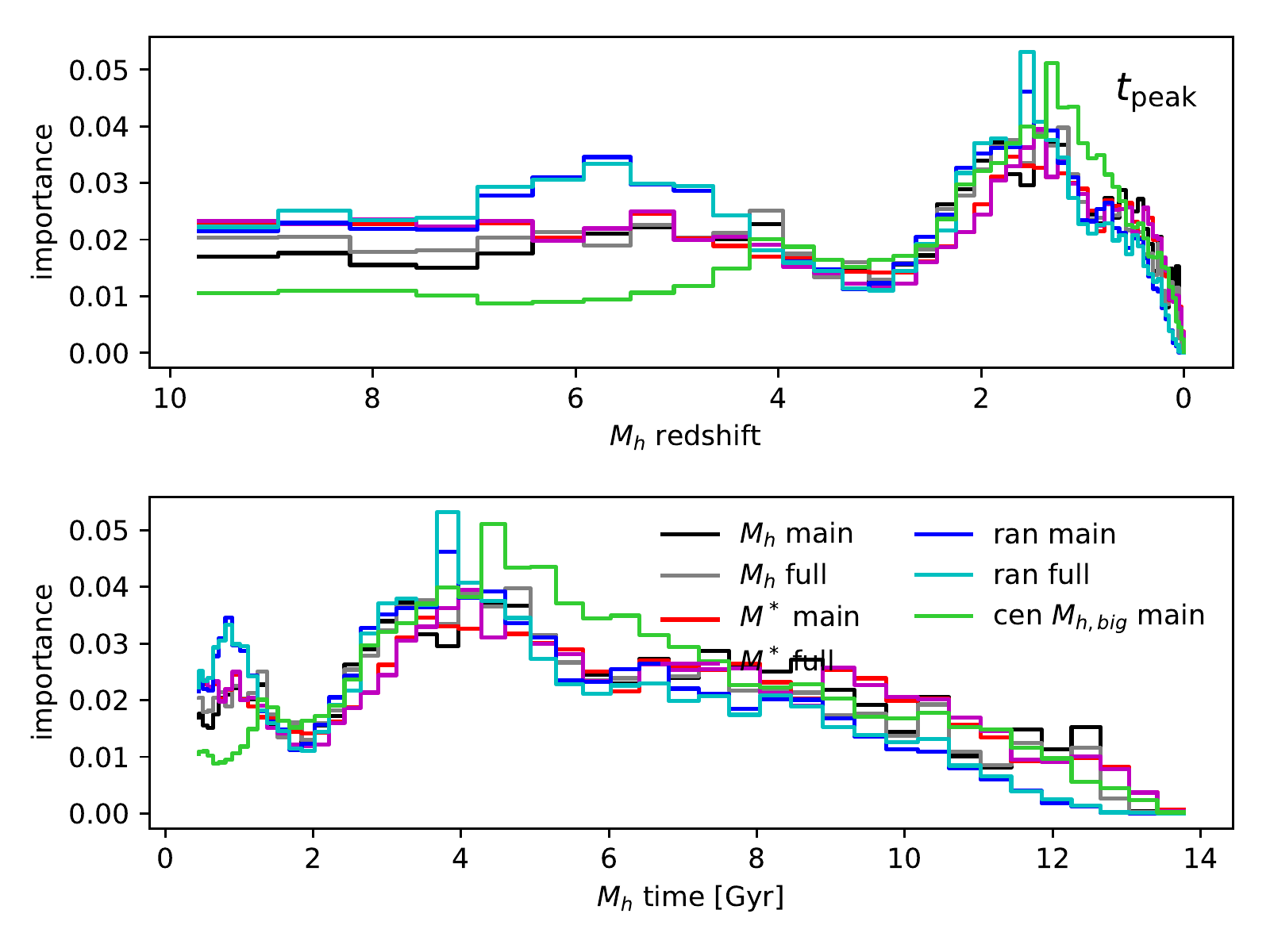}}
\resizebox{3.3in}{!}{\includegraphics{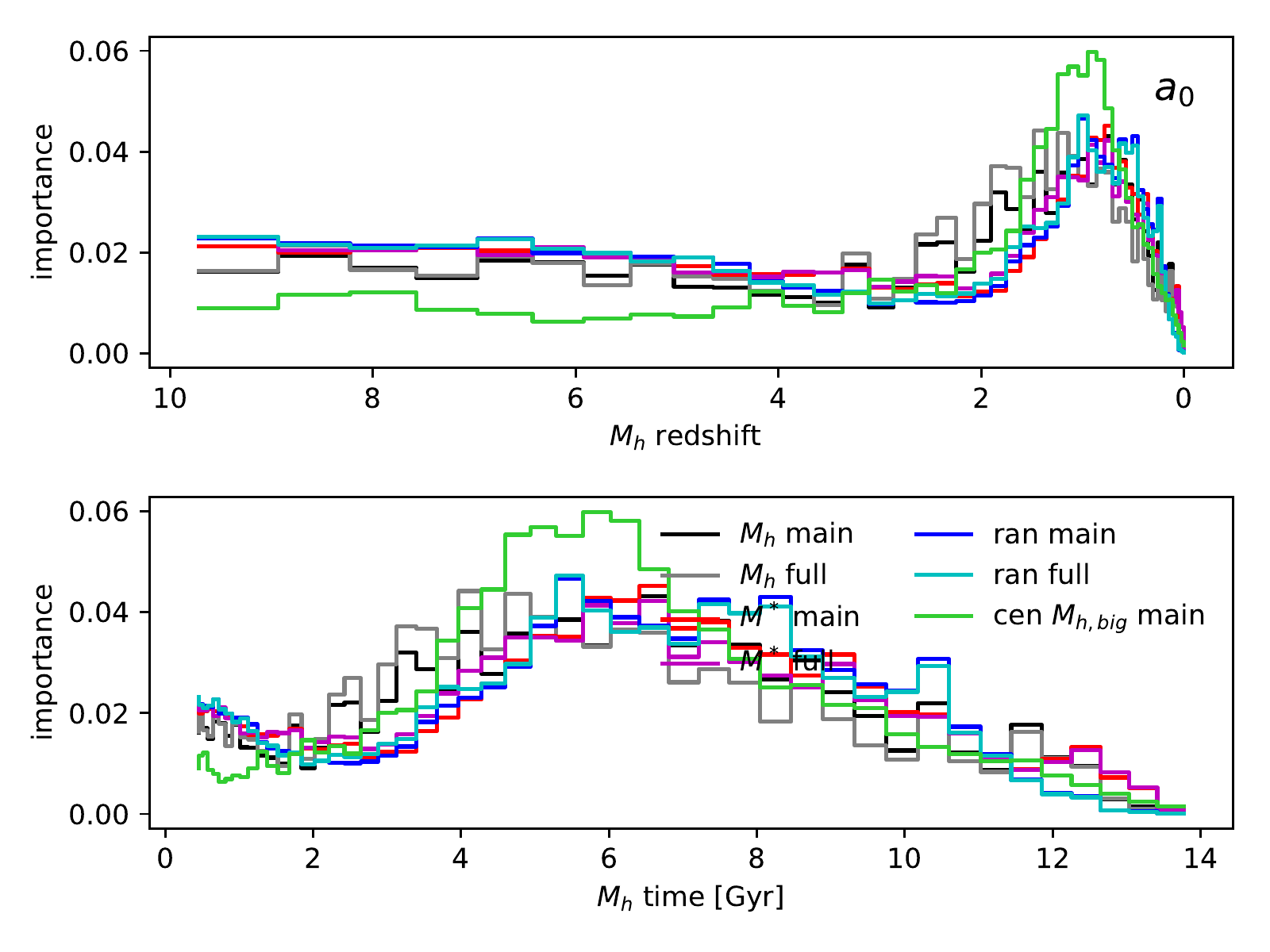}}
\end{center}
\caption{Importance of different
  times in $M_h(t)$, for
  machine learning predictions for $t_{\rm peak}$ and $a_0$ (top,
  redshift space, bottom, time).  Different colored lines are
  different samples.  Values of $M_h(t)$ around 4-5 Gyr seem to
  have the most weight.
}
\label{fig:import} 
\end{figure}

\subsection{Approximating final time star formation rate}
For the final time  $SFR(M^*)$ predictions, Fig.~\ref{fig:predresults}
shows the correlations between true and found for the final $M^*$
machine learning predictions.  Two
other $M^*$ tests not
shown are successful:  the stellar mass to halo
mass relation, considered in \citet{KamTurBru16a} and the stellar mass function (well reproduced except
for losing some galaxies
at the high mass end).   
However, the stellar mass to star formation rate relations are worse
because of the final time star
formation rate discrepancies, shown in Fig.~\ref{fig:sfrhistml}.  This
can be compared to Fig.~\ref{fig:sfrhist} in \S\ref{sec:finaltime};
the black lines give the machine learning predictions, while the
magenta lines show the earlier predictions using the direct fits. 
The machine learning prediction for the
number of galaxies in the green valley decreases for both fits for the
{\tt ran} sample, but behaves differently in other samples.  All samples show some increase
of galaxies with fairly high star formation rates above the number in
the simulation when the rates are found using machine learning fits.

The resulting stellar mass to star formation rates, the machine
learning version of Fig.~\ref{fig:mstarsfr}, are shown in
Fig.~\ref{fig:mstarsfrml}.
Just as in Fig.~\ref{fig:mstarsfr} the simulation
result is at top, and the fits (this time from machine learning) are
below.  The fractions of
negative star formation rates for the PCA fit are again given on the $y$-axis on
the lowest row, after which all galaxies with SFR $\leq 10^{-7}M_\odot
yr^{-1}$ are assigned to the minimal SFR $=10^{-7}M_\odot
yr^{-1}$.  These minimal SFR galaxies are compared in the
simulation and the fits:
the numbers of low star formation rate galaxies
common to the fit and the simulation (``both''), in the
simulation (``true'') and in the fit  (``fit'') are compared in the
ratios  (``both/true''), (``fit/true''), shown in each fit panel.
In the lower panel, again all galaxies with SFR $\leq 10^{-7}M_\odot
yr^{-1}$ are plotted as galaxies with SFR $=10^{-7}M_\odot
yr^{-1}$.  The {\tt ran} sample lognormal fit from machine learning has the smallest number of ``fit/true'' low
star formation rate galaxies, but this number does not go above 0.31
for either fit among any of the samples.

The machine learning fits also provide the ``importance'' of difference components of
the input in producing the final results.  
Examples of this importance for $t_{\rm peak}$ and $a_0$ predictions 
are in Fig.~\ref{fig:import}, in terms of redshift in the top panel, and time on the
bottom panel, and are hard to interpret. Many individual
redshifts between 0 and 2 seem to have more importance than earlier
times, even though many of the galaxies have their peak time well
before redshift 2 (at $\sim 2$ Gyr, Fig.~\ref{fig:import}).  This is
likely related to the fit information available when star formation
declines at these later times.

\section{Separating galaxy histories into many subfamilies}
\label{sec:moresplit}
As discussed in \S\ref{sec:split}, integrated star formation rate
histories can be separated into many different subfamilies, once a
continuous parameter such as $t_{\rm peak}$ is assigned to each
history.  A variety of different splits were tried, with $\Delta
\sigma^2_{\rm tot}$, Eq.~\ref{eq:deltavar}, and the overlaps,
Eq.~\ref{eq:scattersep}, compared for each.  
Subfamilies were divided
according to
quenching time, $t_{\rm peak}$ and $a_0$, 
changing the number of subfamilies and dividing values of the
parameters.
Some splits were uniform, others were based on different regions
visible in figures such as Fig.~\ref{fig:tppc0hex}.
(Uniform splits for $t_{\rm peak}$ grouped all histories with $t_{\rm peak}
\gtrsim 7.5$ Gyr together as they always significantly overlapped.)  
A sampling of how the scatter
changes in different subfamily choices, and 2 examples of galaxy properties in
the different subfamilies are given in this section.  A full exploration of all possible subfamilies is
beyond the scope of this work.
 
\begin{figure} %
\begin{center}
\resizebox{3.5in}{!}{\includegraphics{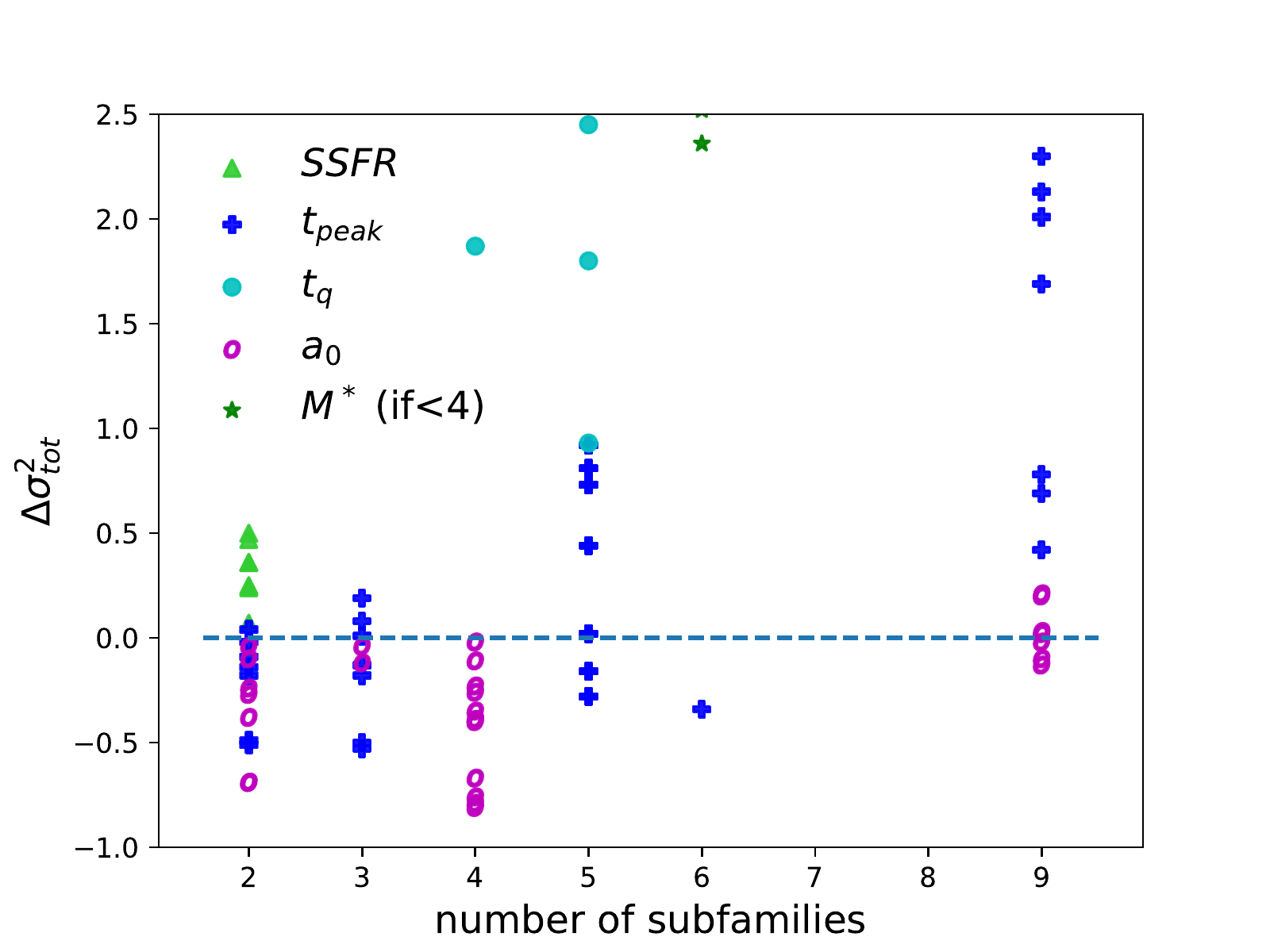}}
\end{center}
\caption{The change in total variance, $\Delta \sigma^2_{\rm tot} = \sum_i
  \sigma^2_{i,{\rm final}} - \sigma^2_{\rm initial}$, for galaxy samples split into
  different numbers of
 subfamilies (numbers noted on the $x$-axis), for different parameter values and
 different samples (i.e., $M_h$, $M^*$, {\tt ran}, cen $M_{h,{\rm big}}$).  The average histories for the splits
 into two subfamilies are in  Fig.~\ref{fig:splitthree}.   Splits
were uniform or guided by features in the individual or joint distributions of
$t_{\rm peak}, a_0, t_q$; the divisions based upon $M^*$ and SSFR
mentioned in the text are also included. 
Subfamilies with $\Delta
\sigma^2_{\rm tot} \geq 4$ are not shown. 
 The more negative $\Delta \sigma^2_{\rm tot}$,
the more the split into subfamilies reduces total scatter.  
For all samples with $\Delta
\sigma^2_{\rm tot}<0$ shown here, 
the scatter between adjacent subfamilies was also less than the
distance between the corresponding average integrated star formation
rate histories, i.e., satisfying Eq.~\ref{eq:scattersep}. }
\label{fig:subscat} 
\end{figure}

For this small set of assorted subfamilies, a chart of $\Delta
\sigma^2_{\rm tot}$ as a function of number of subfamilies is shown in
Fig.~\ref{fig:subscat} (for splits into subfamilies with $\Delta
\sigma^2_{\rm tot}<4$).  The
divisions based upon $M^*$ and SSFR discussed in \S\ref{sec:split} are also included.
In this small
sample, splitting integrated star formation rate histories 
based upon $a_0$ again tended to succeed more often than splitting on the
basis of  $t_{\rm   peak}$ or $t_q$, perhaps because $a_0$ is the
coefficient of the largest fluctuation around the average history.
However, these examples do not preclude better (in terms of $\Delta
\sigma^2_{\rm tot}$ and Eq.~\ref{eq:scattersep}) separations in terms of
$t_{\rm peak}$ or $t_q$.

\begin{figure*} %
\begin{center}
\resizebox{8.4in}{!}{\includegraphics{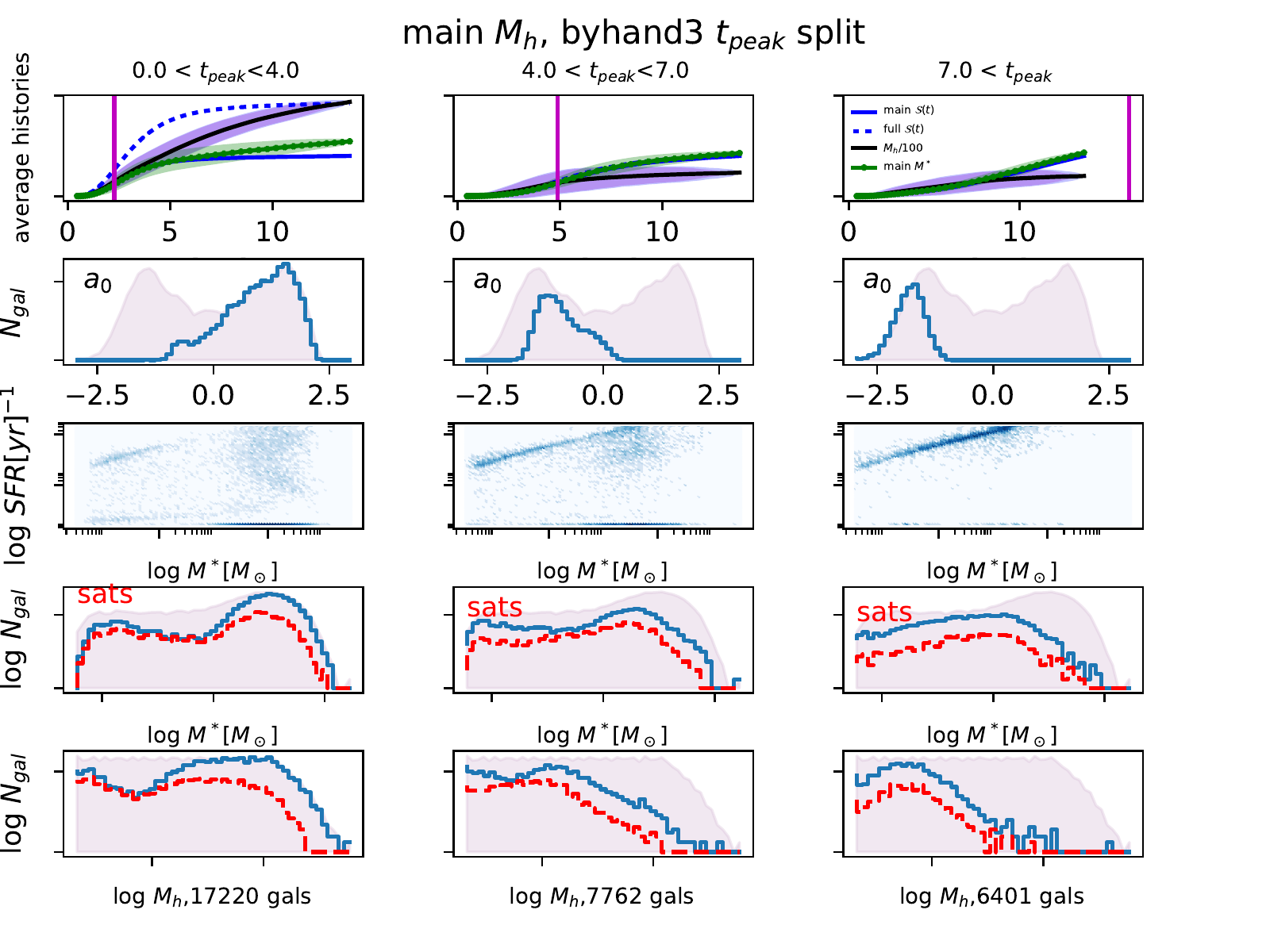}}
\end{center}
\caption{
Properties of the galaxies sorted by the $t_{\rm peak}$ separations 
producing the average integrated and instantaneous star formation rate
histories in the middle row of  Fig.~\ref{fig:samp3}.  Each column
corresponds to one subfamily, with sample galaxies chosen according to 
the $t_{\rm
  peak}$ range listed at top.  The highest row shows the average main integrated
SFR (solid blue), full integrated SFR (dashed blue),
$M_h$ (solid black) and $M^*$ (green stars) histories.  Each type of history is normalized to the
same final value, described in the text. The scatters
are not rescaled with the final values, so sizes and shapes of scatters can be compared
directly across subfamilies, and the vertical axis is shared across all
columns on the top row.  Histories are all as a function of
time, listed in Gyr along the $x$-axis.
The magenta vertical line shows the average peak time for the each subfamily.
In the second row, the solid line shows the  $a_0$
distribution for the subfamily, with light purple shading
showing the full sample distribution.  This breaks down 
the relation between $a_0$ and $t_{\rm peak}$.
The third row shows the logarithm of number of galaxies with a given final
log $M^*$ and SFR.  Galaxies with SFR=0
are set to $SFR = 10^{-7} M_\odot yr^{-1}$.
Solid dark lines in the fourth and fifth rows give the log of the number of galaxies with
a given final $M^*$ and final $M_h$ for each
subfamily, along with the full distribution, shaded in light purple.
The distribution of satellites in the lower two panels is shown as a
dashed red line.  
}
\label{fig:galproptp}
\end{figure*}

\begin{figure*} %
\begin{center}
\resizebox{8.4in}{!}{\includegraphics{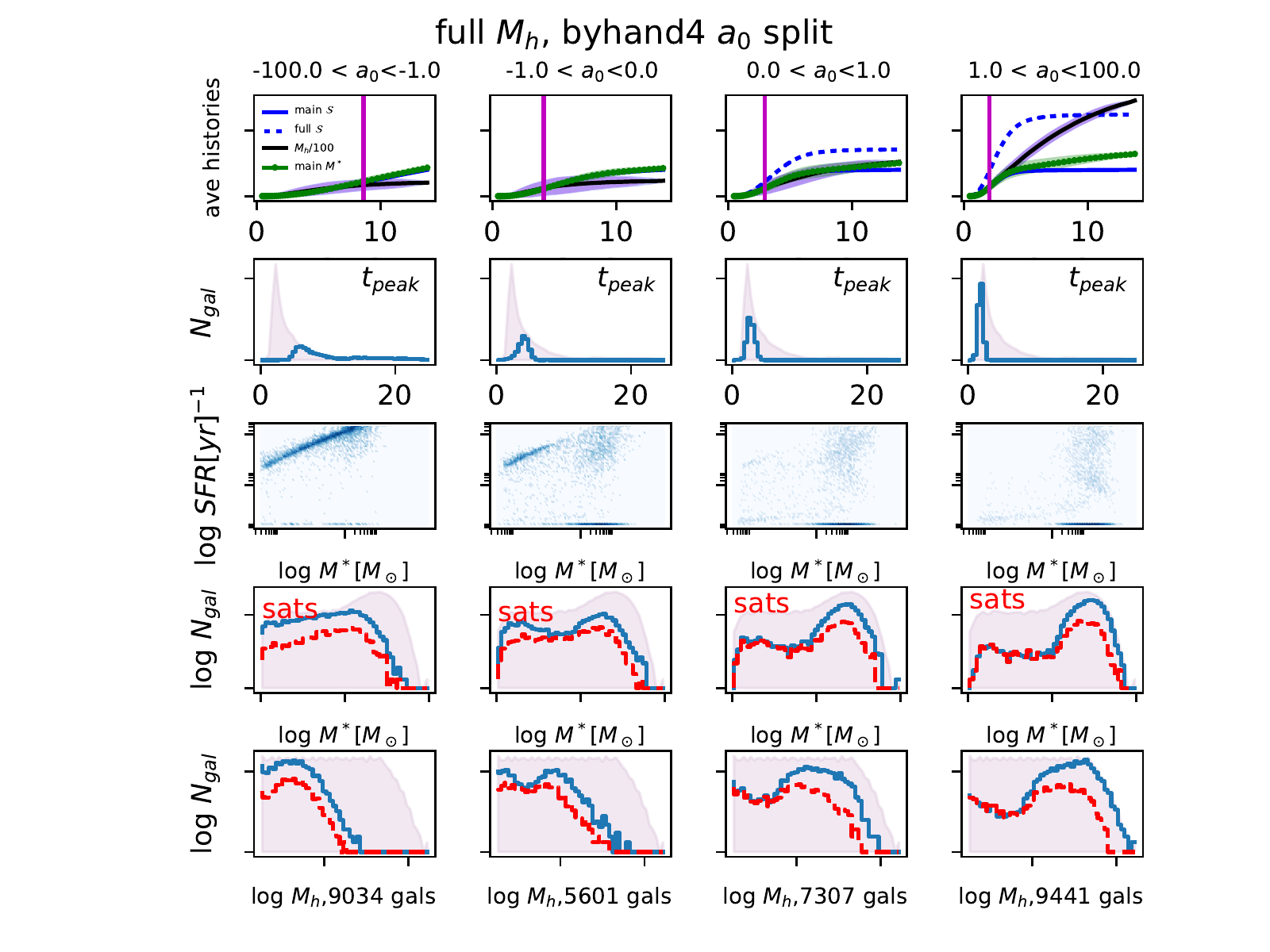}}
\end{center}
\caption{
Final time properties of the galaxies in
subfamilies corresponding to the $a_0$ separation with average
integrated and instantaneous star formation rate histories shown in the bottom row of
Fig.~\ref{fig:samp3}.  Lines and colors are as in
Fig.~\ref{fig:galproptp} above, except for the second row.  In this
case,
as $a_0$ is used to separate the histories into subfamilies,
the $t_{\rm peak}$ distribution of each subfamily is given, with the
full distribution shown as a shaded region.
In the bottom two rows,
the subfamily (2nd column) which is bimodal in $M_h$ and $M^*$
can be further split into two samples by using $a_1$ or $a_2$, both of
which are correlated with high or low final $M^*,M_h$ in this subfamily. 
}
\label{fig:galproppc}
\end{figure*}

The galaxies in the different subfamilies can be compared in terms of their
average history properties (stellar mass, halo mass and main and full
integrated star formation rates), and other properties.  Two examples are shown in
Fig.~\ref{fig:galproptp} and Fig.~\ref{fig:galproppc}.  For these
subfamilies, the
average integrated star formation rate histories,
scatter around these averages, and instantaneous star formation rates
are in the bottom two panels in 
Fig.~\ref{fig:samp3}.  
Fig.~\ref{fig:galproptp} corresponds to properties of the 3
subfamilies separated using
$t_{\rm
  peak}$, for the full $M_h$ sample, and Fig. ~\ref{fig:galproppc} corresponds to the separation of
galaxies into 4 subfamilies using $a_0$, in the main $M_h$ sample.

In each plot,
each column is a different subfamily, and each row focuses the
distribution of the same property or properties.

\begin{itemize}
\item The top row panel shows average histories for $M^*$, $M_h$, and the
full and main integrated star formation rate.  (These are full and
main histories in subfamilies determined by
$a_0$ or $t_{\rm peak}$ from either the main or full integrated star
formation rate histories.)
The main integrated star
formation rate is rescaled to have final value 1.  The other histories
are rescaled to have the same final value, corresponding to the median
values of $\tilde{\cal
  S}_{\rm full}(t_f) /\tilde{\cal
  S}_{\rm main}(t_f) $ for the full integrated star formation
rate histories,
$ 1.868 M^*(t_f)/\tilde{\cal
  S}_{\rm main} (t_f) $ for the stellar mass histories, and
$ M_h(t_f)/(100 \tilde{\cal
  S}_{\rm main} (t_f)) $ for the halo mass histories.  

The prefactor for the
stellar mass history rescaling is to 
take out the stellar ageing; if the stellar mass history and main
integrated star formation rate history roughly coincide, then almost
all the star formation occurs in the main history, i.e. in situ.
This is also seen in the difference between the full (dashed line) and
main (solid line) integrated star formation rate histories; these two lines
overlap if the median full and main histories overlap after this rescaling.  As might be expected,
this overlap
occurs more for low
$a_0$ or high $t_{\rm peak}$, corresponding to the lowest final $M^*$ and
final $M_h$ galaxies.  These galaxies are not expected to gain much
stellar mass through mergers, tending to be small and star forming (or
quenched satellites).
The vertical solid line is the $t_{\rm peak}$ for the average
integrated star formation rate history for each subfamily.
  
The shading around the main halo mass histories and
the main
stellar mass histories are the 3 principal components of scatter
(purple) and the full variance (blue, at edges) around the galaxy
histories.  These scatters are rescaled to histories with a final value of 1, so
that their relative size can be compared directly between different
columns, similarly,
the vertical scales for the top row are identical for all columns.
\item The solid lines in the second rows
show the distribution of $a_0$ (for the sample split
on $t_{\rm peak}$) or $t_{\rm peak}$ (for the sample split on $a_0$),
 to see how well the cut in one quantity corresponds to a cut
in the other.  The shaded region shows the full sample distribution of $a_0$
or $t_{\rm peak}$ respectively.  
\item The next three
rows are final time properties, showing the stellar mass to halo mass,
final stellar mass and final halo mass of galaxies in each
subfamily.  The shaded regions show the full sample distribution of these
quantities.  For final $M^*$ and final $M_h$, the distribution for
satellites is also shown separately. \footnote{
There is some bimodality in final halo masses and stellar masses
for samples with $t_{\rm peak} \sim 5$ Gyr and $-1< a_0< 0$, correlated
with the value of $a_1,a_2$.  This is somewhat perhaps hinted at in
Fig.~\ref{fig:pcaeverything} in the main text.}
\end{itemize}
These figures compare different properties of the subfamilies which
have different integrated star formation rate histories.  Similar
final properties are often spread across several subfamilies.  For
instance, galaxies with moderately high final $M^*$ can lie in 3 of
the 4 subfamilies split using $a_0$ in Fig.~\ref{fig:galproppc}, with different contributions from,
for instance, mergers.  These comparisons may give another angle on
trying to disentangle contributing factors to the formation of
galaxies.  Similarly, galaxies with early peaks in star formation rate can often span a wide
range of final stellar masses.  Again, the hope would be that these
comparisons would help identify cases where some features are shared
and others differ, which could motivate searches for physical causes of
either the differences or the similarities.  These simpler questions
may give useful angles for approaching the wide diversity of
properties and galaxy histories
produced by galaxy formation models.


\begin{thebibliography}{99.}
\bibitem[{{Abramson et al}(2016)}]{Abr16}
Abramson, L.E., Gladders, M.D., Dressler, A., Oemler, A.,Jr,
Poggianti, B., Vulcani, B., 2016, ApJ, 832, 7
\bibitem[{{Agarwal, Dave \& Bassett}(2017)}]{AgaDavBas17}
Agarwal, S., Dave, R., Bassett, B.A., 2017, arXiv:1712.03255
\bibitem[{{Angulo \& White}(2010)}]{AngWhi10}
Angulo, R.E., White, S.D.M.,  2010, MNRAS, 405, 143
\bibitem[{{Angulo \& Hilbert}(2015)}]{AngHil15}
Angulo, R.E., Hilbert, S., 2015, MNRAS, 448, 364

\bibitem[{{Behroozi et al}(2013a)}]{Beh13a}
Behroozi, P.S., Marchesini, D., Wechsler, R.H., Muzzin, A.,
      Papovich, C., Stefanon, M., 2013a, ApJL, 777, 10

\bibitem[{{Behroozi, Wechsler \& Conroy}(2013b)}]{BWCz8}
Behroozi, P.S., Wechsler, R.H., Conroy, C., 2013a, ApJ, 770, 57

\bibitem[{{Bluck et al}(2016)}]{Blu16}
Bluck, A.F.L., et al., 2016, MNRAS, 462, 2559
\bibitem[{{Breiman et al}(1984)}]{Bre84}
Breiman L., Friedman J., Stone C. J., Olshen R. A., 1984, Clas-
sification and regression trees. CRC press
\bibitem[{{Carnall et al}(2017)}]{Car17}
Carnall, A.C., McLure, R.J.,  Dunlop, J.S.,  Dave, R., 2017, arxiv:1712.04452
\bibitem[{{Cohn \& Van de Voort}(2015)}]{CohvdV15}
Cohn, J.D., van de Voort, F., 2015, MNRAS, 446, 3253

\bibitem[{{Dekel et al}(2013)}]{Dek13}
Dekel, A. Zolotov, A., Tweed, D., Cacciato, M., Ceverino, D., Primack,
J.R., 2013, MNRAS, 435, 999

\bibitem[{{Diemer et al}(2017)}]{Die17}
Diemer, B., Sparre, M., Abramson, L.E., Torrey, P., 2017, ApJ, 839, 26

\bibitem[{{Eales et al}(2018)}]{Eal18}
Eales, S., et al, 2018, MNRAS, 473, 3507

\bibitem[{{Fu et al}(2013)}]{Fu13}
Fu, J., et al, 2013, MNRAS, 434, 1531

\bibitem[{{Geurts, Ernst \& Wehenkel}(2006)}]{GeuErnWeh06}
Geurts P., Ernst D., Wehenkel L., 2006, Machine learning, 63, 3
\bibitem[{{Gladders et al}(2013)}]{Gla13}
Gladders, M.D., Oemler, A., Dressler, A., Poggianti, B., Vulcani, B.,
Abramson, L., 2013, ApJ, 770, 64
\bibitem[{{Guo et al}(2010)}]{Guo10}
Guo, Q., White, S., Boylan-Kolchin, M., De Lucia, G., Kauffmann, G.,
Lemson, G., Li, C., Springel, V., Weinmann, S., 2011, MNRAS, 413, 101
\bibitem[{{Guo et al}(2013)}]{Guo13}
Guo, Q., White, S., Angulo, R. E., Henriques, B., Lemson, G.,
Boylan-Kolchin, M., Thomas, P., Short, C., 2013, MNRAS, 428, 1351
\bibitem[{{Hearin \& Watson}(2013)}]{HeaWat13}
Hearin, A.P., Watson, D.F., 2013, MNRAS, 435, 1313
\bibitem[{{Henriques et al}(2015)}]{Hen15}
Henriques, B. M. B., White, S. D. M., Thomas, P. A., Angulo, R., Guo,
Q., Lemson, G., Springel, V., Overzier, R.,2015,
MNRAS, 451, 2663
\bibitem[{{Hopkins, Quataert \& Murray}(2011)}]{HopQuaMur11}
Hopkins, P.F., Quataert, E., Murray, N., 2011, MNRAS, 417, 950 %

\bibitem[{{Kamdar, Turk \& Brunner}(2016a)}]{KamTurBru16a}
Kamdar, H.M., Turk, M.J., Brunner, R.J., 2016a, MNRAS, 455, 642

\bibitem[{{Kamdar, Turk \& Brunner}(2016b)}]{KamTurBru16b}
Kamdar, H.M., Turk, M.J., Brunner, R.J., 2016b, MNRAS, 457, 1162
\bibitem[{{Kelson}(2014)}]{Kel14}
Kelson, D.D., 2014, arxiv:1406.5191

\bibitem[{{Kelson, Benson and Abramson}(2016)}]{Kel16}
Kelson, D.D., Benson, A.J., Abramson, L.E., 2016, arxiv:1610.06566


\bibitem[{{Lemson et al}(2006)}]{Lem06}
Lemson, G., and the Virgo consortium, 2006, arXiv:astro-ph/0608019

\bibitem[{{Martinez-Garcia et al}(2018)}]{Mar18}
Martinez-Garcia, E.E., Bruzual, G., Magris, G.,
Gonzalez-Lopezlira,R.A., 2018, MNRAS, 1862

\bibitem[{{McBride, Fakhouri \& Ma}(2009)}]{McBFakMa09}
McBride, J., Fakhouri, O., Ma, C.-P., 2009, MNRAS, 298, 1858


\bibitem[{{Nadler et al}(2017)}]{Nad17}
Nadler, E.O, Mao, Y.-Y., Wechsler, R.H., Garrison-Kimmel, S., Wetzel,
A., 2017, arxiv:1712.04467


\bibitem[{{Ntampka et al}(2015)}]{Nta15}
Ntampaka M., Trac H., Sutherland D. J., Battaglia N., Poczos
B., Schneider J., 2015, ApJ, 803, 50
\bibitem[{{Pacifici et al}(2013)}]{Pac13}
Pacifici, C., Kassin, S.A., Weiner, B.,Charlot, S., Gardner, J.P.,
2013, ApJL, 762, 15
\bibitem[{{Pacifici et al}(2016)}]{Pac16}
Pacifici, C., Kassin, S.A., Weiner, B.J., et al, 2016, ApJ, 832, 79

\bibitem[{{Rodriguez-Puebla et al}(2017)}]{RodPriAviFab17}
Rodriguez-Puebla, A., Primack, J.R., Avila-Reese, V., Faber, S., 2017,
MNRAS, 470, 651
\bibitem[{{Schaye et al}(2010)}]{Sch09}
Schaye, J., et al, 2010, MNRAS, 402, 1536

\bibitem[{{Shamshiri et al}(2015)}]{Sha15}
Shamshiri, S., Thomas, P. A., Henriques, B. M., Tojeiro, R., Lemson,
G., Oliver, S. J., Wilkins, S.,2015, MNRAS, 451, 2681

\bibitem[{{Sparre et al}(2015)}]{Spa15}
Sparre, M., et al, 2015, MNRAS, 447, 3548

\bibitem[{{Speagle et al}(2014)}]{Spe14}
Speagle, J.S., Steinhardt, C.L., Capak, P.L., Silverman, J.D., 2014,  ApJS, 214, 15

\bibitem[{{Springel et al}(2005)}]{Spr05}
Springel, V., et al, 2005, Nature, 435, 629

\bibitem[{{Tasitsiomi et al}(2004)}]{Tas04}
Tasitsiomi, A., Kravtsov, A.V., Gottlober, S., Klypin, A.A., 2004,
ApJ, 621, 673.

\bibitem[{{Wechsler et al}(2002)}]{Wec02}
Wechsler, R.H., Bullock, J.S., Primack, J.R., Kravtsov, A.V., Allgood,
B., 2006, ApJ, 652, 71

\bibitem[{{Wong and Taylor}(2012)}]{WonTay12}
Wong, A.W.C., Taylor, J.E., 2012, ApJ, 757, 102
\bibitem[{{Xu et al}(2013)}]{Xu13}
Xu X., Ho S., Trac H., Schneider J., Poczos B., Ntampaka M.,
2013, ApJ, 772, 147


\bibitem[{{Zhao et al}(2003)}]{Zha03}
D.H. Zhao, Y.P. Jing, H.J. Mo, G. Boerner, 2003, ApJL, 597,  9
\bibitem[{{Zhao et al}(2009)}]{Zha09}
D.H. Zhao, Y.P. Jing, H.J. Mo, G. Boerner, 2009, ApJ, 707,  354
\end{thebibliography}
\end{document}